\RequirePackage{etex}

\documentclass[AMA,STIX1COL]{WileyNJD-v2-arxiv}

\articletype{ORCID: \quad \href{https://orcid.org/0000-0001-6444-0361}{A. Borsotti} \quad \href{https://orcid.org/0000-0001-5294-6840}{L. Breveglieri} \quad \href{https://orcid.org/0000-0001-5061-7402}{S. Crespi Reghizzi} \quad \href{https://orcid.org/0000-0002-6469-2929}{A. Morzenti}}

\usepackage{etex}               
\usepackage{amscd}              
\usepackage{amsmath}            
\usepackage{stmaryrd}           
\usepackage{mathrsfs}           
\usepackage{times}              
\usepackage{latexsym}           
\usepackage{hhline}             
\usepackage{multicol}           
\usepackage{multirow}           
\usepackage{rotating}           
\usepackage[latin1]{inputenc}   
\usepackage{calc}               
\usepackage{graphicx}           
\usepackage{ifthen}             
\usepackage{colortbl}           
\usepackage{array}              
\usepackage{booktabs}           
\usepackage{enumitem}           
\usepackage{leftidx}            
\usepackage{longtable}          
\usepackage{mathtools}          
\usepackage{framed}             
\usepackage{setspace}           
\usepackage{srcltx}             
\usepackage{epstopdf}           
\usepackage{changebar}          
\usepackage{fancybox}           
\usepackage{nicematrix}         
\usepackage{hyperref}           

\usepackage[algo2e, plain, vlined]{algorithm2e}         

\usepackage{pst-all}            
\usepackage{pst-poly}           
\usepackage{pst-node}			
\usepackage{pstricks}			
\usepackage{multido}            
\usepackage[off]{auto-pst-pdf}       

\usepackage{tikz}
\usetikzlibrary{shapes}
\usetikzlibrary{arrows}
\usetikzlibrary{trees}
\usetikzlibrary{matrix}
\usetikzlibrary{automata}
\usetikzlibrary{plotmarks}
\usetikzlibrary{external}

\usepackage{pgfplots}
\usepackage{pgfplotstable}
\pgfplotsset{compat=newest}

\usepackage{caption}
\usepackage{subcaption}
\usepackage{extarrows}

\def\REc{\url{https://re2c.org}}

\def\bible{\url{https://www.gliscritti.it/dchiesa/bibbia_cei08/indice.htm}}

\def\fasta{\url{https://open.oregonstate.education/computationalbiology/chapter/patterns-regular-expressions}}

\def\traffic{\url{https://zenodo.org/record/5789064\#.ZCHHQ9LP0eM}}

\def\AMDEPYC{\url{https://www.amd.com/en/products/embedded/epyc/epyc-7001-series.html}}

\def\RePar{\url{https://github.com/FLC-project/RE-Parser}}

\def\REgen{\url{https://github.com/FLC-project/REgen}}

\def\parRErectool{\url{https://zenodo.org/records/14219357}}

\newcommand{\set}[1]{\left\{ #1 \right\}}

\newcommand{\asgn}{\vcentcolon =}

\begin{document}

\title{A parallel parser for regular expressions}

\author[1]{Angelo Borsotti}

\author[1]{Luca Breveglieri}

\author[1,2]{Stefano Crespi Reghizzi}

\author[1]{Angelo Morzenti}

\authormark{A parallel parser for regular expressions \hspace{6.25cm} A. Borsotti \quad L. Breveglieri \quad S. Crespi Reghizzi \quad A. Morzenti}

\address[1]{\orgdiv{\href{https://www.deib.polimi.it/eng/home-page}{DEIB}}, \orgname{Politecnico di Milano}, \orgaddress{\country{Italy}}}

\address[2]{\orgdiv{\href{https://www.cnr.it/en/institute/029}{IEIIT}}, \orgname{CNR}, \orgaddress{\country{Italy}}}

\corres{\href{https://orcid.org/0000-0001-5294-6840}{Luca Breveglieri} \email{\href{mailto:luca.breveglieri@polimi.it}{luca.breveglieri@polimi.it}}}

\presentaddress{\href{https://www.polimi.it/en}{Politecnico di Milano}, Piazza Leonardo Da Vinci $32$, $20133$ Milano, Italy}

\abstract[SUMMARY]{
Regular expression (RE)	matching is a very common functionality that scans a text to find   occurrences of  patterns specified by an RE; it includes the simpler function of RE recognition.  Here we address  RE parsing, which subsumes matching by providing not just the pattern positions in the text, but also the syntactic structure  of each pattern occurrence, in the form of a tree representing how the RE operators  produced the patterns.  RE parsing increases the selectivity of matching, yet avoiding the complications of context-free grammar parsers. Our parser manages ambiguous REs and texts by returning the set of all  syntax trees, compressed into a Shared-Packed-Parse-Forest data-structure.
We initially convert the RE into a serial parser, which simulates a finite automaton (FA) so that the states the automaton passes through encode the syntax tree of the input.
On long texts, serial matching and parsing may be too slow for time-constrained applications. Therefore, we present a novel efficient parallel parser for multi-processor computing platforms; its  speed-up over the serial algorithm  scales well with the text length. We innovatively apply to RE parsing the  approach  typical of parallel RE matchers / recognizers, where the text is split  into   chunks to be parsed in parallel and then joined together. Such an approach suffers from the so-called speculation overhead, due to the lack of knowledge by a chunk processor about the state reached at the end of the preceding chunk;  this forces each chunk processor to speculatively start in all its states. We introduce a novel technique that minimizes the speculation overhead.
The multi-threaded parser program, written in Java, has been validated and its performance has been measured on a commodity multi-core computer, using public and synthetic RE benchmarks. The speed-up over serial parsing, parsing times, and parser construction times are reported.}


\maketitle

\section{Introduction} \label{sec:introduction}
This paper is about text processing using Regular Expressions (RE). The wide world of REs is populated with many algorithms and is supported by many libraries. In increasing order of complexity,   the simplest algorithm, the RE \emph{recognizer}, tells if a text matches an RE. The RE submatch addressing algorithm, for short \emph{matcher},  tells if a string matching an RE occurs in a larger text, and returns either all the  occurrences, or (\emph{disambiguating matcher}) just the one that meets a disambiguation policy, e.g.,
Posix standard or greedy. At the highest level, the RE \emph{parser}  returns not just a yes\,/\,no answer or a set of  matching positions,
but a set of matches with their internal structure. RE parsers subsume recognizers and matchers, but are currently less common  than the others; their
greater precision for text searching tasks is discussed below.
\par
Recognizer and matcher algorithms are classically implemented as  a finite-state automaton (FA) derived from the RE. If the FA is deterministic (DFA), the time complexity of recognizers and matchers is linear. Besides the widely available serial algorithms, also parallel algorithms have been proposed  to reduce the execution time by exploiting parallel hardware, in particular the multi-core computers of interest for us. This paper presents a \emph{parallel} parser algorithm that accepts \emph{any} RE, returns \emph{all} the \emph{parses} and operates in \emph{linear time}.  It is a new accomplishment as, to our knowledge,  none is available
that supports all these features.
\par
To  situate our project with respect to the existing ones, we first  establish  terminology. The set of strings defined by an RE\footnote{We denote by RE the mathematical formalism  that defines regular languages, i.e., finite-state, while the term RegExp refers to the notation from the POSIX standard that is typically assumed by libraries such as RE2c, see \REc.}  is a formal language of the \emph{regular} family. Given an input string over a terminal alphabet $\Sigma$, a \emph{recognizer}  is a decision function that  accepts the strings valid (legal) according to the RE and rejects the invalid (illegal) ones. The recognizer is typically organized as a finite-state automaton (FA), deterministic (DFA) or nondeterministic (NFA), constructed starting from the RE by means of one of the classical methods: Gluskhov / McNaughton-Yamada (GMY)~\cite{VMGlushkov_1961, DBLP:journals/tc/McNaughtonY60}, Thompson~\cite{DBLP:journals/cacm/Thompson68}, and derived ones. Given an RE $e$, a \emph{matcher} is essentially  a recognizer of language $\Sigma^\ast \cdot e \cdot \Sigma^\ast$, but that also returns the position(s) where a valid string for $e$ occurs.
\par
\begin{figure}[ht]
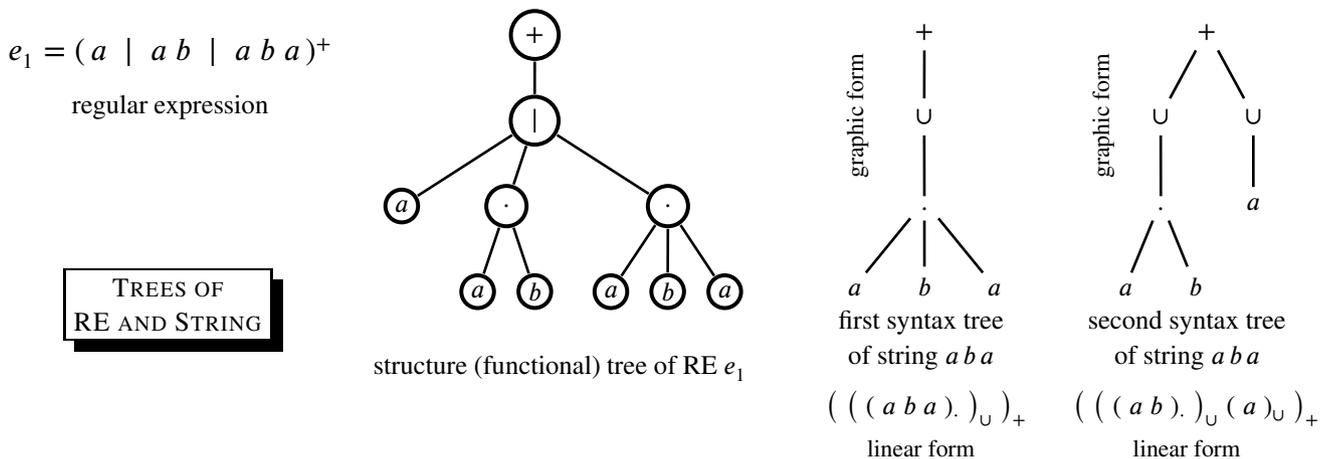

 	\begin{center}
 		\rnode{structure}{
 			\tabcolsep=0.0cm\def\arraystretch{1.25}
 			\begin{tabular}{c@{\hspace{0.0cm}}c@{\hspace{0.5cm}}c@{\hspace{0.75cm}}c}
 				\tabcolsep=0.0cm\def\arraystretch{1.5}
 				\begin{tabular}{c}
 					\large $e_1  =  ( \, a \; \mid \; a \; b \; \mid \; a \; b \; a \, )^+$
 					\\ regular expression
 					\\ \vspace{1.5cm}
 				\end{tabular}
 				
 				&
 				
 				\tabcolsep=0.0cm\def\arraystretch{0}
 				\psset{arrows=-, linestyle=solid, linewidth=1pt, levelsep=1.125cm, treesep=0.75cm, nodesep=0pt, border=0.0cm, treefit=loose, nodealign=true}
 				\begin{tabular}{c}
 					\pstree[treesep=0.5cm] {\TCircle[linewidth=1.5pt, radius=0.35] {$+$}} {
 						\pstree {\TCircle[linewidth=1.5pt, radius=0.35] {$\mid$}} {
 							\TCircle[linewidth=1.5pt, radius=0.25] {$a$}
 							\pstree[treesep=0.25cm]{\TCircle[linewidth=1.5pt, radius=0.3] {$.$}} {
 								\TCircle[linewidth=1.5pt, radius=0.25] {$a$}
 								\TCircle[linewidth=1.5pt, radius=0.25] {$b$}
 							}
 							\pstree[treesep=0.25cm] {\TCircle[linewidth=1.5pt, radius=0.3] {$.$}} {
 								\TCircle[linewidth=1.5pt, radius=0.25] {$a$}
 								\TCircle[linewidth=1.5pt, radius=0.25] {$b$}
 								\TCircle[linewidth=1.5pt, radius=0.25] {$a$}
 							}
 						}
 					}
 				\end{tabular}
 				
 				&
 				
                \tabcolsep=0.0cm\def\arraystretch{0}
 				\psset{arrows=-, linestyle=solid, linewidth=1pt, levelsep=1.125cm, treesep=1.0cm, nodesep=0pt, border=0.0cm, treefit=loose, nodealign=true}
 				\begin{tabular}{c}
 					\psset{treesep=1.0cm, nodesep=4pt, treefit=tight, nodealign=true}
 					\pstree[name=tree1]{\TR{$+$}} {
 						\pstree{\TR{$\cup$}} {
 							\pstree[treesep=0.75cm]{\TR{$.$}} {
 								\TR{$a$}
 								\TR{$b$}
 								\TR{$a$}
 							}
 						}
 					}
 				\end{tabular}
 				
 				&
 				
 				\tabcolsep=0.0cm\def\arraystretch{0}
 				\psset{arrows=-, linestyle=solid, linewidth=1pt, levelsep=1.125cm, treesep=1.0cm, nodesep=0pt, border=0.0cm, treefit=loose, nodealign=true}
 				\begin{tabular}{c}
 					\psset{treesep=1.0cm, nodesep=4pt, treefit=tight, nodealign=true}
 					\pstree[name=tree2]{\TR{$+$}} {
 						\pstree{\TR{$\cup$}} {
 							\pstree[treesep=0.75cm]{\TR{$.$}} {
 								\TR{$a$}
 								\TR{$b$}
 							}
 						}
 						\pstree{\TR{$\cup$}} {
 							\TR{$a$}
 						}
 					}
 				\end{tabular}
 				
 				\\[0.5cm]
 				
 				& \parbox{6.0cm}{\centering structure (functional) tree of RE $e_1$} & \parbox{2.5cm}{\centering first syntax tree \par of string $a\,b\,a$ \par \vspace{0.2cm} $\big( \; \big( \; ( \; a \; b \; a \; )_\cdot \; \big)_\cup \; \big)_+$} & \parbox{3.0cm}{\centering second syntax tree \par of string $a\,b\,a$ \par \vspace{0.2cm} $\big( \; \big( \; ( \; a \; b \; )_\cdot \; \big)_\cup \; ( \; a \; )_\cup \; \big)_+$} \\ & & \scriptsize linear form & \scriptsize linear form
 		\end{tabular}}
 		\nput[labelsep=-3.875cm] {-170} {structure} {\shadowbox{\parbox{2.5cm}{\centering \textsc{Trees of} \par \textsc{RE and String}}}}
        \nput[labelsep=0.625cm, rot=90, origin=c] {-180} {tree1} {\scriptsize graphic form}
        \nput[labelsep=0.5cm, rot=90, origin=c] {-180} {tree2} {\scriptsize graphic form}
 	\end{center}
\caption{The RE $e_1$ with its functional (structure) tree and the two syntax trees of the valid string $a\,b\,a$ that are produced by a parser. The two syntax trees are represented as graphs and linearly as parenthesized strings.} \label{fig:structuretree}
\end{figure}

A \emph{parser}  additionally assigns  syntactic structure(s) to any   valid string. The notion of syntactic structure is central for context-free (CF) languages, i.e., BNF grammars.  All compilers include a  parser that returns the syntax  tree(s) of the source text, e.g., see in~\cite{AhoLamSethiUl2006, DBLP:series/txcs/ReghizziBM19}.
It is well-known that any regular language can be also specified by a simple kind of CF grammar and be parsed by a CF parser that simulates  a pushdown automaton -- a type of abstract machine more powerful than an FA -- such that the pushdown store is limited in size. But a vast number of applications prefer to  avoid the complexity of the grammar and pushdown approach, and turn to REs, which are quite suitable for many pattern matching tasks; e.g., to  find patterns in  JSON data files,  RE matchers are often used instead of CF parsers.  In such cases, we propose to use an RE parser, therefore the parse structure(s)  assigned to the input are not specified by a grammar, but by the functional structure of the RE. We clarify this important concept by means of an example in Fig.~\ref{fig:structuretree}. We  show an RE with its functional \emph{structure tree}, and a valid input string that has two distinct \emph{syntax trees}, i.e., it is ambiguously matched in two ways.
\par\noindent
As said, RE parsers may operate differently  with respect to ambiguity:  a \emph{disambiguating} parser returns only  one  tree,  chosen according to specific criteria, while more complete parsers, including ours, return all the syntax trees, i.e., a forest. The  forest is packed in a  data-structure that we call a \emph{shared linearized parse forest} (SLPF). Clearly, disambiguating parsers  make a selection that may be appropriate to or independent of the pattern matching applications.
To contrast the parser and recognizer for the same RE, we  anticipate that their respective FAs are formally equivalent in the classical sense since they accept the same language, but the former needs more states  to  preserve information on the RE functional structure.
\paragraph{Parsing is more precise than matching}
We explain by means of a typical case  how a parser may answer user queries more selectively and comfortably than a matcher does.
Imagine our text is a file \verb!examplemail.txt! of e-mail messages in, say, the MIME format. Disregarding the details and simplifying, let us assume that each message orderly contains the following fields:
\[
\overbracket[0.75pt]{\, \texttt{MIME: \ldots} \, }^\text{version} \qquad
\overbracket[0.75pt]{\, \texttt{Date: \ldots} \, }^\text{date} \qquad
\overbracket[0.75pt]{\, \texttt{Subject: \ldots} \, }^\text{subject} \qquad
\overbracket[0.75pt]{\, \texttt{From: \ldots} \, }^\text{sender} \qquad
\overbracket[0.75pt]{\, \texttt{To: \ldots} \, }^\text{recipients} \qquad
\overbracket[0.75pt]{\, \texttt{Content: \ldots} \, }^\text{contents}
\]
which could be easily formalized by  an RE. A user query could be: make a list of all the recipients of all messages.
We compare the queries and answers of a traditional Linux \emph{grep} style against the same
done with our parser tool.
\par
A \emph{grep}-like query could be: \verb!grep "To:" examplemail.txt!. Unfortunately, the query also returns false positives because the contents field of an e-mail messsage may contain
the "\verb!To:!" string. Moreover, the answer is formatted as lines containing several recipients separated by commas.
To overcome the latter problem, the user should pipe the result of \emph{grep} to some other utility, such as \emph{awk}, to extract the individual recipients.
\par
On the other hand, the same query, done using an RE parser, cancels the risk of false positives and directly executes in  one
step without post-processing to edit the recipient lists. The query uses \emph{regrep}, a proof-of-concept utility that comes together with our tool: \verb!regrep mail.re examplemail.txt tree-query!. It accepts three arguments: an RE \verb!mail.re! describing e-mail messages with the structure and marks of the MIME format as shown above, a text file \verb!examplemail.txt! and a tree-query similar to an XPath selection. This utility, like \emph{grep}, seeks in the text all the occurrences of the specified RE, builds
a syntax tree for each of them and then delivers the selection of elements as specified in the tree-query.
\paragraph{Short introduction to related work}
We anticipate some elements from Sect.~\ref{sec:relatedwork} about the existing serial and parallel recognizers, matchers and parsers, to help understanding the contributions of our paper.  We do not consider past efforts that only work for restricted types of RE, e.g., for string matching using the Aho-Corasick algorithm~\cite{DBLP:journals/cacm/AhoC75}. While many serial  and some parallel recognizers and matchers  exist  (discussed  in Sect.~\ref{sec:relatedwork}), very  few existing  RE parsers are known to us, and none of them is targeted to parallel architectures.
\par
Two leading examples that return just one syntax tree are the parser~\cite{DBLP:conf/wia/OkuiS10,DBLP:conf/flops/SulzmannL14} that chooses the prior tree according to POSIX, and the other~\cite{DBLP:conf/icalp/FrischC04} that chooses by the greedy criterion. Two parsers that return all the trees are:
Sulzmann~\cite{DBLP:journals/ijfcs/SulzmannL17}, which enumerates sequentially all the parse trees, and our Berry-Sethi Parser (BSP)~\cite{DBLP:journals/acta/BorsottiBCM21}, which instead returns the forest representing all the syntax trees and is the baseline of the present work. To construct  BSP, the RE is directly converted  by  the Berry-Sethi algorithm~\cite{BerstelPin1996} to a non-minimal deterministic FA, such that its states are linked to the positions of the terminals and operators that occur in the RE formula. In this way the parser, while scanning the text,  incrementally builds a linear representation of the syntax tree(s) organized as a directed acyclic graph (DAG).
\paragraph{Parallelization}
The idea of reducing the time for recognition by executing FA transitions in parallel  is very old, but  here it suffices to recall the influential presentation~\cite{DBLP:conf/wia/HolubS09} that sets a standard followed and also improved by  subsequent developments (some discussed in the Sect.~\ref{sec:relatedwork} on related work), including the present one. Such a \emph{standard approach} is a data-parallel algorithm that operates in three  phases: (\emph{i}) splitting the input into as many chunks as available processors, (\emph{ii})  parallel scanning of each chunk by means of a chunk automaton FA that speculatively starts in each one of its states and returns  the corresponding last state reached at the chunk end, and (\emph{iii}) orderly joining  the pairs  $\langle \, \emph{start-state}, \, \emph{last-state} \, \rangle$  for each pair of adjacent chunks, thus obtaining the successful run(s) for the whole input.
\par
The drawback of the standard approach is that each chunk automaton, although identical to the original FA,  is used as a nondeterministic machine  since it has multiple initial states, except for the first chunk automaton. Each chunk is  scanned in parallel by a different processor, therefore all FAs, from the second chunk onwards, must \emph{speculatively} start in \emph{any} of their states, since the last state reached by the preceding chunk is still unknown. In practice, most speculative attempts will do useless work and fail after some transition steps.  Therefore,  many later R\&D efforts have addressed such a major cause of overhead, which may reduce and even neutralize the benefit of parallelization.
\paragraph{Our contribution}
Our main contribution is both theoretical and practical: (\emph{i}) an  RE parsing algorithm, derived from~\cite{DBLP:journals/acta/BorsottiBCM21}, suitable to parallelization by the standard approach~\cite{DBLP:conf/wia/HolubS09}, and (\emph{ii}) a novel effective optimization of the chunk FA that reduces the number of starting states and therefore  the speculation overhead. The parser-generator and parser are available at the GitHub repository \RePar, as Java programs, to ensure the reproducibility of measurements. We report the absolute parsing speed and especially the speed-up achieved by using more processors up to $64$. For the benchmarks considered we have obtained  speed-up values  in the range $9$-$24$, i.e., at least one order of magnitude over the serial parser. The time to generate the parser from a given RE is short enough to be viable in practice.
\par
Our measurements  show that the speed-up obtainable strongly depends on the characteristics of the RE. This tool  may help future investigation to understand how specific optimizations may suit the classes of REs typical of certain areas. The addition of this parser to the vast existing collection of RE tools will hopefully allow the users to experiment the  friendliness of the finer analysis supported by parsing. Moreover, faster execution on parallel architectures will make larger data-sets viable to processing.
However this tool is not a fully engineered software product, and improvements in processing speed and user interface are still possible, e.g., to exploit advanced thread management techniques on multi-core computers.
\paragraph{Paper organization}
The presentation is articulated into four major sections, plus related work, conclusion and three appendices. Section~\ref{sec:serialParser} -- \emph{Serial parsing} --  introduces the basic concepts for structure and syntax trees, and presents the  parsing automaton we use for the serial parser algorithm. Section~\ref{sec:parallelparser} -- \emph{Parallel parser} -- recalls the standard (speculative) approach,  and presents the construction of the automaton used for chunk parsing and the organization  of our parallel parser algorithm with a complete example. Section~\ref{sec:tool} -- \emph{Tool design} -- describes the functions and structure of the software tool, and the concurrent organization in threads. Section~\ref{sec:experimentation} -- \emph{Experimentation} -- describes the benchmarks used with related measurements, and reports the results obtained in terms of absolute timing, speed-up of many vs one thread, and  memory space consumption. Then Section~\ref{sec:relatedwork} on related work discusses pertinent research on the standard parallel approach and the developments for reducing  speculation overhead, and compares to existing RE parsing algorithms. Section~\ref{sec:conclusion} concludes. Appendix~\ref{app:otherfeatures} describes  additional features of the parser tool, which are useful for real-life. Appendix~\ref{app:compareSLPFandSPPF} compares the SLPF representation of syntax trees with the standard one used by  context-free parsers, and  Appendix~\ref{app:SLPFoptimization} presents  memory optimizations for the SLPF.
\section{Serial parsing} \label{sec:serialParser}
This section includes what is relevant for understanding the serial parsing algorithm, presented at the section end and then incorporated in the parallel parser of Sect.~\ref{sec:parallelparser}. Although most concepts considered are not new in the theory of REs and their translation to finite automata (FA), our presentation differs from the traditional ones and is especially suited to our needs. For the convenience of the reader, we collect in Tab.~\ref{tab:tableterms1} the technical terms introduced.
\begin{table}[h]
\begin{center}
\tabcolsep=0.0cm
\begin{tabular}{l@{\hspace{1.0cm}}l}	
\textbf{term} & \textbf{denotation / description}
\\ \toprule
regular expression& RE -- standard regular expression $e$ with concatenation, union, Kleene star and cross \\ \midrule
finite-state automaton & FA -- deterministic or nondeterministic finite automaton, in short DFA or NFA
\\ \midrule
metasymbol & operators (``\,$\cdot$\,'', ``\,$\mid$\,'' or ``\,$\cup$\,'', ``\,$\ast$\,'' and ``\,$+$\,''), parentheses ``\,$( \ )$\,'' and empty string $\varepsilon$
\\ \midrule
numbered regular expression & an RE $e_\#$ where terminals, empty strings $\varepsilon$ and parentheses are distinctly numbered
\\ \midrule
structure tree & functional tree of an RE, where leaf and inner nodes are terminals and operators, resp.
\\ \midrule
linearized syntax tree & LST -- parenthesized expression that represents a syntax tree
\\ \midrule
shared linearized parse forest & SLPF -- compact data-structure  that represents all the LSTs of an ambiguous text
\\ \midrule
shared packed parse forest & SPPF -- compact data-structure of the syntax trees of an ambiguous context-free text
\\ \midrule
segment & string of parentheses and $\varepsilon$'s that ends with a single terminal, all symbols numbered
\\ \midrule
follower set  & $\mathit{Fol} \, (\ldots)$ -- set of numbered symbols that may follow another numbered symbol
\\ \midrule
follower segment set & $\mathit{FolSeg} \, (\ldots)$ -- same as before but for segments
\\ \bottomrule
\end{tabular}
\end{center}
\caption{Terms and acronyms -- Part I.} \label{tab:tableterms1}
\end{table}
\subsection{Regular expressions and syntax trees} \label{subsec:regexp}
This subsection contains the relevant aspects of RE parsing, starting from the difference between parsing and mere validity checking. We discuss ambiguity and we show how to mark the positions in the RE in order to obtain a linear representation of the parse or \emph{syntax trees}.
\paragraph{What is RE parsing}
We omit the standard definition of RE and we just introduce our terminology.
A \emph{regular expression} (RE) is a formula with \emph{terminals}, operators and parentheses. The basic operators  are
\emph{concatenation}, \emph{union} (both can be binary, ternary, etc.), \emph{Kleene star} and \emph{cross}, the latter two called \emph{iterators}. The operators are respectively denoted  by ``\,$\cdot$\,'' (often understood), ``\,$\mid$\,'' or ``\,$\cup$\,'', ``\,$\ast$\,'' and ``\,$+$\,''. The \emph{empty} (or \emph{null}) \emph{string} is denoted by $\varepsilon$ (epsilon). The  default precedence between  operators is: first compute the iterators, second concatenation, and last union; parentheses may be used to override the default. For convenience, we classify the signs occurring in REs as \emph{terminals} and \emph{metasymbols} (operators, parentheses and $\varepsilon$). We illustrate with an example.
\paragraph{Example $1$ -- RE string and trees}
We represent an RE formula as a \emph{structure tree}, where  the terminals and the empty string $\varepsilon$ occur as leaf nodes and the operators (metasymbols) as internal nodes; see the RE $e_1$ with its structure tree in Fig.~\ref{fig:structuretree}. An RE generates a \emph{regular language}, i.e., a set of strings  over the terminal alphabet, e.g., $\set{ \, a, \; a\,b, \; a\,b\,a, \; a\,a, \; \ldots \, }$ for RE $e_1$. The way a string is generated is visualized as a \emph{syntax tree}, which is modeled on the RE structure tree, as follows: for each concatenation node orderly take (from left to right) all the child subtrees, for each union node choose exactly one of the child subtrees, and for each iterator node, say ``\,$\ast$\,'', take zero, i.e., $\varepsilon$, one or more instances of the child subtree. The left-to-right sequence of leaf nodes of the syntax tree is the generated string, e.g., $a\,b\,a$  for the first (leftmost) syntax tree in Fig.~\ref{fig:structuretree}; notice that  the cross iterates once. A string  may have more than one syntax tree, e.g., the second (rightmost) syntax tree of $a\,b\,a$ for $e_1$ in Fig.~\ref{fig:structuretree}, where the cross iterates twice. \qed
\par
When it has two or more syntax trees, the string, as well as the RE, is \emph{ambiguous}. In some cases, a string may even have infinitely many syntax trees. For example, the RE $( \, a \; \mid \; \varepsilon \, )^\ast$ is infinitely ambiguous.  For brevity we do not discuss the REs that are infinitely ambiguous. Our approach for dealing with them is similar to the one in~\cite{DBLP:journals/acta/BorsottiBCM21} and is  hinted in App.~\ref{app:otherfeatures}.
\par
Given an RE and an input string,  the \emph{parser job} precisely consists of returning all the \emph{syntax trees} of the string in a suitable representation. In recent years a few serial parsing algorithms have been published (references are in Sect.~\ref{sec:relatedwork}) that are based on some kind of parsing automaton. Their construction and operation is later described in view of the developments introduced in  Sect.~\ref{sec:parallelparser} to obtain a parallel parser. Comparing string parsing with string recognition and matching, it is obvious that the former produces a richer information at a higher computational cost. Moreover, parsing enhances matching, in that it finds the inner structure of the matched substring(s), i.e., their syntax tree(s). To represent syntax trees we introduce a linearized form.
\paragraph{Linearized syntax tree (LST)}
It is well-known that a syntax tree can be encoded in the form of a well-balanced parenthesized string called \emph{linearized syntax tree} (LST). In such a parenthesized string, each pair of matching parentheses represents an internal node of the tree, and the substring contained in the pair represents the subtree rooted at that node; each node operator is appended (as a subscript) to the associated closed parenthesis; and each terminal (or $\varepsilon$) represents a leaf node; e.g., see the  two  graphic trees  and their linearized encoding in Fig.~\ref{fig:structuretree}. Therefore, the alphabet of the LST includes terminals, the empty string $\varepsilon$ and (subscripted) parentheses.
\subsection{Numbered regular expressions} \label{subsec:numberedRE}
Our parser  returns the packed LSTs of the input string; it  generalizes the classic Glushkov / McNaughton-Yamada (GMY) construction~\cite{VMGlushkov_1961,DBLP:journals/tc/McNaughtonY60} of the RE recognizer. We recall that the GMY method belongs to the class of \textit{positional} RE-to-FA translators because it distinctly numbers the RE positions where a terminal occurs.
Then GMY constructs a nondeterministic finite automaton (NFA) that has one state per numbered terminal.
\par
We generalize the GMY  technique by numbering also the occurrences of the RE operators. Since each operator occurrence is associated to a subtree, i.e., to a parenthesized substring of the LST, we move the operator number from the operator to the parenthesis pair that corresponds to  it. The \emph{numbered} RE, denoted by  $e_\#$, associated to an RE $e$ is obtained by distinctly numbering each terminal (and epsilon, always understood), i.e., $a$ becomes $a_i$ (as in GMY). Moreover, the scope of each operator is surrounded by distinctly numbered parentheses, i.e., $_i( \ )_i$. A fresh integer subscript $i$ is used each time; e.g., see the numbered RE  $e_{1\#}$ of  RE $e_1$  in Eq.~\eqref{eq:numberedRE}, alongside with the correspondence table between subscripts and operators:
\begin{equation} \label{eq:numberedRE}
\vspace{0.125cm}
\tabcolsep=0.0cm\def\arraystretch{3}
\begin{tabular}[c]{c@{\hspace{1.0cm}}c}
\tabcolsep=0.0cm\def\arraystretch{0.5}
\begin{tabular}[c]{c}
$e_{1\#} = \, _1( \; \; _2( \; \, a_3 \; \mid \; _4( \; \; a_5 \; b_6 \; \; )_4 \; \mid \; _7( \; \; a_8 \; b_9 \; a_{10} \; \; )_7 \, \; )_2 \, \; {)_1}^+$ \\
\end{tabular}
&
\tabcolsep=0.0cm\def\arraystretch{0.75}
\begin{tabular}[c]{c@{\hspace{0.375cm}}l@{\hspace{0.375cm}}c@{\hspace{0.375cm}}l}
\emph{num} & RE \emph{operator} & \emph{symbol} & \emph{arity} \\ \toprule
$1$ & Kleene cross & $+$ & unary \\ \midrule
$2$ & union & $\mid$ or $\cup$ & ternary \\ \midrule
$4$ & concatenation & $\cdot$ & binary \\ \midrule
$7$ & concatenation & $\cdot$ & ternary \\ \bottomrule
\end{tabular}
\end{tabular}
\vspace{0.125cm}
\end{equation}
The numbered RE $e_\#$ has many more terminals than RE $e$; e.g., for $e_{1\#}$ the alphabet includes $a_3$, $a_5$, \ldots, $b_9$, $a_{10}$, and   ``\,$_1($\,'', ``\,$_2($\,'', \ldots, ``\,$)_2$\,'', ``\,$)_1$\,''. Strictly speaking, the formula thus obtained is not an RE because the original parentheses, after their subscripting, are viewed as terminals. To restore their role as metasymbols for grouping a subformula, we should re-introduce some kind of  brackets, e.g., square, that duplicate the subscripted parentheses, as shown below:
\[
e_{1\#} = \, _1( \; \Big[ \; \underbracket[0.75pt]{_2( \; \big[ \, \underbracket[0.75pt]{a_3 \; \mid \; _4( \; [ \; \underbracket[0.75pt]{a_5 \; b_6}_\text{concat.} \; ] \; )_4 \; \mid \; _7( \; [ \; \underbracket[0.75pt]{a_8 \; b_9 \; a_{10}}_\text{concat.} \; ] \; )_7}_\text{scope of the union operator} \, \big] \; )_2}_\text{scope of the cross operator} \, \Big]^+ \; )_1
\]
This formula is now formally correct but cumbersome, and from now on for simplicity the square brackets will be omitted. The reader should interpret  the (sub)expression $_1( \; \ldots \; {)_1}^+$  as $_1( \; [ \, \ldots \, ]^+ \; )_1$, by virtually reintroducing the metasymbols $[ \ ]^+$ that delimit the scope of the cross operator, and similarly $_2( \; \ldots \; )_2$ as $_2( \;  [ \, \ldots \, ] \; )_2$, where the metasymbols $[ \ ]$ delimit the scope of the union operator; similarly for all the other numbered parenthesis pairs. In fact, the parser-generator only needs the numbered parentheses since it knows where they fit in the syntactic structure of the RE.
\par
\begin{figure}[ht]
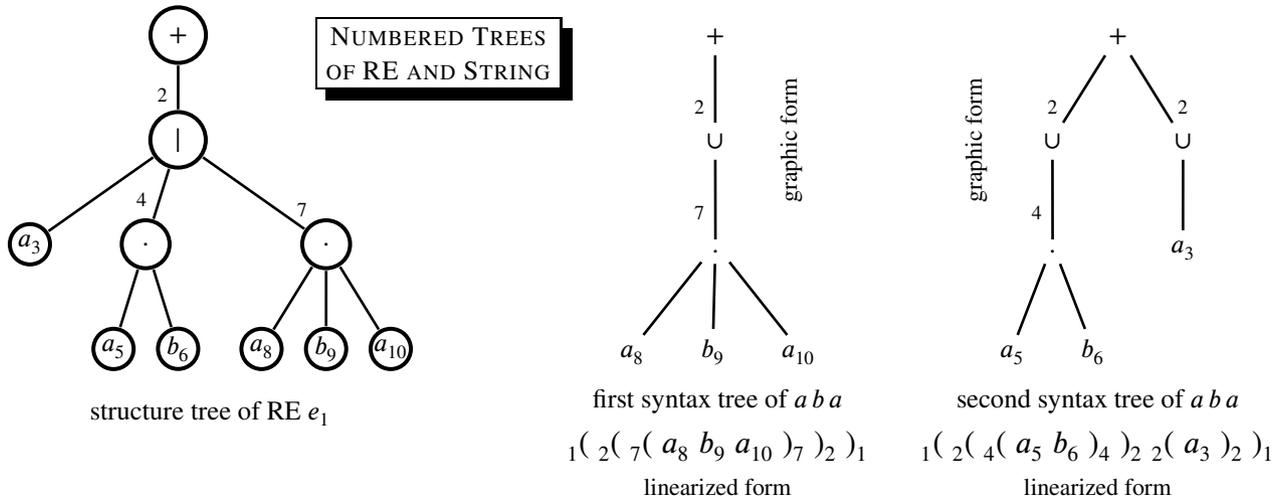

\begin{center}
\rnode {numberedstructure} {
\psset{arrows=-, linestyle=solid, linewidth=1pt, levelsep=1.375cm, treesep=1.0cm, nodesep=0pt, labelsep=3pt, treefit=loose, nodealign=true, tpos=0.55}
\tabcolsep=0.0cm\def\arraystretch{0.5}
\begin{tabular}{c@{\hspace{1.5cm}}cc}
\pstree[treesep=0.5cm]{\TCircle[linewidth=1.5pt, radius=0.40]{$+$}} {
    \pstree{\TCircle[linewidth=1.5pt, radius=0.40]{$\mid$} \tlput{\footnotesize 2}} {
        \TCircle[linewidth=1.5pt, radius=0.30]{$a_3$}
        \pstree[treesep=0.25cm, tpos=0.6]{\TCircle[linewidth=1.5pt, radius=0.35]{$.$} \tlput{\footnotesize 4}} {
            \TCircle[linewidth=1.5pt, radius=0.30]{$a_5$}
            \TCircle[linewidth=1.5pt, radius=0.30]{$b_6$}
        }
        \pstree[treesep=0.25cm, tpos=0.6]{\TCircle[linewidth=1.5pt, radius=0.35]{$.$} \trput[tpos=0.65]{\footnotesize 7}} {
            \TCircle[linewidth=1.5pt, radius=0.30]{$a_8$}
            \TCircle[linewidth=1.5pt, radius=0.30]{$b_9$}
            \TCircle[linewidth=1.5pt, radius=0.30]{$a_{10}$}
        }
    }
}

&

\psset{arrows=-, linestyle=solid, linewidth=1pt, levelsep=1.375cm, treesep=1.0cm, nodesep=4pt, labelsep=3pt, treefit=tight, nodealign=true, tpos=0.675}
\pstree[name=syntree1]{\TR{$+$}} {
    \pstree{\TR{$\cup$} \tlput{\footnotesize 2}} {
            \pstree[treesep=0.75cm]{\TR{$.$} \tlput{\footnotesize 7}} {
                \TR{$a_8$}
                \TR{$b_9$}
                \TR{$a_{10}$}
            }
    }
}

&

\psset{arrows=-, linestyle=solid, linewidth=1pt, levelsep=1.375cm, treesep=1.0cm, nodesep=4pt, labelsep=3pt, treefit=tight, nodealign=true, tpos=0.675}
\pstree[name=syntree2, treesep=1.5cm]{\TR{$+$}} {
    \pstree{\TR{$\cup$} \tlput{\footnotesize 2}} {
            \pstree[treesep=0.75cm]{\TR{$.$} \tlput{\footnotesize 4}} {
                \TR{$a_5$}
                \TR{$b_6$}
            }
    }
    \pstree{\TR{$\cup$} \trput{\footnotesize 2}} {
                \TR{$a_3$}
    }
}
\\ \\
\multirow{2}{*}{\parbox{5.0cm}{\centering structure tree of RE $e_1$}} & \parbox{5.0cm}{\centering first syntax tree of $a\,b\,a$} & \parbox{5.0cm}{\centering second syntax tree of $a\,b\,a$} \\ \\
& \large $_1( \; _2( \; _7( \; a_8 \; b_9 \; a_{10} \; )_7 \; )_2 \; )_1$ & \large $_1( \; _2( \; _4( \; a_5 \; b_6 \; )_4 \; )_2 \; _2( \; a_3 \; )_2 \; )_1$ \\ \\
& \scriptsize linearized form & \scriptsize linearized form
\end{tabular}}
\nput[labelsep=-1.25cm] {125} {numberedstructure} {\shadowbox{\parbox{3.0cm}{\centering \textsc{Numbered Trees} \par \textsc{of RE and String}}}}
\nput[labelsep=0.75cm, rot=90, origin=c] {0} {syntree1} {\scriptsize graphic form}
\nput[labelsep=0.75cm, rot=90, origin=c] {180} {syntree2} {\scriptsize graphic form}
\end{center}
\caption{Structure tree of RE $e_1$, and syntax trees of string $a\,b\,a$ in graphic and linearized form (the root subscript is omitted).} \label{fig:linearizedtrees}
\end{figure}

Since the REs $e$ and $e_\#$  have the same tree structure, we can transfer the numbering from $e_\#$  to the structure tree and syntax trees, as illustrated in Fig.~\ref{fig:linearizedtrees}. For better reading, in the trees the (round) parentheses are omitted but their subscripts are shown as edge labels, near to the child nodes they refer to. The subscript $1$ for the tree root node is redundant and is omitted. Notice that  also the terminals are numbered.
We can then encode the structure tree of $e$ and the syntax tree(s) of a string produced by $e$, in the form of \emph{linearized syntax trees} (LSTs) over the \emph{terminal alphabet of $e_\#$}; e.g., see the LSTs shown in Fig.~\ref{fig:linearizedtrees}.
\par\noindent
The  order chosen to number an RE is irrelevant, and we adhere to a simple option:  to proceed left to right. Such an order  corresponds to the left-to-right preorder visit of the graphic tree. From the preceding discussion, the next  statement (Prop.~\ref{prop:regulartreelang}) immediately follows,  which  is the foundation of the parser automaton construction in Sect.~\ref{subsec:parserautomaton}.
\begin{proposition}[language of LST] \label{prop:regulartreelang} The (regular) language generated by the numbered RE $e_\#$ is the set of the linearized syntax trees (LSTs) of the original RE $e$. \qed
\end{proposition}
Coming back to ambiguity, since the number of syntax trees of an ambiguous string may rapidly increase  with its length, it is necessary to compactly represent all such trees.  To this end a well-known approach relies on  a \emph{forest} graph called \emph{shared packed parse forest} (SPPF), e.g., in~\cite{DBLP:journals/toplas/ScottJ06}. The SPPF model was originally developed for grammars, but can be used also for REs. We later develop our representation, called \emph{shared linearized parse forest} (SLPF), designed to encode LSTs. A comparison between SPPF and SLPF is in App.~\ref{app:compareSLPFandSPPF}.
\subsection{Parser automaton} \label{subsec:parserautomaton}
As said, the construction of the serial parser resembles the GMY technique that constructs a nondeterministic FA recognizer  from an RE where all positions have been numbered. The essential difference is  that the parser NFA returns not just a yes\,/\,no, but also the string(s) representing the LST(s). The LST elements are the terminals of the  numbered RE $e_\#$ and comprise not just the numbered terminals of RE $e$, but also the numbered parentheses. The next example (Sect.~\ref{subsubsec:initial-ex-serial-parser}) intuitively presents  the whole construction from  numbered RE to the NFA (parser). The construction is then formalized in Sect.s~\ref{subsubsec:foundationsRE2NFA} and \ref{subsubsec:segment-computation}, and is further developed into a complete algorithm in Sect.~\ref{subsubsec:constrParserNFA}. To complete the construction, Sect.~\ref{subsubsec:case-ambiguousRE} discusses the case of ambiguous REs  and deals with the construction of  the SLPF forest of syntax trees. At the end, Sect.~\ref{subsec:serialalgorithmVectorial} reformulates  the serial parser in a terser matrix notation.
\subsubsection{Initial example} \label{subsubsec:initial-ex-serial-parser}
\paragraph{Example $2$ -- recognizing and then parsing a non-ambiguous RE}
In the GMY approach, the terminals of  RE $e_2 =  ( \, a \, b \; \mid \; a \, )^\ast$ are numbered as in $e_\text{$2$N} = ( \, a_4 \, b_5 \; \mid \; a_6 \, )^\ast$
(numbering  from $4$ is for consistency with later examples). The states of the classic GMY NFA (shown below, left) of $e_2$ are obtained from $e_\text{$2$N}$ and are the numbered terminals $\set{a_4, \, b_5, \, a_6}$ plus the end-mark $\dashv$, which for convenience we assume to terminate each string and is the final state (marked with the exit arrow). The three initial states are $a_4$, $a_6$ and $\dashv$ (marked with the entry arrows):
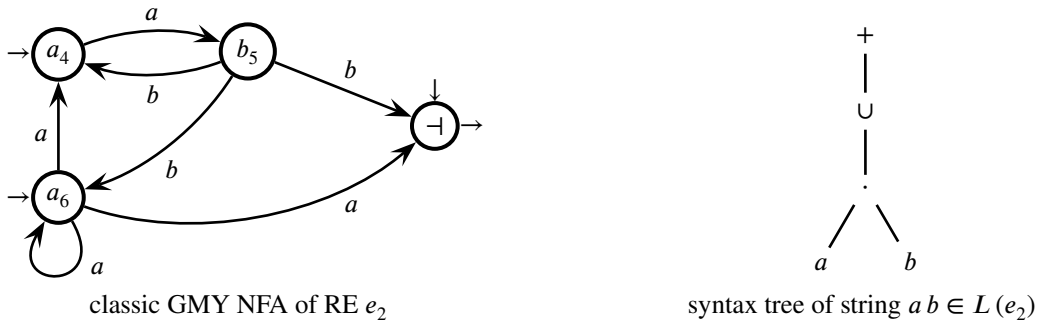
\begin{center}
\vspace{0.375cm}
\tabcolsep=0.0cm\def\arraystretch{0}
\begin{tabular}{c@{\hspace{3.0cm}}c}
\tabcolsep=0.0cm\def\arraystretch{0}
\begin{tabular}[c]{c}
\scalebox{1.0}{
	\psset{nodesep=0.0cm, arrows=->, linewidth=1.5pt, arrowscale=2, rowsep=0.25cm, colsep=1.75cm, border=0.00cm}
	\begin{psmatrix}
		\circlenode 4 {$a_4$} & \circlenode 5 {$b_5$} \\
		& & \circlenode {dashv} {$\dashv$} \\
		\circlenode 6 {$a_6$} \\[0.30cm]
		
		\psset{labelsep=0pt, linewidth=1pt}
		
		\nput {180} 4 {$\to$}
		
		\nput {180} 6 {$\to$}
		
		\nput {90} {dashv} {$\downarrow$}
		
		\nput {0} {dashv} {$\to$}
		
		\psset{labelsep=3pt}
				
		\psset{labelsep=3pt, linewidth=1pt, border=0.00cm}
		
		\ncarc[arcangle=20] 4 5 \naput{$a$}
		
		\ncarc[arcangle=20] 5 4 \naput{$b$}
		
		\ncline 6 4 \naput[npos=0.4]{$a$}
		
		\nccurve[angleA=-60, angleB=-120, ncurv=6] 6 6 \naput[npos=0.25]{$a$}
		
		\ncarc[arcangle=-30] 6 {dashv} \nbput[npos=0.75]{$a$}
		
		\ncarc[arcangle=20] 5 6 \naput[npos=0.55]{$b$}
		
		\ncline 5 {dashv} \naput{$b$}
	\end{psmatrix}}
\end{tabular}
&
\tabcolsep=0.0cm\def\arraystretch{0}
\begin{tabular}[c]{c}
	\psset{nodesep=4pt, linewidth=1pt, arrows=-, levelsep=1.0cm, treesep=1.0cm}
	\pstree{\TR{$+$}} {
		\pstree{\TR{$\cup$}} {
			\pstree{\TR{$\cdot$}} {
				\TR{$a$}
				\TR{$b$}
			}
		}
	}
\end{tabular}
\\ \\[0.25cm]
classic GMY NFA of RE $e_2$ & syntax tree of string $a\,b \in L \, (e_2)$
\end{tabular}
\vspace{0.25cm}
\end{center}
The run $a_4 \xrightarrow a b_5 \xrightarrow b \; \dashv$ recognizes the input $a\,b \dashv$. Now we depart from the pure GMY recognizer and move on to the parser. From  the states traversed, i.e., $a_4$, $b_5$ and $\dashv$, it is easy to construct the syntax tree above (right) of $a\,b$. Assuming the numbered RE to be $\,e_{2\#} =  \, _1( \; _2( \; _3( \; a_4 \; b_5 \; )_3 \; \mid \; a_6 \; )_2 \; {)_1}^\ast\, $, the tree representation of $a\,b$ as an LST is the string ``$_1( \, _2( \, _3( \, a_4 \, b_5 \, )_3 \, )_2 \, )_1$''. By analogy with the behaviour of the GMY automaton, we should expect the parser to simulate some sort of NFA such that the sequence of states traversed by a run on a given input string coincides with the LST of the string. Such an NFA is named a \emph{parser automaton} and is shown in  Fig.~\ref{fig:NFA-RE2}.
\begin{figure}[ht]
	\begin{center}
		\rnode {NFA1} {
			\psset{nodesep=0.0cm, arrows=->, linewidth=1.5pt, arrowscale=2, rowsep=1.25cm, colsep=2.5cm, border=0.00cm}
			\begin{psmatrix}
				& \ovalnode {32234} {$)_3 \, )_2 \, _2( \, _3( \, a_4$} \\[-0.25cm]
				\ovalnode {1234} {$_1( \, _2( \, _3( \, a_4$} & \ovalnode {5} {$ \ b_5 \ $} & \ovalnode {321dashv} {$)_3 \, )_2 \, )_1 \dashv$} \\
				\ovalnode {11dashv} {$_1( \ )_1 \dashv$} & \ovalnode {2234} {$)_2 \, _2( \, _3( \, a_4$} & \ovalnode {3226} {$)_3 \, )_2 \, _2( \, a_6$} \\
				\ovalnode {126} {$_1( \, _2( \, a_6$} & \ovalnode {226} {$)_2 \, _2( \, a_6$} & \ovalnode {21dashv} {$)_2 \, )_1 \dashv$}
				
				\psset{labelsep=3pt}
				
				\nput {135} {11dashv} {\large $q_1$}
				
				\nput {135} {1234} {\large $q_2$}
				
				\nput {135} {126} {\large $q_3$}
				
				\nput {-135} {5} {\large $q_4$}
				
				\nput[labelsep=4pt] {-150} {32234} {\large $q_5$}
				
				\nput {45} {3226} {\large $q_6$}
				
				\nput {150} {2234} {\large $q_7$}
				
				\nput {-135} {226} {\large $q_8$}
				
				\nput {45} {321dashv} {\large $q_9$}
				
				\nput {45} {21dashv} {\large $q_{10}$}
				
				\psset{labelsep=0pt, linewidth=1pt}
				
				\nput {180} {11dashv} {$\to$}
				
				\nput {0} {11dashv} {$\to$}
				
				\nput {180} {1234} {$\to$}
				
				\nput {180} {126} {$\to$}
				
				\nput {0} {321dashv} {$\to$}
				
				\nput {0} {21dashv} {$\to$}
				
				\psset{labelsep=3pt, linewidth=1pt, border=0.00cm}
				
				\ncline {1234} {5} \naput[npos=0.4]{$a$}
				
				\ncarc[arcangle=-25] {5} {32234} \nbput{$b$}
				
				\ncarc[arcangle=-25] {32234} {5} \nbput{$a$}
				
				\ncline {5} {3226} \naput{$b$}
				
				\ncline {5} {321dashv} \naput[npos=0.575]{$b$}
				
				\ncarc[arcangle=10] {126} {2234} \naput[npos=0.6]{$a$}
				
				\ncline {126} {226} \naput{$a$}
				
				\ncarc[arcangle=-20] {126} {21dashv} \nbput[npos=0.775]{$a$}
				
				\ncline {226} {21dashv} \naput[npos=0.475]{$a$}
				
				\ncarc[arcangle=-20] {226} {2234} \nbput{$a$}
				
				\ncline {3226} {226} \nbput[npos=0.475]{$a$}
				
				\ncline {3226} {2234} \nbput[npos=0.525]{$a$}
				
				\ncline {3226} {21dashv} \naput{$a$}
				
				\ncline {2234} {5} \naput[npos=0.4]{$a$}
				
				\nccurve[angleA=110, angleB=150, ncurv=5.5] {226} {226} \nbput[npos=0.575]{$a$}
			\end{psmatrix}
			\hspace{4.5cm}}
		\nput[labelsep=-2.5cm] {-8} {NFA1} {\shadowbox{\Large \textsc{NFA $e_2$}}}
		\nput[labelsep=-1.825cm] {19} {NFA1} {
			\rnode {classicNFA1} {
				\psframebox[framesep=0.375cm, linewidth=1pt]{
					\scalebox{0.75}{
						\psset{nodesep=0.0cm, arrows=->, linewidth=1.5pt, arrowscale=2, rowsep=0.0cm, colsep=1.0cm, border=0.00cm}
						\begin{psmatrix}
							\circlenode 4 {$a_4$} & \circlenode 5 {$b_5$} \\
							& & \circlenode {dashv} {$\dashv$} \\
							\circlenode 6 {$a_6$} \\[0.30cm]
							
							\psset{labelsep=0pt, linewidth=1pt}
							
							\nput {180} 4 {$\to$}
							
							\nput {180} 6 {$\to$}
							
							\nput {90} {dashv} {$\downarrow$}
							
							\nput {0} {dashv} {$\to$}
							
							\psset{labelsep=3pt}
							
							\nput[rot=30] {120} 4 {\footnotesize $q_2\, q_5\, q_7$}
							
							\nput {60} 5 {\footnotesize $q_4$}
							
							\nput[rot=-40] {-40} 6 {\footnotesize $q_3\, q_6\, q_8$}
							
							\nput[rot=30] {-60} {dashv} {\footnotesize $q_1\, q_9\, q_{10}$}
							
							\psset{labelsep=3pt, linewidth=1pt, border=0.00cm}
							
							\ncarc[arcangle=15] 4 5 \naput{$a$}
							
							\ncarc[arcangle=15] 5 4 \naput{$b$}
							
							\ncline 6 4 \naput[npos=0.4]{$a$}
							
							\nccurve[angleA=-60, angleB=-120, ncurv=5] 6 6 \naput[npos=0.25]{$a$}
							
							\ncarc[arcangle=-30] 6 {dashv} \nbput[npos=0.75]{$a$}
							
							\ncarc[arcangle=20] 5 6 \naput[npos=0.55]{$b$}
							
							\ncline 5 {dashv} \naput{$b$}
						\end{psmatrix}
						\hspace{0.05cm}
		}}}}
		\nput[labelsep=-0.35cm] {-70} {classicNFA1} {\scriptsize classic GMY NFA}
	\end{center}
	\vspace{0.5cm}
	\caption{The parser NFA for RE $e_2$ of Ex. $2$. The  NFA  encodes in its states the language of the LSTs of RE $e_2$ (to be detailed in  Tab.~\ref{tab:strings}). The upper right box  reproduces the classic GMY NFA that  recognizes (but does not parse) the language  $L \, (e_2)$; its states correspond to subsets of the parser NFA states. } \label{fig:NFA-RE2}
\end{figure}
Clearly the parser  NFA  and classic  GMY NFA are equivalent, i.e., both recognize the same language.  The former contains the states of the latter  and its states $q_1$, \ldots, $q_{10}$ carry the information needed to output the LST.
\par
Looking at Fig.~\ref{fig:NFA-RE2}, we see that each state  contains  a string over the numbered alphabet, e.g., ``\,$)_3 \, )_2 \, _2( \, _3( \, a_4$'' in the state $q_5$. We name such a string a \emph{segment} because the LST of any valid  string is the concatenation of one or more segments.
The reader may also notice that the segments associated to the states are distinct, and that each segment consists of a (possibly empty) prefix made of numbered  parentheses followed by a numbered terminal or by the end-mark. Thus exactly one numbered terminal is present in a segment and always at the end.
Returning to parsing, the states  traversed by the parser  NFA in the run  that recognizes the string $a\,b$ are $q_2$, $q_4$ and $q_9$, and the concatenation of the three associated segments is exactly the LST:
\[
\overbracket[0.75pt]{\, _1( \, _2( \, _3( \, a_4 \,}^\text{segment $q_2$} \quad \overbracket[0.75pt]{\hspace{0.2cm} b_5 \hspace{0.2cm}}^\text{seg. $q_4$} \quad \overbracket[0.75pt]{\, )_3 \, )_2 \, )_1 \; \dashv \,}^\text{segment $q_9$} \quad = \quad \overbracket[0.75pt]{\, _1( \, _2( \, _3( \, a_4 \, b_5 \, )_3 \, )_2 \, )_1 \; \dashv \,}^\text{linearized syntax tree (LST)}
\]
We may disregard the end-mark, which is present in the final states of parser NFA just for technical reasons. \qed
\par\noindent
To intuitively explain how  the parser NFA is built from the numbered RE, we  make explicit the analogy with the GMY construction:
 \begin{itemize}[leftmargin=*]
 	\item  The states of the GMY NFA are  the numbered terminals occurring in RE $e_\text{$2$N}$ and those of the parser NFA are the numbered segments occurring in RE $e_{2\#}$; for brevity let us call them  state \emph{contents}.
 	\item A state is initial if  its contents may occur as prefix in a valid string or in the LST of a valid string.
 	\item A state is final if its contents consist of or terminate by the end-marker $\dashv$.
 	\item There is a state transition $q_j \xrightarrow a q_k$ if the contents of states $q_j$ and $q_k$ may orderly occur next to each other in a valid string or in the LST of a valid string.
 \end{itemize}
The details of the preceding definitions and constructions are developed in the next sections.
\subsubsection{Definition of segment} \label{subsubsec:foundationsRE2NFA}
We have seen that the states of the parser NFA  are the segments of the numbered RE $e_\# \dashv$. We define the segments more precisely\footnote{For brevity we resume and simplify the presentation in~\cite{DBLP:journals/acta/BorsottiBCM21}, where the properties of the segments are formally stated and proved.}: a \emph{segment} is a string of the form  $\mu \, a$, where part $\mu$ is a possibly empty string exclusively made of numbered  parentheses and empty strings $\varepsilon$, and part $a$ is a numbered terminal or the end-mark $\dashv$. The two parts $\mu$ and $a$ are called \emph{meta-prefix} and \emph{end-letter}, respectively.  A string  $\mu \, a$ is a segment of an RE $e$ if it is a \emph{maximal} substring of an LST generated by the associated numbered RE $e_\#$ (more precisely by $e_\# \dashv$). For instance, looking back to the RE $e_1$ of Ex.~$1$ and Fig.~\ref{fig:linearizedtrees}, both string ``\,$_1( \, _2( \, _4( \, a_5$\,'', which has meta-prefix $\mu = \text{``\,$_1( \, _2( \, _4($\,''}$ and end-letter $a = a_5$, and string ``\,$)_4 \, )_2 \, _2( \, a_3$\,'' are segments of $e_1$. On the other hand, the string ``\,$_2( \, a_3$\,'' is not a segment of $e_1$, because it is not maximal.
\par
Given a valid input string   $a_1 \,  a_2 \, \ldots \, a_{l} \dashv$, with $l \geq 0$ (for $l = 0$ the string is $\varepsilon\dashv$),  the corresponding LST can be factored into segments in a unique way:  $\mu_1 \, a_1 \, \mu_2 \, a_2 \, \ldots \, \mu_l \, a_l \, \mu_{l + 1} \dashv$ (or just $\mu_1 \dashv$ if $l = 0$).
Notice that the last end-letter is the end-mark $\dashv$, and that the meta-prefixes $\mu$ may be empty. The segments that occur at the beginning or at the end of some LSTs are called \emph{initial} or \emph{final}, respectively, and the others \emph{internal}.
\subsubsection{Computation of segments} \label{subsubsec:segment-computation}
\begin{figure}[ht]
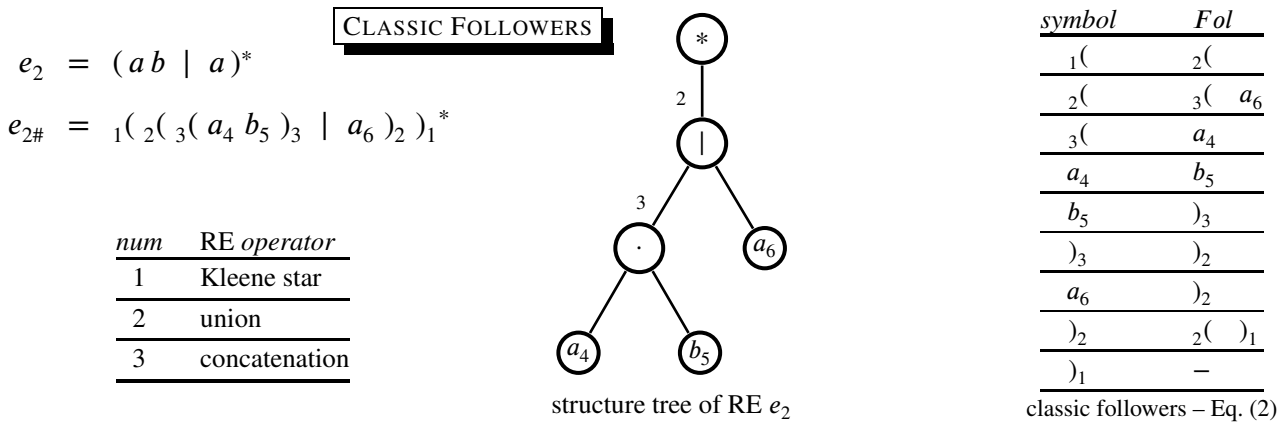

	\begin{center}
		\tabcolsep=0.0cm\def\arraystretch{0}
		\begin{tabular}{c@{\hspace{1.0cm}}c@{\hspace{3.0cm}}c}
			\begingroup
			\tabcolsep=0.0cm\def\arraystretch{1.25}
			\begin{tabular}{c}
				\begingroup
				\large
				$\arraycolsep=0.15cm\def\arraystretch{0}
				\begin{array}{rcl}
					e_2 & = & ( \, a \, b \; \mid \; a \, )^\ast \\[0.5cm]
					e_{2\#} & = & _1( \; _2( \; _3( \; a_4 \; b_5 \; )_3 \; \mid \; a_6 \; )_2 \; {)_1}^\ast
				\end{array}$
				\endgroup
				\\[1.5cm]
				\begingroup
				\tabcolsep=0.0cm\def\arraystretch{0.75}
				\begin{tabular}{c@{\hspace{0.5cm}}l}
					\emph{num} & RE \emph{operator} \\ \toprule
					$1$ & Kleene star \\ \midrule
					$2$ & union \\ \midrule
					$3$ & concatenation \\ \bottomrule
				\end{tabular}
				\endgroup
			\end{tabular}
			\endgroup
			&
			\tabcolsep=0.0cm\def\arraystretch{0}
			\rnode {tree1} {
				\begin{tabular}{c}
					\psset{arrows=-, linestyle=solid, linewidth=1pt, levelsep=1.375cm, treesep=1.0cm, nodesep=0pt, border=0.0cm, treefit=tight, tpos=0.55}
					\pstree{\TCircle[linewidth=1.5pt, radius=0.35]{$ \ast$}}{
						\pstree{\TCircle[linewidth=1.5pt, radius=0.35]{$\mid$} \tlput{\footnotesize 2}}{
							\pstree{\TCircle[linewidth=1.5pt, radius=0.35]{$.$} \tlput{\footnotesize 3}}{
								\TCircle[linewidth=1.5pt, radius=0.30]{$a_4$}
								\TCircle[linewidth=1.5pt, radius=0.30]{$b_5$}
							}
							\TCircle[linewidth=1.5pt, radius=0.30]{$a_6$}
						}
					}
					\\ \\[0.25cm]
					structure tree of RE $e_2$
			\end{tabular}}
			&
			\tabcolsep=0.0cm\def\arraystretch{0.25}
			\begin{tabular}{c}
				$\arraycolsep=0.0cm\def\arraystretch{0.25}
				\begin{array}{c@{\hspace{1.0cm}}l}
					\emph{symbol} & \mathit{Fol} \\ \toprule
					_1( & _2( \\ \midrule
					_2( & _3( \quad a_6 \\ \midrule
					_3( & a_4 \\ \midrule
					a_4 & b_5 \\ \midrule
					b_5 & )_3 \\ \midrule
					)_3 & )_2 \\ \midrule
					a_6 & )_2 \\ \midrule
					)_2 & _2( \quad )_1 \\ \midrule
					)_1 & - \\ \bottomrule
				\end{array}$ \\ \\
				\text{\scriptsize classic followers -- Eq.~\eqref{eq:classicfollowers}}
			\end{tabular}
		\end{tabular}
		\nput[labelsep=-0.25cm] {123} {tree1} {\shadowbox{\textsc{Classic Followers}}}
	\end{center}
	\caption{Non-ambiguous RE $e_2$ with its table of operators and their numbering, structure tree, and list of all classic followers. The end-mark is omitted.} \label{fig:nonambiguousREclassicFollowers}
\end{figure}

For a simple RE $e$ the segments can be found by visual inspection of $e_\#$. Here we describe the algorithm we use to compute the segments for any RE. We recall the well-known notion of \emph{follower set}, for a terminal symbol $r$ of an RE $e$. Such a set, denoted $\mathit{Fol} \, (e, \, r)$,   contains all the terminal symbols $s$ that are consecutive to $r$ in some string generated by $e$, in formula:
\begin{equation} \label{eq:classicfollowers}
	\mathit{Fol} \, (e, \, r) = \big\{ \, s \; \vert \quad \text{$s$ is a terminal and there exist strings $u$, $v$ such that $u\,r\,s\,v \in L \, (e)$} \, \big\}
\end{equation}
There are well-known algorithms, e.g., in~\cite{DBLP:series/txcs/ReghizziBM19}, for computing the follower sets of regular languages.
Of course, the definition of $\mathit{Fol}$ applies also to  the set of LSTs, i.e.,  language $L \, (e_\# \dashv)$, which is regular  by Prop.~\ref{prop:regulartreelang}. Thus, we can compute the classic follower set $\mathit{Fol} \, (e_\# \dashv, \, s)$ for any numbered symbol $s$. This is illustrated for the RE $e_2$ in Fig.~\ref{fig:nonambiguousREclassicFollowers}.
\par\noindent
A recursive algorithm  for computing the set $Q$ of all the segments for  RE $e$ is listed in Fig.~\ref{fig:segmentalgo}. It also returns  the subsets $I, \,F \subseteq Q$ of initial and final segments. Symbols $a$ denotes a numbered terminal (not $\varepsilon$).
\begin{figure}[ht]
	\begin{center}
		\definecolor{lightgray}{rgb}{0.875,0.875,0.875}
		\vspace{0.125cm}
		\tabcolsep=0.0cm
		\def\arraystretch{0.5}
		\begin{tabular}{p{0.475\textwidth}@{\hspace{0.5cm}}p{0.475\textwidth}} \toprule
			\cellcolor{lightgray}
			\begin{minipage}[t]{0.475\textwidth}
				\begin{algorithm2e}[H]
					\setstretch{1.0}
					\SetAlgorithmName{algorithm}{}{}
					\SetKwInOut{Input}{input}
					\SetKwInOut{Output}{output}
					\SetKwInOut{LocVars}{variable}
					\DontPrintSemicolon
					\TitleOfAlgo{\emph{segment computation}}
					\hrule height 0.25pt
					\BlankLine
					\Input{numbered end-marked RE $e_\# \dashv$ for RE $e$}
					\Output{sets $Q$, $I$ and $F$ of segments for RE $e$}
					\LocVars{numbered terminal $a$}
					\BlankLine
					\hrule height 0.25pt
					\BlankLine
					$Q \asgn \emptyset$ \tcp*[r]{initialization}
					\ForEach (\tcp*[f]{scan end-letters}) {\text{\rm end-letter $a$}} {
						$\mathit{MetaPrefix} \, (a, \, a)$ \tcp*[r]{set $Q$}
					}
					$I \asgn \set{ \, \sigma \; \vert \quad \text{\rm $\sigma \in Q$ begins with ``\,$_1($\,''} \, }$ \tcp*[r]{set $I$}
					$F \asgn \set{ \, \sigma \; \vert \quad \text{\rm $\sigma \in Q$ ends with $\dashv$} \, }$ \tcp*[r]{set $F$}
				\end{algorithm2e}
			\end{minipage}
			&
			\cellcolor{lightgray}
			\begin{minipage}[t]{0.475\textwidth}
				\begin{algorithm2e}[H]
					\setstretch{1.205}
					\SetKwInOut{LocVars}{variable}
					\DontPrintSemicolon
					\textbf{procedure} $\mathit{MetaPrefix} \, (\text{num. symb. $s$}, \, \text{segment $\sigma$})$ \\
					\LocVars{numbered symbol $r$}
					\lIf (\tcp*[f]{init symb}) {s = \text{\rm ``\,$_1($\,''}} {$Q \asgn Q \, \cup \, \set{ \, \sigma \, }$}
					\Else (\tcp*[f]{scan all symbols}) {
						\ForEach {\text{\rm symbol $r$ such that $\mathit{Fol} \, (e_\# \dashv, \, r) \ni s$}} {
							\lIf {\text{\rm $r$ is an end-letter}} {
								$Q \asgn Q \, \cup \, \set{ \, \sigma \, }$
							} \lElse (\tcp*[f]{extend $\sigma$}) {
								$\mathit{MetaPrefix} \, (r, \, r \cdot \sigma)$
							}
						}
					}
					\textbf{end procedure} \tcp*[r]{all symbols done}
				\end{algorithm2e}
			\end{minipage} \\ \bottomrule
		\end{tabular}
	\end{center}
	\caption{A recursive algorithm for computing all the segments for an RE, with the initial and final ones distinguished.} \label{fig:segmentalgo}
\end{figure}
\par\noindent
The algorithm computes each segment $\sigma = \mu \, a$ by starting from the end-letter $a$, or from the end-marker $\dashv$, and by extending the meta-prefix $\mu$ from right to left, one parenthesis  at a time, in all valid ways. A parenthesis  $r$ is prepended to $\sigma$ if symbol $r$ admits the leftmost symbol $s$ of $\mu$ as (classic) follower (initially $s = a$ and $\mu = \varepsilon$). Such an extension goes on until the end-letter of the preceding segment or the initial  parenthesis ``\,$_1($\,'' are found. At this point, segment $\sigma$ is complete and is stored into the segment set $Q$. The initial segments are those that have a meta-prefix starting with ``\,$_1($\,'', and the final ones those with the end-letter $\dashv$. The sets $I$ and $F$ are filtered out from set $Q$ at the end of the algorithm.
\par
We prove termination. We start from the observation that, since the RE is not infinitely ambiguous, a metasymbolic prefix may not contain two identical metasymbols subscripted with the same number. Therefore the number of meta-prefixes is upper bounded (Prop.~\ref{prop:segmentfiniteness}) and the algorithm terminates.
\begin{proposition}[segment finiteness] \label{prop:segmentfiniteness}
For any non-infinitely ambiguous RE $e$, the number of segments is finite. \qed
\end{proposition}
Note that the complexity of the parser NFA, i.e., its number of states, is the same as the number of segments. A loose upper bound to the number of segments is the number of permutations (without repetitions) of the $k$ elements of the alphabet of the numbered RE $e_\#$, i.e., $k!$. But quite often the number of segments is by far smaller (as shown in Ex.~$5$ of Sect.~\ref{subsec:deterministicFA} and in Fig.~\ref{fig:segments-vs-REsize} of Sect.~\ref{subsec:results})\footnote{Remember that we are excluding the case of infinitely ambiguous REs. Otherwise,  an LST of an infinitely ambiguous string may contain meta-prefixes of arbitrary length. In fact, infinite ambiguity stems exactly from the existence in the RE of an iterator (star or cross) with a nullable argument, which generates LST segments that consist of numbered parentheses repeated arbitrarily many times.}. For brevity we do not simulate the algorithm of Fig.~\ref{fig:segmentalgo} and we proceed informally by means of examples for unambiguous and then ambiguous REs.
  \begin{table}[h]
 	\begin{center}
 		\tabcolsep=0.0cm
 		\begin{tabular}{l@{\hspace{1.0cm}}r@{\hspace{0.875cm}}r@{\hspace{0.875cm}}r@{\hspace{0.875cm}}r@{\hspace{0.875cm}}r}
 			\textbf{type} & \multicolumn{5}{l}{\textbf{list of segments} \quad (total $10$ segments -- segment $_1( \ )_1 \dashv$ is both initial and final)} \\ \toprule
 			initial & $_1( \ )_1 \dashv \colon 1$ & $_1( \, _2( \, _3( \, a_4 \colon 2$ & $_1( \, _2( \, a_6 \colon 3$ \\ \midrule
 			internal & $b_5 \colon 4$ & $)_3 \, )_2 \, _2( \, _3( \, a_4 \colon 5$ & $)_3 \, )_2 \, _2( \, a_6 \colon 6$ & $)_2 \, _2( \, _3( \, a_4 \colon 7$ & $)_2 \, _2( \, a_6 \colon 8$ \\ \midrule
 			final & $_1( \ )_1 \dashv \colon 1$ & $)_3 \, )_2 \, )_1 \dashv \colon 9$ & $)_2 \, )_1 \dashv \colon 10 \hspace{-0.18cm}$ \\ \bottomrule
 		\end{tabular}
 	\end{center}
 	\caption{List of all the segments  (numbered $1$ \ldots $10$ for reference) of RE $e_2$ (symbols are numbered  as in $e_{2\#}$).} \label{tab:segments}
 \end{table}
 \begin{table}[ht]
 	\begin{center}
 		\scalebox{0.975}{
 			\tabcolsep=0.0cm\def\arraystretch{0}
 			\begin{tabular}{c@{\hspace{0.5cm}}c@{\hspace{1.0cm}}c}
 				\tabcolsep=0.0cm\def\arraystretch{1.125}
 				\begin{tabular}{c@{\hspace{0.25cm}}l@{\hspace{0.375cm}}l}
 					$\#$ & \textbf{string} & \textbf{linearized syntax tree} -- LST \\ \toprule
 					$1$ & $\varepsilon$ & $\underbracket[0.75pt]{_1( \ )_1 \dashv}_1$ \\ \midrule
 					$2$ & $a \, b$ & $\underbracket[0.75pt]{_1( \, _2( \, _3( \, a_4}_2 \, b_5 \, )_3 \, )_2 \, )_1 \dashv$ \\ \midrule
 					$3$ & $a$ & $\underbracket[0.75pt]{_1( \, _2( \, a_6}_3 \, )_2 \, )_1 \dashv$ \\ \midrule
 					$4$ & $a \, b$ & $_1( \, _2( \, _3( \, a_4 \, \underbracket[0.75pt]{b_5}_4 \, )_3 \, )_2 \, )_1 \dashv$ \\ \midrule
 					$5$ & $a \, b \, a \, b$ & $_1( \, _2( \, _3( \, a_4 \, b_5 \, \underbracket[0.75pt]{)_3 \, )_2 \, _2( \, _3( \, a_4}_5 \, b_5 \, )_3 \, )_2 \, )_1 \dashv$ \\ \bottomrule
 					\multicolumn{3}{c}{\scriptsize list continues on the right}
 				\end{tabular}
 				&
 				\tabcolsep=0.0cm\def\arraystretch{1.125}
 				\begin{tabular}{c@{\hspace{0.25cm}}l@{\hspace{0.375cm}}l}
 					$\#$ & \textbf{string} & \textbf{linearized syntax tree} -- LST \\ \toprule
 					$6$ & $a \, b \, a$ & $_1( \, _2( \, _3( \, a_4 \, b_5 \, \underbracket[0.75pt]{)_3 \, )_2 \, _2( \, a_6}_6 \, )_2 \, )_1 \dashv$ \\ \midrule
 					$7$ & $a \, a \, b$ & $_1( \,  _2( \, a_6 \, \underbracket[0.75pt]{)_2 \, _2( \, _3( \, a_4}_7 \, b_5 \, )_3 \, )_2 \, )_1 \dashv$ \\ \midrule
 					$8$ & $a \, a$ & $_1( \, _2( \, a_6 \, \underbracket[0.75pt]{)_2 \, _2( \, a_6}_8 \, )_2 \, )_1 \dashv$ \\ \midrule
 					$9$ & $a \, b$ & $_1( \, _2( \, _3( \, a_4 \, b_5 \, \underbracket[0.75pt]{)_3 \, )_2 \, )_1 \dashv}_9$ \\ \midrule
 					$10$ & $a$ & $_1( \, _2( \, a_6 \, \underbracket[0.75pt]{)_2 \, )_1 \dashv}_{10}$ \\ \bottomrule
 					\multicolumn{3}{c}{\scriptsize end of the list}
 				\end{tabular}
 				&
 				\tabcolsep=0.0cm\def\arraystretch{0.25}
 				\begin{tabular}{c}
 					$\arraycolsep=0.0cm\def\arraystretch{0.775}
 					\begin{array}{c@{\hspace{0.5cm}}l}
 						\emph{segment} & \mathit{FolSeg} \\ \toprule
 						1 & - \\ \midrule
 						2 & 4 \\ \midrule
 						3 & 7 \; \; 8 \; \; 10 \\ \midrule
 						4 & 5 \; \; 6 \; \; 9 \\ \midrule
 						5 & 4 \\ \midrule
 						6 & 7 \; \; 8 \; \; 10 \\ \midrule
 						7 & 4 \\ \midrule
 						8 & 7 \; \; 8 \; \; 10 \\ \midrule
 						9 & - \\ \midrule
 						10 & - \\ \bottomrule
 					\end{array}$ \\ \\
 					\text{\scriptsize follower seg.s -- Eq.~\eqref{eq:folsegdef}}
 				\end{tabular}
 		\end{tabular}}
 	\end{center}
 	\caption{Shortest strings and LSTs where each segment of RE $e_2$ (as in Tab.~\ref{tab:segments}) makes its appearance, and follower segments.} \label{tab:strings}
 \end{table}
\subsubsection{Construction of the parser NFA} \label{subsubsec:constrParserNFA}
To construct the parser NFA , it is necessary to generalize the notion of follower from symbols to segments. Given an RE $e$, for every segment $\rho \in Q$, define the set $\mathit{FolSeg}$ of the \emph{follower segments} of $\rho$ as:
\begin{equation} \label{eq:folsegdef}
	\mathit{FolSeg} \, (e_\# \dashv, \, \rho) = \set{ \, \sigma \; \vert \quad \text{$\sigma \in Q$ \quad and \quad $(\text{first symbol of $\sigma$}) \in \mathit{Fol} \, (e_\# \dashv, \, \text{end-letter of $\rho$})$} \, }
\end{equation}
In practice,  Eq.~\eqref{eq:folsegdef} says that segment $\sigma$ is a follower of $\rho$, i.e., the segment sequence $\rho \, \sigma$ occurs in an LST if the initial symbol of the meta-prefix of $\sigma$ (or the end-letter of $\sigma$ if the meta-prefix is empty) follows (in the classical sense) the end-letter of $\rho$. After computing the segment set $Q$ for an RE $e$, it is a simple matter to compute all the sets $\mathit{FolSeg}$ for $e$ as of Eq.~\eqref{eq:folsegdef}, by reapplying the classic follower $\mathit{Fol}$ approach as of Eq.~\eqref{eq:classicfollowers}, and we omit the obvious algorithm.
\par
The non-ambiguous RE $e_2$, numbered  $e_{2\#}$,  is shown in Fig.~\ref{fig:nonambiguousREclassicFollowers}, along with the structure tree, and the (classic) followers as of Eq.~\eqref{eq:classicfollowers}. All the segments for $e_2$ are listed in Tab.~\ref{tab:segments}. There are three initial, three final (one is both initial and final) and five internal segments. Tab.~\ref{tab:strings} identifies the segments  in the LSTs, and tabulates  their followers defined by  Eq.~\eqref{eq:folsegdef}. Tab.~\ref{tab:strings} lists the ten shortest  strings of $L \, (e_2)$ and their LSTs (exactly one per string since RE $e_2$ is not ambiguous) where each segment shows up.
\par
\begin{figure}[ht]
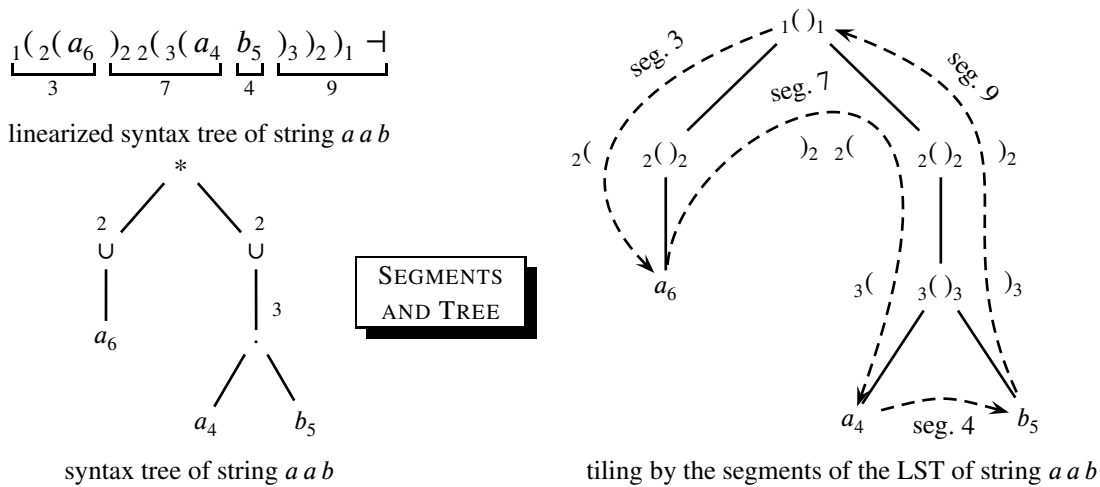

	\begin{center}
		\tabcolsep=0.0cm\def\arraystretch{1.75}
		\begin{tabular}{c@{\hspace{2.5cm}}c}
			\begin{tabular}{c}
				\large $\underbracket[0.75pt]{_1( \, _2( \, a_6}_3 \; \underbracket[0.75pt]{)_2 \, _2( \, _3( \, a_4}_7 \; \underbracket[0.75pt]{b_5}_4 \; \underbracket[0.75pt]{)_3 \, )_2 \, )_1 \dashv}_9$ \\ linearized syntax tree of string $a\,a\,b$ \\
				\tabcolsep=0.0cm\def\arraystretch{0}
				\rnode {tree2} {
					\begin{tabular}{c}
						\psset{arrows=-, linestyle=solid, linewidth=1pt, levelsep=1.125cm, treesep=1.0cm, nodesep=4pt, treefit=loose, tpos=0.675}
						\pstree{\TR{$\ast$}} {
							\pstree{\TR{$\cup$} \tlput{\footnotesize 2}}{
								\TR{$a_6$}
							}
							\pstree{\TR{$\cup$} \trput{\footnotesize 2}}{
								\pstree{\TR{$.$} \trput{\footnotesize 3}}{
									\TR{$a_4$}
									\TR{$b_5$}
								}
							}
						}
				\end{tabular}}
				\\
				syntax tree of string $a\,a\,b$
			\end{tabular}
			&
			\tabcolsep=0.0cm\def\arraystretch{1.875}
			\begin{tabular}{c}
				\tabcolsep=0.0cm\def\arraystretch{0}
				\begin{tabular}{c}
					\psset{arrows=-, linestyle=solid, linewidth=1pt, levelsep=1.75cm, treesep=2.0cm, nodesep=4pt, treefit=loose}
					\pstree{\TR[name=0]{$_1( \ )_1$}}{
						\pstree{\TR[name=1]{$_2( \ )_2$}}{
							\TR[name=2]{$a_6$}
						}
						\pstree{\TR[name=3]{$_2( \ )_2$}}{
							\pstree{\TR[name=4]{$_3( \ )_3$}}{
								\TR[name=5]{$a_4$}
								\TR[name=6]{$b_5$}
							}
						}
					}
					\nccurve[arrows=->, arrowscale=1.5, linestyle=dashed, linewidth=1pt, nodesepA=3pt, nodesepB=3pt, angleA=-160, angleB=135, ncurvA=1.25, ncurvB=1.0] 0 2 \nbput[nrot=:D,  npos=0.25]{seg.~$3$} \nbput[npos=0.59]{$_2($}
					\nccurve[arrows=->, arrowscale=1.5, linestyle=dashed, linewidth=1pt, nodesepA=6pt, nodesepB=3pt, angleA=85, angleB=70, ncurvA=1.25, ncurvB=4.0] 2 5 \naput[nrot=:U, npos=0.45]{seg.~$7$} \nbput[npos=0.4375]{$)_2$} \nbput[npos=0.57]{$_2($} \nbput[npos=0.9005]{$_3($}
					\nccurve[arrows=->, arrowscale=1.5, linestyle=dashed, linewidth=1pt, nodesepA=6pt, nodesepB=3pt, angleA=15, angleB=165, ncurvA=1.0, ncurvB=1.0] 5 6 \nbput[nrot=:U]{seg.~$4$}
					\nccurve[arrows=->, arrowscale=1.5, linestyle=dashed, linewidth=1pt, nodesepA=6pt, nodesepB=3pt, angleA=115, angleB=-20, ncurvA=1.0, ncurvB=1.25] 6 0 \nbput[nrot=:D,  npos=0.75]{seg.~$9$} \nbput[npos=0.2115]{$)_3$} \nbput[npos=0.55]{$)_2$}
				\end{tabular}
				\\
				tiling by the segments of the LST of string $a\,a\,b$
			\end{tabular}
		\end{tabular}
		\nput[labelsep=0.5cm] {0} {tree2} {\shadowbox{\parbox{2.0cm} {\centering \textsc{Segments} \par \textsc{and Tree}}}}
	\end{center}
	\caption{Sample segmented syntax tree for the valid string $a\,a\,b$ of RE $e_2$ of Ex.~$2$ (the end-marker $\dashv$ is omitted).} \label{fig:segtree}
\end{figure}

We anticipate that the segment sets $Q$, $I$ and $F$  respectively denote the sets of all the states, of the initial states and of the final states for the parser NFA, while the follower segment sets $\mathit{FolSeg}$  provide all the state transitions.
More insight on the relationship between  syntax trees and segments is provided in Fig.~\ref{fig:segtree}. An LST is shown for $e_2$ (left), factored into four segments from Tab.~\ref{tab:segments}. The syntax tree is shown on the left and also on the right where each inner node is  labeled with the numbered parenthesis of the corresponding operator.  The dashed lines display the preorder tree visit path with segments as label.
\paragraph{Local properties of LSTs}
To understand why the parser NFA construction works, it is important to observe that the language of the LSTs  is \emph{local}, since  it is exactly defined by the relation between each numbered symbol and the symbols that may follow it~\cite{DBLP:series/txcs/ReghizziBM19}, i.e., by the sets $\mathit{Fol}$. Through a process of abstraction, we may view an LST as a string over the finite alphabet of segments, rather than over the alphabet of numbered symbols, e.g., for RE $e_2$ the alphabet comprising the ten segments  listed in Tab.~\ref{tab:strings}. In this way, the language of the LSTs is exactly defined by the relation between each segment and its follower segments,  i.e., by the sets $\mathit{FolSeg}$ as of Eq.~\eqref{eq:folsegdef}, plus the initial and final segments; hence it is a local language  also at this level of abstraction. The rationale is the same underlying the classic GMY  RE-to-NFA construction: each numbered symbol occurs in the RE exactly once, in a precise position.
\par
We  proceed to build the parser NFA for RE $e_2$. Its states  coincide with the segments of the  numbered RE, i.e., with the set $Q$ computed by the algorithm of Fig.~\ref{fig:segmentalgo}. The initial and final states of the NFA are the initial and final segments in sets $I$ and $F$, respectively, computed by the algorithm. Two states are connected by a transition if the segment of the source state is followed by the segment of the target state,  as specified by the set $\mathit{FolSeg}$ of Eq.~\eqref{eq:folsegdef}. The label on each arc is a terminal  to be read by the NFA; it coincides with the end-letter, without numbering, of the source segment.
\par
Adopting a standard terminology for automata, we say that a segment $q$, i.e., a state, is \emph{reachable} from a segment $p$ if there is a path from $p$ to $q$; a segment is \emph{accessible} if it is reachable from an initial one; it is \emph{post-accessible} if a final segment can be reached from it;  a segment is \emph{useful} if it is both accessible and post-accessible, otherwise it is \emph{useless}. The parser NFA for RE $e_2$  is shown in Fig.~\ref{fig:NFA-RE2}.
\begin{figure}[ht]
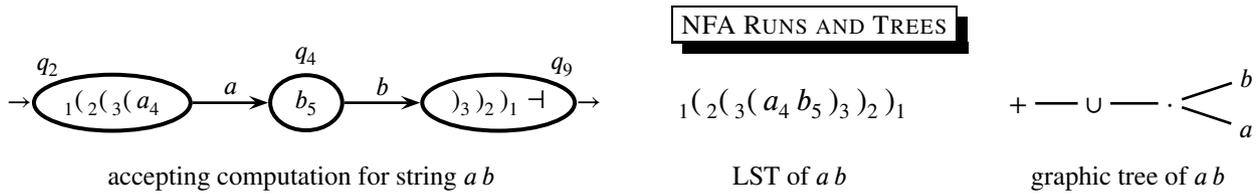

	\begin{center}
		\tabcolsep=0.0cm\def\arraystretch{0}
		\begin{tabular}[c]{c@{\hspace{1.25cm}}c@{\hspace{1.25cm}}c}
			& \multicolumn{2}{l}{\shadowbox{\textsc{NFA Runs and Trees}}} \\[0.2cm]
			\tabcolsep=0.0cm\def\arraystretch{0}
			\begin{tabular}[c]{c}
				\psset{nodesep=0pt, linewidth=1.5pt, colsep=1.0cm, nodealign=true, linecolor=black, labelsep=2pt, arrows=->, arrowscale=1.5, npos=0.5}
				\begin{psmatrix}
					\ovalnode {T1} {$_1( \, _2( \, _3( \, a_4$} & \ovalnode {T2} {$ \ b_5 \ $} & \ovalnode {T3} {$)_3 \, )_2 \, )_1 \dashv$}
					
					\ncline[linewidth=1pt] {T1} {T2} \naput{$a$}
					
					\ncline[linewidth=1pt] {T2} {T3} \naput{$b$}
					
					\nput[labelsep=0pt] {180} {T1} {$\to$}
					
					\nput[labelsep=0pt] {0} {T3} {$\to$}
					
					\nput[labelsep=3pt] {150} {T1} {$q_2$}
					
					\nput[labelsep=3pt] {90} {T2} {$q_4$}
					
					\nput[labelsep=3pt] {30} {T3} {$q_9$}
				\end{psmatrix}
			\end{tabular}
			&
			\tabcolsep=0.0cm\def\arraystretch{0}
			\begin{tabular}[c]{c}
				\large $_1( \, _2( \, _3( \, a_4 \, b_5 \, )_3 \, )_2 \, )_1$
			\end{tabular}
			&
			\tabcolsep=0.0cm\def\arraystretch{0}
			\begin{tabular}[c]{c}
				\psset{treemode=R, nodesep=3pt, linewidth=1pt, arrows=-, levelsep=1.0cm, treesep=0.5cm}
				\pstree{\TR{$+$}} {
					\pstree{\TR{$\cup$}} {
						\pstree{\TR{$\cdot$}} {
							\TR{$b$}
							\TR{$a$}
						}
					}
				}
			\end{tabular}
			\\[0.75cm]
			accepting computation for string $a\,b$
			&
			LST of $a\,b$
			&
			graphic tree of $a\,b$
		\end{tabular}
	\end{center}
	\caption{Accepting run of the parser NFA for RE $e_2$, and corresponding LST and graphic tree for string $a\,b$.} \label{fig:NFAcomputation}
\end{figure}
\par\noindent
Clearly, the device acts as a \emph{parser}: the sequence of segments traversed by an accepting run over an  input text is the LST. For instance the run in Fig.~\ref{fig:NFAcomputation} reads  the input string $a\,b$ of RE $e_2$ and yields the LST $_1( \, _2( \, _3( \, a_4 \, b_5 \, )_3 \, )_2 \, )_1$; see also Tab.~\ref{tab:strings}. Since by construction all and only the LSTs of the RE are encoded in the accepting runs, the  NFA  is correct. We observe that, being the device nondeterministic, an input string may have two or more accepting runs, hence two or more LSTs. Such a case may only arise when the RE is ambiguous and is now considered.
\subsubsection{The case of ambiguous RE} \label{subsubsec:case-ambiguousRE}
\begin{figure}[ht]
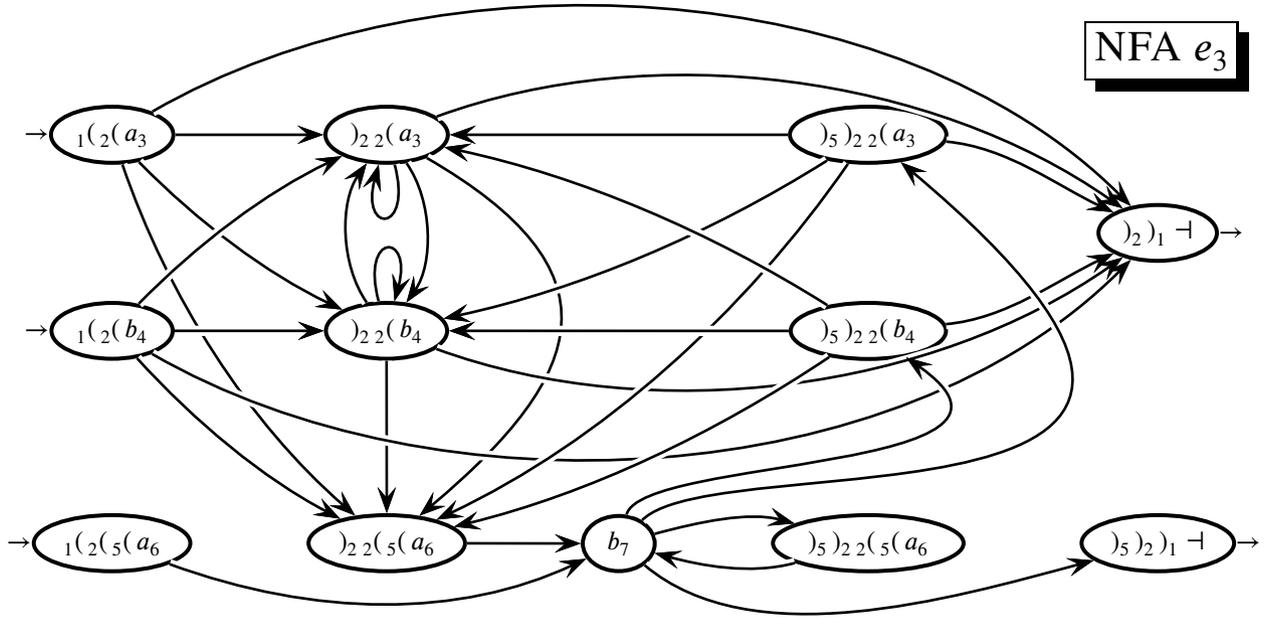

	\vspace{1.5cm}
	\begin{center}
		\rnode {NFA2} {
			\psset{nodesep=0.0cm, arrows=->, linewidth=1.5pt, arrowscale=2, rowsep=0.5cm, colsep=1.5cm, border=0.00cm}
			\begin{psmatrix}
				\ovalnode {123} {$_1( \, _2( \, a_3$} & \ovalnode {223} {$)_2 \, _2( \, a_3$} && \ovalnode {5223} {$)_5 \, )_2 \, _2( \, a_3$} \\
				
				& & & & \ovalnode {21dashv} {$)_2 \, )_1 \dashv$} \\
				
				\ovalnode {124} {$_1( \, _2( \, b_4$} & \ovalnode {224} {$)_2 \, _2( \, b_4$} && \ovalnode {5224} {$)_5 \, )_2 \, _2( \, b_4$} \\[1.5cm]
				
				\ovalnode {1256} {$_1( \, _2( \,_5( \, a_6$} & \ovalnode {2256} {$)_2 \, _2( \, _5( \, a_6$} & \ovalnode {7} {$ \ b_7 \ $} & \ovalnode {52256} {$)_5 \, )_2 \, _2( \,_5( \, a_6$} & \ovalnode {521dashv} {$)_5 \, )_2 \, )_1 \dashv$}
				
				\psset{labelsep=0pt}
				
				\nput {180} {123} {$\to$}
				
				\nput {180} {124} {$\to$}
				
				\nput {180} {1256} {$\to$}
				
				\nput {0} {21dashv} {$\to$}
				
				\nput {0} {521dashv} {$\to$}
				
				\psset{labelsep=3pt, arcangle=20, linewidth=1pt}
				
				\psset{border=0.05cm}
				
				\nccurve[angleA=30, angleB=135] {123} {21dashv}
				
				\nccurve[angleA=20, angleB=145] {223} {21dashv}
				
				\ncline {123} {223}
				
				\ncarc[arcangle=-10] {123} {224}
				
				\nccurve[angleA=-30, angleB=40, ncurv=1.25] {223} {2256}
				
				\ncline {224} {2256}
				
				\ncarc[arcangle=-15] {123} {2256}
				
				\ncarc[arcangle=10] {124} {223}
				
				\ncline {124} {224}
				
				\ncarc[arcangle=-10] {124} {2256}
				
				\nccurve[angleA=-30, angleB=-135] {124} {21dashv}
				
				\nccurve[angleA=-20, angleB=-145] {224} {21dashv}
				
				\ncarc[arcangle=35] {223} {224}
				
				\ncarc[arcangle=35] {224} {223}
				
				\nccurve[angleA=-75, angleB=-105, ncurv=9] {223} {223}
				
				\nccurve[angleA=105, angleB=75, ncurv=9] {224} {224}
				
				\ncline {2256} {7}
				
				\nccurve[angleA=-20, angleB=-155, ncurvA=0.75] {1256} {7}
				
				\ncarc[arcangle=15] {7} {52256}
				
				\ncarc[arcangle=15] {52256} {7}
				
				\nccurve[angleA=-40, angleB=-165, ncurvA=0.625] {7} {521dashv}
				
				\nccurve[angleA=-5, angleB=155] {5223} {21dashv}
				
				\nccurve[angleA=5, angleB=-155] {5224} {21dashv}
				
				\ncline {5223} {223}
				
				\ncarc[arcangle=10] {5223} {224}
				
				\ncarc[arcangle=15] {5223} {2256}
				
				\ncarc[arcangle=-10] {5224} {223}
				
				\ncline {5224} {224}
				
				\ncarc[arcangle=10] {5224} {2256}
				
				\nccurve[angleA=45, angleB=-40, ncurvA=0.5, ncurvB=2.75] {7} {5223}
				
				\nccurve[angleA=75, angleB=-35, ncurvA=0.375, ncurvB=1.25] {7} {5224}
		\end{psmatrix}}
		\nput[labelsep=0.375cm] {30} {NFA2} {\shadowbox{\Large \textsc{NFA $e_3$}}}
	\end{center}
	\vspace{0.5cm}
	\caption{Parser NFA  of RE $e_3$. Transition labels are redundant as they coincide with the end-letter of the source-state segment  (without numbering):   e.g., the  transition  ``$_1( \, _2( \, b_4\;\to \;)_2 \, _2( \, a_3$'' reads letter $b$.}
	\label{fig:NFA3}
\end{figure}

The above construction of the parser NFA works also for  finitely ambiguous REs, with the significant difference that the parser has to return more than one LST for the input text.  We focus on this aspect and explain by means of an example how the multiple LSTs are represented in the  \emph{shared linearized parse forest} (SLPF).
\paragraph{Example $3$ -- parsing an ambiguous RE}
Given the  ambiguous RE $e_3$, in Fig.~\ref{fig:NFA3} we show the parser NFA, and in Fig.~\ref{fig:SLPFe3} the LSTs and their SLPF for the input $x = a\,b\,a\,b$:
\begin{center}
	\vspace{0.125cm}
	\tabcolsep=0.0cm\def\arraystretch{0}
	\begin{tabular}{c@{\hspace{2.0cm}}c}
		\tabcolsep=0.15cm\def\arraystretch{1.25}
		\begin{tabular}{rcl}
			\large $e_3$ & \large $=$ & \large $( \, a \; \mid \; b \; \mid \; a \, b \, )^+$
			\\
			\large $e_{3\#}$ & \large $=$ & \large $_1( \; _2( \; a_3 \; \mid \; b_4 \; \mid \; _5( \; a_6 \; b_7 \; )_5 \; )_2 \; {)_1}^+$
		\end{tabular}
		&
		\tabcolsep=0.0cm\def\arraystretch{0.75}
		\begin{tabular}[c]{c@{\hspace{0.5cm}}l}
			\emph{num} & RE \emph{operator} \\ \toprule
			$1$ & Kleene cross \\ \midrule
			$2$ & union (ternary) \\ \midrule
			$5$ & concatenation \\ \bottomrule
		\end{tabular}
	\end{tabular}
\end{center}
The SLPF is a directed acyclic graph (DAG) that embeds all the NFA computations for the input string. The complete SLPF of string $x = a\,b\,a\,b$ is shown in Fig.~\ref{fig:SLPFe3}; it represents exactly four LSTs, also listed at the bottom.  The SLPF nodes are layered in a series $C_0, \ldots, C_4$ of columns,  each one containing a set of NFA states (segments). Each path from a node in the initial column $C_0$ to one in the final column $C_4$ represents an LST. The solid line part of this SLPF is \emph{clean} in the sense that  all its nodes are useful since they lie on at least one accepting NFA run. Just for explanation, two useless NFA nodes are shown (dashed) in Fig.~\ref{fig:SLPFe3}: the left one is accessible but not post-accessible, and conversely the right one is post-accessible but not accessible. \qed
\par
\begin{figure}[ht]
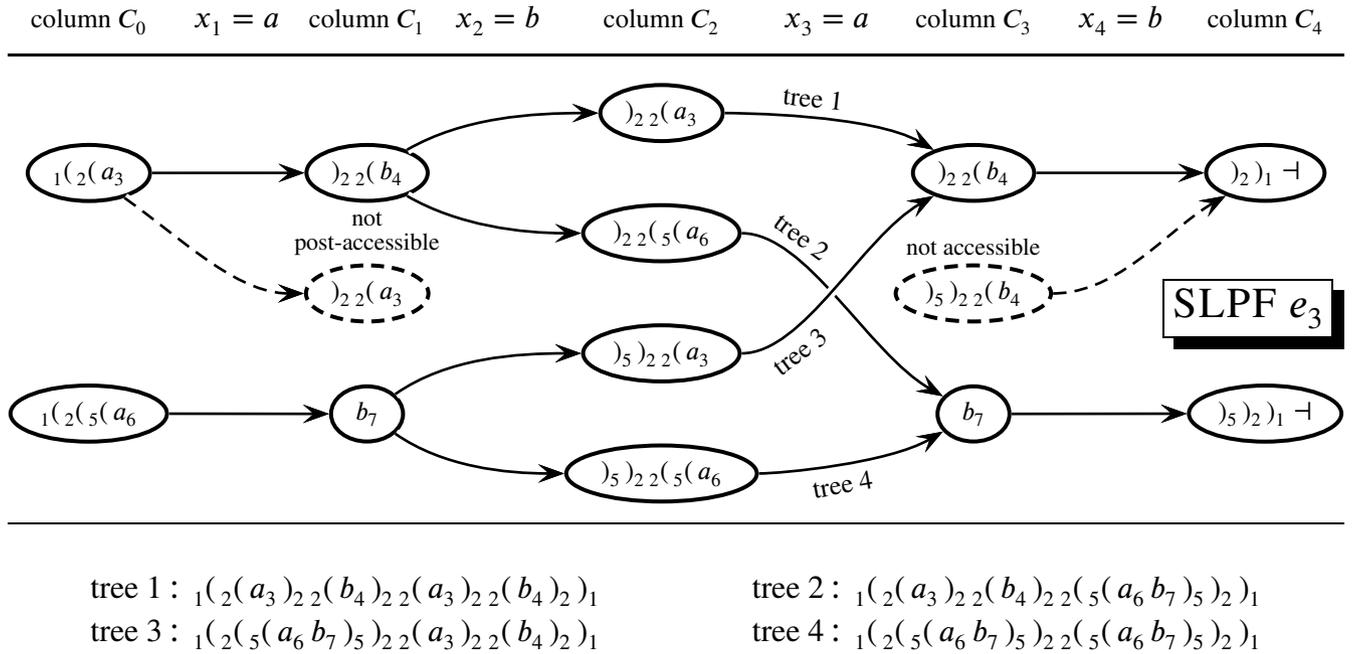

	\vspace{0.5cm}
	\begin{center}
		\rnode{SLPF} {
			\psset{nodesep=0.0cm, arrows=->, linewidth=1.5pt, arrowscale=2, rowsep=0.0cm, colsep=0.325cm, border=0.00cm}
			\begin{psmatrix}
				column $C_0$ & \large $x_1 = a$ & column $C_1$ & \large $x_2 = b$ & column $C_2$ & \large $x_3 = a$ & column $C_3$ & \large $x_4 = b$ & column $C_4$ \\[0.25cm] \hline \\[0.25cm]
				
				& & & & \ovalnode {223ab} {$)_2 \, _2( \, a_3$} & & & & \\
				
				\ovalnode {123i} {$_1( \, _2( \, a_3$} & & \ovalnode {224a} {$)_2 \, _2( \, b_4$} & & & & \ovalnode {224aba} {$)_2 \, _2( \, b_4$} & & \ovalnode {21dashvabab} {$)_2 \, )_1 \dashv$} \\
				
				& & & & \ovalnode {2256ab} {$)_2 \, _2( \, _5( \, a_6$} & & & & \\
				
				& & \ovalnode[linestyle=dashed] {undefined} {$)_2 \, _2( \, a_3$} & & & & \ovalnode[linestyle=dashed] {unreachable} {$)_5 \, )_2 \, _2( \, b_4$} & & \\
				
				& & & & \ovalnode {5223ab} {$)_5 \, )_2 \, _2( \, a_3$} & & & & \\
				
				\ovalnode {1256i} {$_1( \, _2( \,_5( \, a_6$} & & \ovalnode {7a} {$ \ b_7 \ $} & & & & \ovalnode {7aba} {$ \ b_7 \ $} & & \ovalnode {521dashvabab} {$)_5 \, )_2 \, )_1 \dashv$} \\
				
				& & & & \ovalnode {52256ab} {$)_5 \, )_2 \, _2( \,_5( \, a_6$} & & & & \\[0.25cm] \hline \\
				
				\psset{linewidth=1pt, border=0.05cm}
				
				\ncline {123i} {224a}
				
				\ncline {1256i} {7a}
				
				\nccurve[angleA=30, angleB=-180] {224a} {223ab}
				
				\nccurve[angleA=-30, angleB=180] {224a} {2256ab}
				
				\nccurve[angleA=35, angleB=-180] {7a} {5223ab}
				
				\nccurve[angleA=-35, angleB=180] {7a} {52256ab}
				
				\nccurve[angleA=-0, angleB=150, ncurvA=0.125] {223ab} {224aba} \naput[labelsep=3pt, nrot=:U]{tree $1$}
				
				\nccurve[angleA=-0, angleB=150, ncurvA=0.5] {2256ab} {7aba} \naput[labelsep=3pt, nrot=:U, npos=0.25]{tree $2$}
				
				\nccurve[angleA=0, angleB=-150, ncurvA=0.5] {5223ab} {224aba} \nbput[labelsep=3pt, nrot=:U, npos=0.25]{tree $3$}
				
				\nccurve[angleA=0, angleB=-150, ncurvA=0.25] {52256ab} {7aba} \nbput[labelsep=3pt, nrot=:U]{tree $4$}
				
				\ncline {224aba} {21dashvabab}
				
				\ncline {7aba} {521dashvabab}
				
				\psset{linestyle=dashed, linewidth=1pt, border=0.05cm}
				
				\nccurve[angleA=-35, angleB=180] {123i} {undefined}
				
				\nccurve[angleA=0, angleB=-150] {unreachable} {21dashvabab}
				
				\psset{labelsep=3pt}
				
				\nput[nrot=:U] {90} {undefined} {\parbox{2.0cm}{\centering \scriptsize not post-accessible}}
				
				\nput[nrot=:U] {90} {unreachable} {\parbox{2.0cm}{\centering \scriptsize not accessible}}
		\end{psmatrix}}
		\nput[labelsep=-2.375cm] {-3} {SLPF} {\shadowbox{\Large \textsc{SLPF $e_3$}}}
		\tabcolsep=0.0cm\def\arraystretch{1.25}
		\begin{tabular}{l@{\hspace{2.0cm}}l}
			\large tree $1 \colon _1( \, _2( \, a_3 \, )_2 \, _2( \, b_4 \, )_2 \, _2( \, a_3 \, )_2 \, _2( \, b_4 \, )_2 \, )_1$
			&
			\large tree $2 \colon _1( \, _2( \, a_3 \, )_2 \, _2( \, b_4 \, )_2 \, _2( \, _5( \, a_6 \, b_7 \, )_5 \, )_2 \, )_1$
			\\
			\large tree $3 \colon _1( \, _2( \, _5( \, a_6 \, b_7 \, )_5 \, )_2 \, _2( \, a_3 \, )_2 \, _2( \, b_4 \, )_2 \, )_1$
			&
			\large tree $4 \colon _1( \, _2( \, _5( \, a_6 \, b_7 \, )_5 \, )_2 \, _2( \, _5( \, a_6 \, b_7 \, )_5 \, )_2 \, )_1$
		\end{tabular}
	\end{center}
	\vspace{0.0cm}
	\caption{SLPF of the ambiguous string $x = a\,b\,a\,b$ of RE $e_3$. The final  clean form (solid nodes and arcs)  encodes four LSTs. Two non-valid NFA nodes (among many others omitted)  are  shown (dashed), which are not on a complete LST path.}
	\label{fig:SLPFe3}
\end{figure}
\noindent
For convenience of the readers we sum up the aspects that are relevant for parsing:
\begin{itemize}[leftmargin=*]
	\item The states of the parser NFA and the RE segments  are in one-to-one correspondence.
	\item  Each  accepting run  defines an LST as the concatenation of the segments traversed.
	\item   The SLPF forest of a valid  text $x = x_1 \, \ldots \, x_n$ or $x = \varepsilon$, of length $\vert \, x \, \vert = n \geq 0$, is a DAG, the nodes of which are segments; the segments are placed in a series $C_0, C_1, \ldots, C_n$ of   columns, and each column contains a non-empty set of segments.
	\item Columns $C_0$ and $C_n$ contain only initial and final segments, respectively.  If $x = \varepsilon$ there is only one column $C_0$.
	\item Each arc of the SLPF connects a segment in  column $C_r$ to one in  column $C_{r+1}$ (arcs may not span   distant  columns).
	\item The clean SLPF of $x$ has an arc from a \emph{source} segment in $C_r$ to a \emph{target} one in $C_{r+1}$ if, and only if, the accepting run of $x$ has an arc with label $x_{r + 1}$ from the source to the target segment. If $x = \varepsilon$, each initial-final segment in $C_0$ encodes an LST for $\varepsilon$.
	\item  The \emph{clean} SLPF of a valid text  encodes all and only the LSTs of the text as segment paths from initial  to final column.
\end{itemize}
\subsection{Serial parsing algorithm} \label{subsec:serialalgorithmVectorial}
We show how to  build the straightforward (but inefficient) serial RE parser  driven by the parser NFA, to be  used as baseline for the parallel parser construction.  To this end, we formalize the NFA state-machine operation by using the classical matrix notation; e.g., see in~\cite{Sakarovitch2009, DBLP:conf/asplos/MytkowiczMS14} for a similar use in parallel DFA recognizers. Given an RE $e$ over an alphabet $\Sigma$ and the associated numbered RE $e_\#$, the  definitions for the simple serial parser follow:
\begin{description}[leftmargin=1.75cm, style=multiline]
	\item[set $Q$] The non-empty set of all the segments of the numbered RE $e_\#$ associated to the RE $e$:
	\[
	Q = \set{ \, q \; \vert \quad \text{$q$ is a segment of $e$} \, }
	\]
	The number of segments is $\ell = \vert \, Q \, \vert \geq 1$ and each segment is denoted $q_j$ or simply $j$ (with $1 \leq j \leq \ell$).
	The non-empty sets $I, F \subseteq Q$ store the initial and final segments, respectively. The powerset of $Q$ is denoted $2^Q$.
	\item[NFA] The nondeterministic  FA of RE $e$, obtained through the generalized GMY method applied to RE $e_\#$ (Sect.~\ref{subsubsec:constrParserNFA}):
	\[
	\text{NFA} = \big( \, Q, \, \Sigma, \, I, \, F, \delta_\text{NFA} \, \big) \qquad \text{with $\delta_\text{NFA} \colon Q \times \Sigma \to 2^Q$}
	\]
	where $Q$ is the state set, $\Sigma$ is the alphabet, $I$ and $F$ are  initial and final states, respectively, and $\delta_\text{NFA}$ is the (nondeterministic) transition function.
	\item[input text] A string $x = x_1 \ldots x_n$, $n \geq 1$, to be parsed, where $x_r$, $1 \leq r \leq n$, is the $r$-th character (terminal) of $x$. The trivial case $x = \varepsilon$ is omitted.
	\item[SLPF]
	The DAG that encodes the LSTs of the  text $x$. It is represented as a series of $n + 1$ columns $C= C_0 \, C_1 \, \ldots \ C_n$, where $C_r \subseteq Q$ is a segment set ($0 \leq r \leq n$). Column $C_r$ (with $r \not = 0$ and $r \not = n$), stores the segments that belong to an LST of the input $x$ and are located between characters $x_r$ and $x_{r+1}$. Columns $C_0$ and $C_n$  contain the segments at the input start and end, respectively; see also Fig.~\ref{fig:SLPFe3}. The arcs from a segment to the next one of the LST are not represented in the DAG, as they are already available in the NFA graph.
\end{description}
In the matrix notation, a subset $V$ of the state set $Q$  is  a Boolean column vector $\mathbf V$ of size $\ell = \vert \, Q \, \vert $: it holds $\mathbf V \, [j] = 1$ iff $q_j \in V$. The transition function of the NFA is represented by a collection of $\vert \, \Sigma \, \vert $ Boolean $\ell \times \ell$ (square) connection matrices $\mathbf N_a$, indexed on the rows and columns by the segments, each matrix being associated to a terminal $a \in \Sigma$. The matrix case $\mathbf N_a \, [\mathit{row}, \, \mathit{col}]$, with  $1 \leq \mathit{row}, \, \mathit{col} \leq \ell$, is equal to $1$ if and only if $q_\mathit{row} \in \delta_\text{NFA} \, (q_\mathit{col}, \, a)$, i.e., the NFA has an arc labeled  $a$ from segment $q_\mathit{col}$ to segment $q_\mathit{row}$. The following data-structures are used by the serial parser:
\begin{description}[leftmargin=2.125cm, style=multiline]
	\item[columns $\mathbf I$, $\mathbf F$] The two Boolean column vectors that represent the sets $I$ and $F$ of initial and final segments of the NFA.
	\item[matrix $\mathbf N$] The above described collection of $\vert \, \Sigma \, \vert$ Boolean connection matrices $\mathbf N_a$ ($a \in \Sigma$) that represent NFA transitions.
	\item[column $\mathbf C$] The above described series of $n + 1$ Boolean column vectors $\mathbf C_r$ ($0 \leq r \leq n$) that represent the SLPF graph.
\end{description}
The construction of the SLPF for  text $x$ is then immediately expressed in Eq.~\eqref{eq:matrixmul}, left, where the operation ``\,$\times$\,'' is Boolean matrix multiplication, and the computation scans the text from left to right:
\begin{equation} \label{eq:matrixmul}
	\underbracket[0.75pt]{\mathbf C_0 = \mathbf I \quad \text{and} \quad \mathbf C_r = \mathbf N_{x_r} \, \times \, \mathbf C_{r - 1}}_\text{scan left-to-right (forwards)} \hspace{0.5cm} 1 \leq r \leq n
	\hspace{1.5cm}
	\underbracket[0.75pt]{\mathbf{\widehat C}_n = \mathbf F \quad \text{and} \quad \mathbf{\widehat C}_r  = \mathbf{\widehat N}_{x_{r + 1}} \, \times \, \mathbf{\widehat C}_{r + 1}}_\text{scan right-to-left (backwards)} \hspace{0.5cm} n - 1 \geq r \geq 0
\end{equation}
\paragraph{Need of a reverse NFA}
The so-obtained column series $C$ represents the SLPF segments that are accessible, but not necessarily post-accessible. In other words, the SLPF  encodes all the LSTs of text $x$, but in general it is not clean. To obtain a clean SLPF, a similar construction is carried out by using the \emph{reverse} NFA, denoted by $\widehat{\text{NFA}}$:
\begin{equation}\label{eq:reverseNFAdef}
\text{$\widehat{\text{NFA}}$ with connection matrices for all $a \in \Sigma \colon \mathbf{\widehat N}_a = \mathbf N_a^T$ \qquad i.e., the transpose of $\mathbf N_a$}
\end{equation}
The initial and final segments of the  NFA are switched in the $\widehat{\text{NFA}}$. Then automaton $\widehat{\text{NFA}}$ computes a  \emph{backward column series} $\widehat C = \widehat C_0 \, \widehat C_1 \, \ldots \ \widehat C_n$ by scanning the text $x$ backwards, i.e., from right to left, starting from the final segments; see Eq.~\eqref{eq:matrixmul}, right. Upon termination of the backward scan,  the post-accessible SLPF segments are known. At last the columns of the clean SLPF are  obtained by intersecting the corresponding columns of the forward and backward series, i.e., $C_r \, \cap \, \widehat C_r$ (for $0 \leq r \leq n$). In fact, such columns refer to the same segments accessed or post-accessed between consecutive characters, or at the text endpoints.
\par The pseudo-code of the serial parser algorithm is listed in Fig.~\ref{fig:serialparser}: it parses the text two-way, forwards and backwards, then intersects the columns and outputs the clean SLPF.
\begin{figure}[ht]
	\begin{center}
		\definecolor{lightgray}{rgb}{0.875,0.875,0.875}
		\vspace{0.125cm}
		\tabcolsep=0.0cm
		\def\arraystretch{0.5}
		\begin{tabular}{p{0.405\textwidth}@{\hspace{0.5cm}}p{0.545\textwidth}} \toprule
			\cellcolor{lightgray}
			\begin{minipage}[t]{0.405\textwidth}
				\begin{algorithm2e}[H]
					\setstretch{1.125}
					\SetAlgorithmName{algorithm}{}{}
					\SetKwInOut{Input}{input}
					\SetKwInOut{Output}{output}
					\SetKwInOut{LocVars}{variable}
					\DontPrintSemicolon
					\TitleOfAlgo{\emph{serial parser}}
					\hrule height 0.25pt
					\BlankLine
					\Input{input text $x$, state sets $\mathbf I$ and $\mathbf F$,
						and transition matrix series $\mathbf N_a$ and $\mathbf{\widehat N}_a$}
					\Output{the clean SLPF of the input text}
					\LocVars{column series $\mathbf C$ and $\mathbf{\widehat C}$, index $r$}
				\end{algorithm2e}
			\end{minipage}
			&
			\cellcolor{lightgray}
			\begin{minipage}[t]{0.545\textwidth}
				\begin{algorithm2e}[H]
					\setstretch{1.125}
					\SetKwInOut{LocVars}{variable}
					\DontPrintSemicolon
					$\mathbf C \, [0] \asgn \mathbf I$ \tcp*[r]{initialize and parse forwards}
					\lFor {\text{\rm $r = 1$ \textbf{upto} $n$}} {
						$\mathbf C \, [r] \asgn \mathbf N \, \big[ x \, [r] \big] \, \times \, \mathbf C \, [r - 1]$
					}
					$\mathbf{\widehat C} \, [n] \asgn \mathbf F$ \tcp*[r]{initialize and parse backwards}
					\lFor {\text{\rm $r = n - 1$ \textbf{downto} $0$}} {
						$\mathbf{\widehat C} \, [r] \asgn \mathbf{\widehat N} \, \big[ x \, [r + 1] \big] \, \times \, \mathbf{\widehat C} \, [r + 1]$
					}
					\lFor (\tcp*[f]{intersect}) {\text{\rm $r = 0$ \textbf{upto} $n$}} {
						$\textbf{output} \, \big( \mathbf C \, [r] \, \cap \, \mathbf{\widehat C} \, [r] \big)$
					}
				\end{algorithm2e}
			\end{minipage} \\ \bottomrule
		\end{tabular}
	\end{center}
	\caption{ Pseudo-code of the serial parser algorithm (NFA-based), which outputs the clean SLPF of the input text.} \label{fig:serialparser}
\end{figure}
\par
\paragraph{Example $4$ -- serial NFA-based parser for RE $e_2$}
Consider the RE $e_2$ and its NFA in Fig.~\ref{fig:NFA-RE2}, with ten segments $q_1 \ldots q_{10}$ (see Tab.~\ref{tab:segments}). The matrix representation of the NFA is:
\begin{center}
\scriptsize
$\NiceMatrixOptions{cell-space-limits=1pt}
\mathbf{I} = \begin{bNiceMatrix}[first-col]
	1 \phantom{a} & 1 \\
	2 \phantom{a} & 1 \\
	3 \phantom{a} & 1 \\
	4 \phantom{a} & 0 \\
	5 \phantom{a} & 0 \\
	6 \phantom{a} & 0 \\
	7 \phantom{a} & 0 \\
	8 \phantom{a} & 0 \\
	9 \phantom{a} & 0 \\
	10 \phantom{a} & 0
\end{bNiceMatrix}
\hspace{1.5cm}
\mathbf{F} = \begin{bNiceMatrix}[first-col]
	1 \phantom{a} & 1 \\
	2 \phantom{a} & 0 \\
	3 \phantom{a} & 0 \\
	4 \phantom{a} & 0 \\
	5 \phantom{a} & 0 \\
	6 \phantom{a} & 0 \\
	7 \phantom{a} & 0 \\
	8 \phantom{a} & 0 \\
	9 \phantom{a} & 1 \\
	10 \phantom{a} & 1
\end{bNiceMatrix}
\hspace{1.5cm}
\mathbf{N}_a = \begin{bNiceMatrix}[first-row, first-col]
	\phantom{a}  & 1 & 2 & 3 & 4 & 5 & 6 & 7 & 8 & 9 & 10 \\
	1 \phantom{a}  & 0 & 0 & 0 & 0 & 0 & 0 & 0 & 0 & 0 & 0 \\
	2 \phantom{a}  & 0 & 0 & 0 & 0 & 0 & 0 & 0 & 0 & 0 & 0 \\
	3 \phantom{a}  & 0 & 0 & 0 & 0 & 0 & 0 & 0 & 0 & 0 & 0 \\
	4 \phantom{a}  & 0 & 1 & 0 & 0 & 1 & 0 & 1 & 0 & 0 & 0 \\
	5 \phantom{a}  & 0 & 0 & 0 & 0 & 0 & 0 & 0 & 0 & 0 & 0 \\
	6 \phantom{a}  & 0 & 0 & 0 & 0 & 0 & 0 & 0 & 0 & 0 & 0 \\
	7 \phantom{a}  & 0 & 0 & 1 & 0 & 0 & 1 & 0 & 1 & 0 & 0 \\
	8 \phantom{a}  & 0 & 0 & 1 & 0 & 0 & 1 & 0 & 1 & 0 & 0 \\
	9 \phantom{a}  & 0 & 0 & 0 & 0 & 0 & 0 & 0 & 0 & 0 & 0 \\
	10 \phantom{a} & 0 & 0 & 1 & 0 & 0 & 1 & 0 & 1 & 0 & 0
\end{bNiceMatrix}
\hspace{1.5cm}
\mathbf{N}_b = \begin{bNiceMatrix}[first-row, first-col]
	\phantom{a}  & 1 & 2 & 3 & 4 & 5 & 6 & 7 & 8 & 9 & 10 \\
	1 \phantom{a}  & 0 & 0 & 0 & 0 & 0 & 0 & 0 & 0 & 0 & 0 \\
	2 \phantom{a}  & 0 & 0 & 0 & 0 & 0 & 0 & 0 & 0 & 0 & 0 \\
	3 \phantom{a}  & 0 & 0 & 0 & 0 & 0 & 0 & 0 & 0 & 0 & 0 \\
	4 \phantom{a}  & 0 & 0 & 0 & 0 & 0 & 0 & 0 & 0 & 0 & 0 \\
	5 \phantom{a}  & 0 & 0 & 0 & 1 & 0 & 0 & 0 & 0 & 0 & 0 \\
	6 \phantom{a}  & 0 & 0 & 0 & 1 & 0 & 0 & 0 & 0 & 0 & 0 \\
	7 \phantom{a}  & 0 & 0 & 0 & 0 & 0 & 0 & 0 & 0 & 0 & 0 \\
	8 \phantom{a}  & 0 & 0 & 0 & 0 & 0 & 0 & 0 & 0 & 0 & 0 \\
	9 \phantom{a}  & 0 & 0 & 0 & 1 & 0 & 0 & 0 & 0 & 0 & 0 \\
	10 \phantom{a}  & 0 & 0 & 0 & 0 & 0 & 0 & 0 & 0 & 0 & 0
\end{bNiceMatrix}$
\end{center}
For the input text $x = a \, b$ of length $n = 2$, the computation (forwards) is listed below:
\begin{center}
\scriptsize
$\NiceMatrixOptions{cell-space-limits=1pt}
\mathbf C_0 = \mathbf I = \begin{bNiceMatrix}[last-col=2]
	1 & \phantom{a} 1 \\ 1 & \phantom{a} 2 \\ 1 & \phantom{a} 3 \\ 0 \\ 0 \\ 0 \\ 0 \\ 0 \\ 0 \\ 0
\end{bNiceMatrix}
\hspace{0.625cm}
\mathbf{C}_1 = \mathbf{N}_a \times \mathbf{C}_0 =
\begin{bNiceMatrix}
	0 & 0 & 0 & 0 & 0 & 0 & 0 & 0 & 0 & 0 \\
	0 & 0 & 0 & 0 & 0 & 0 & 0 & 0 & 0 & 0 \\
	0 & 0 & 0 & 0 & 0 & 0 & 0 & 0 & 0 & 0 \\
	0 & 1 & 0 & 0 & 1 & 0 & 1 & 0 & 0 & 0 \\
	0 & 0 & 0 & 0 & 0 & 0 & 0 & 0 & 0 & 0 \\
	0 & 0 & 0 & 0 & 0 & 0 & 0 & 0 & 0 & 0 \\
	0 & 0 & 1 & 0 & 0 & 1 & 0 & 1 & 0 & 0 \\
	0 & 0 & 1 & 0 & 0 & 1 & 0 & 1 & 0 & 0 \\
	0 & 0 & 0 & 0 & 0 & 0 & 0 & 0 & 0 & 0 \\
	0 & 0 & 1 & 0 & 0 & 1 & 0 & 1 & 0 & 0
\end{bNiceMatrix}
\times
\begin{bNiceMatrix}
	1 \\ 1 \\ 1 \\ 0 \\ 0 \\ 0 \\ 0 \\ 0 \\ 0 \\ 0
\end{bNiceMatrix}
=
\begin{bNiceMatrix}[last-col=2]
	0 \\ 0 \\ 0 \\ 1 & \phantom{a} 4 \\ 0 \\ 0 \\ 1 & \phantom{a} 7 \\ 1 & \phantom{a} 8 \\ 0 \\ 1 & \phantom{a} 10
\end{bNiceMatrix}
\hspace{0.625cm}
\mathbf{C}_2 = \mathbf{N}_b \times \mathbf{C}_1 =
\begin{bNiceMatrix}
	0 & 0 & 0 & 0 & 0 & 0 & 0 & 0 & 0 & 0 \\
	0 & 0 & 0 & 0 & 0 & 0 & 0 & 0 & 0 & 0 \\
	0 & 0 & 0 & 0 & 0 & 0 & 0 & 0 & 0 & 0 \\
	0 & 0 & 0 & 0 & 0 & 0 & 0 & 0 & 0 & 0 \\
	0 & 0 & 0 & 1 & 0 & 0 & 0 & 0 & 0 & 0 \\
	0 & 0 & 0 & 1 & 0 & 0 & 0 & 0 & 0 & 0 \\
	0 & 0 & 0 & 0 & 0 & 0 & 0 & 0 & 0 & 0 \\
	0 & 0 & 0 & 0 & 0 & 0 & 0 & 0 & 0 & 0 \\
	0 & 0 & 0 & 1 & 0 & 0 & 0 & 0 & 0 & 0 \\
	0 & 0 & 0 & 0 & 0 & 0 & 0 & 0 & 0 & 0
\end{bNiceMatrix}
\times
\begin{bNiceMatrix}
	0 \\ 0 \\ 0 \\ 1 \\ 0 \\ 0 \\ 1 \\ 1 \\ 0 \\ 1
\end{bNiceMatrix}
=
\begin{bNiceMatrix}[last-col=2]
	0 \\ 0 \\ 0 \\ 0 \\ 1 & \phantom{a} 5 \\ 1 & \phantom{a} 6 \\ 0 \\ 0 \\ 1 & \phantom{a} 9 \\ 0
\end{bNiceMatrix}$
\end{center}
The (non-clean) SLPF $C_0 \, C_1 \, C_2$ is shown below, where the columns are represented as sets of segments numbered as in Tab.~\ref{tab:segments} and the arcs are visible on the NFA in Fig.~\ref{fig:NFA-RE2}, with several non-post-accessible segments, e.g., $7$, $8$ and $10$ in column $C_1$:
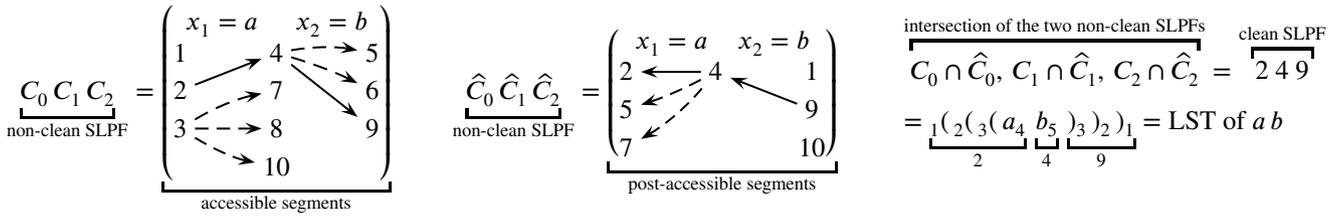
\[
\NiceMatrixOptions{cell-space-limits=1pt}
\psset{rowsep=0.25cm, colsep=1.0cm, arrows=->, linestyle=dashed, arrowscale=1.5, linewidth=0.75pt, nodesep=3pt}
\underbracket[0.75pt]{C_0 \, C_1 \,C_2}_\text{non-clean SLPF} =
\underbracket[0.75pt]{
	\begin{pNiceMatrix}
		x_1 = a \hspace{0.5cm} x_2 = b \\
		\begin{psmatrix}
			1 & 4  & 5 \\
			2 & 7  & 6 \\
			3 & 8  & 9 \\
			& 10 &
			\ncline[linestyle=solid] {2,1} {1,2}
			\ncarc[arcangle=10] {3,1} {2,2}
			\ncline {3,1} {3,2}
			\ncarc[arcangle=-10] {3,1} {4,2}
			\ncarc[arcangle=10] {1,2} {1,3}
			\ncarc[arcangle=5] {1,2} {2,3}
			\ncline[linestyle=solid] {1,2} {3,3}
		\end{psmatrix}
\end{pNiceMatrix}}_\text{accessible segments}
\hspace{0.75cm}
\underbracket[0.75pt]{\widehat C_0 \, \widehat C_1 \, \widehat C_2}_\text{non-clean SLPF} =
\underbracket[0.75pt]{
	\begin{pNiceMatrix}
		x_1 = a \hspace{0.4cm} x_2 = b \\
		\begin{psmatrix}
			2 & 4 & 1 \\
			5 &   & 9 \\
			7 &   & 10
			\ncline[linestyle=solid] {2,3} {1,2}
			\ncline[linestyle=solid] {1,2} {1,1}
			\ncarc[arcangle=5] {1,2} {2,1}
			\ncarc[arcangle=10] {1,2} {3,1}
		\end{psmatrix}
\end{pNiceMatrix}}_\text{post-accessible segments}
\hspace{0.75cm}
\begin{array}{l}
	\overbracket[0.75pt]{C_0 \cap \widehat C_0, \, C_1 \cap \widehat C_1, \, C_2 \cap \widehat C_2}^\text{intersection of the two non-clean SLPFs} = \overbracket[0.75pt]{ \, 2 \ 4 \ 9 \, }^\text{clean SLPF} \\[0.2cm]
	= \underbracket[0.75pt]{_1( \, _2( \, _3( \, a_4}_2 \, \underbracket[0.75pt]{b_5}_4 \, \underbracket[0.75pt]{)_3 \, )_2 \, )_1}_9 = \text{LST of $a\,b$}
\end{array}
\]
The backward computation (not shown in detail) yields the (non-clean) SLPF $\widehat C_0 \, \widehat C_1 \, \widehat C_2$ above, with a few non-accessible segments, e.g., $5$ and $7$ in column $\widehat C_0$ (as they are not initial). Finally the column intersection $C_r \cap \widehat C_r$ yields the clean SLPF, with just one LST  (as string $a\,b$  is not ambiguous). For an ambiguous text the clean SLPF would encode multiple LSTs. \qed
\par
The memory consumption of the serial NFA-based parser consists of the data-structures $I$, $F$ and $N_a$ (for every terminal $a \in \Sigma$) for the NFA, and data-structure $C$ for the SLPF, the latter being by far dominating since its size is proportional to the text length $n$. Note that the backward computation of $\widehat C$ and the intersection of columns $C$ and $\widehat C$ do not need additional memory, as the data-structure $C$ can be simply overwritten on-the-fly during the backward computation, to obtain the result of the intersection. With such an optimization, the algorithm in Fig.~\ref{fig:serialparser} would allocate only the data-structure $C$ and write it twice, forwards and backwards (to be further discussed in Sect.~\ref{subsec:scheduling} and Fig.~\ref{fig:builders&merger}), thus saving memory.
\par
This simple serial parser may be quite slow for large NFAs on long inputs. Among the various optimizations   proposed in the past,  but only for RE recognition and matching, we mention the following: implementing efficient sparse matrix multiplication (as most connection matrices  are very sparse), and fine-tuning on the instruction set architecture (ISA), in particular to exploit single-instruction-multiple-data architectures. Our project uses a DFA instead of an NFA, to the advantage that,  instead of matrix multiplication, a state-transition function is used, i.e., a faster look-up table. DFAs have also been used by most past projects of parallel RE recognizers, e.g., in~\cite{DBLP:conf/asplos/MytkowiczMS14}. But these projects  do not perform parsing, and are unable to produce the syntax trees and their forest representation.  To improve the time performance of parsing,  a novel most attractive  strategy is parallelization with a combined use of nondeterministic and deterministic state-machines. This is  the subject of the next section.
\section{Parallel parser} \label{sec:parallelparser}
The baseline consists of the nondeterministic FAs used as forward  and backward parsers (NFA and $\widehat{\text{NFA}}$) in Sect.~\ref{subsec:serialalgorithmVectorial}.
To exploit parallelism, we uniformly cut the input text into up to as many chunks as the available processors, so that each chunk is assigned a processor for parsing. We  mainly discuss the forward parser, because the other one is similar. Since the starting \emph{segment}, i.e., state, is unknown for all chunks but the first, each chunk parser must speculatively execute multiple runs  starting from  each segment, and for each run it may scan the whole chunk. The \emph{first phase} of the parallel parser, called \emph{reach}, aims at finding the segments reached by the NFA at the end of each chunk, to be called \emph{edge segments}.
Only such edge segments, and not the intermediate ones, are stored, because it would be wasteful to save the many runs that are not part of any accepting computations. Then, in the \emph{second phase}, called \emph{join}, the edge segments of each chunk are selected to become those from where to start parsing the next chunk. Symmetrically, the reverse automaton $\widehat{\text{NFA}}$ is used to execute a right-to-left reach-join phase, so that after the two forward and backward passes, the two sets of starting segments for each chunk in both directions are known.
\par
The \emph{third phase}, called \emph{build}, refines parsing. It  starts from the segments at the edge of each chunk that were identified by the join phase, and stores the entire computation in view of the construction of the SLPF. However, by carrying out the parsing in only one direction, e.g., left-to-right, some non-accepting computations would be found and stored, thus building a graph that includes useless nodes. To avoid this, for each chunk two  opposite runs are executed by the direct and reverse NFA: one in the forward direction starting from the segments at the beginning of the chunk, and the other one in the backward direction starting from the segments at the chunk end. Then the intersection of the segment sets reached at each string position, i.e., input character, by the two runs is computed, to obtain a clean graph, where all the nodes are on a path from an initial segment to a final one. The overall DAG that represents the SLPF for the whole text is then obtained by concatenating the results of the build phase for all chunks.
\par
To simplify the serial parser presentation, we have so far glossed over a major improvement  that is next presented in Sect.~\ref{subsec:deterministicFA} and was prompted by early experimentation: the use of a deterministic automaton (DFA) that  preserves all the information on the segments  needed to compute the SLPF forest. In the DFA each state is associated to a set of segments that is computed from the NFA by means of the standard powerset construction.
\par
We directly use the DFA  only in the build phase. For each  chunk, parsing starts from the DFA state that represents the segment set that was selected by the join phase. On the other hand, it would be wasteful to use such a DFA in the reach phase, since for each chunk too many  parsing runs would have to be executed because the ratio DFA \emph{state number} / NFA \emph{state number} may be large and at worst exponential. We use a more efficient variant of the DFA, called \emph{multi-entry} DFA (ME-DFA), that reduces the amount of speculative runs.
\par
For each  segment, the ME-DFA has an entry state from where a deterministic run begins that embeds all the NFA runs that start from that segment. It follows that any reasoning in Sect.~\ref{subsec:parserautomaton} for proving the correctness of the parser NFA carries over to the parallel parser, since each NFA parsing run that was executed by the reacher processor starting from one of the segments, is now executed deterministically by the ME-DFA starting from the entry state that corresponds to that segment. For the reader's convenience, we collect in Tab.~\ref{tab:tableterms2} the technical terms introduced.
\begin{table}[h]
	\begin{center}
        \tabcolsep=0.0cm
		\begin{tabular}{l@{\hspace{1.0cm}}l}	
			\textbf{term} & \textbf{denotation / description}
			\\ \toprule
            parser NFA (or $\widehat{\text{NFA}}$) & baseline NFA (or reverse NFA), see Sect.~\ref{subsec:serialalgorithmVectorial}, from which to build the parallel parser \\ \midrule
			multi-entry DFA (or $\widehat{\text{DFA}}$) & ME-DFA (or $\widehat{\text{ME-DFA}}$) -- a DFA (or reverse DFA) with more than one initial state
			\\ \midrule
			text chunk (or simply chunk) & substring resulting from a uniform partition of the input text
		\\ \midrule
		reach phase (direct or reverse) & finds the  segments reached at the (right or left) end of a chunk (edge segments)
		\\ \midrule
		join phase (direct or reverse) & selects the (left or right) starting segments of a chunk
		\\ \midrule
		build phase (direct or reverse) & refines parsing (either direction) and eventually computes the (clean) SLPF forest
		\\ \midrule
		edge segment & segment reached by the parser NFA (or $\widehat{\text{NFA}}$) after scanning a whole chunk
			\\ \bottomrule
		\end{tabular}
	\end{center}
	\caption{Terms and acronyms -- Part II.} \label{tab:tableterms2}
\end{table}
\subsection{Deterministic parser automaton} \label{subsec:deterministicFA}
\begin{figure}[ht]
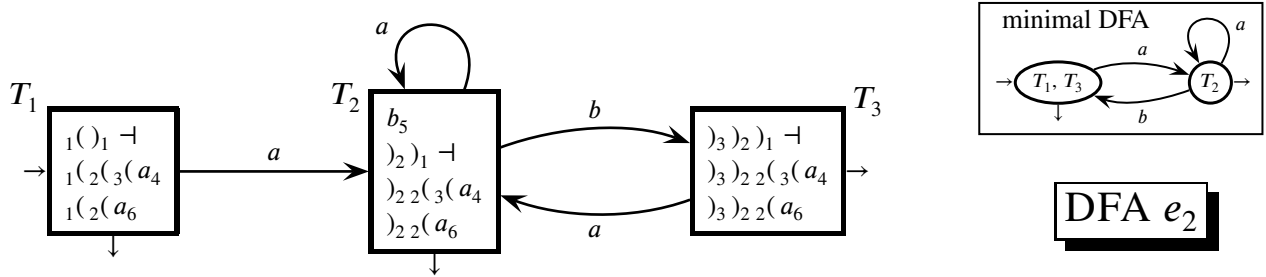

	\begin{center}
		\vspace{1.0cm}
		\tabcolsep=0.0cm
		\begin{tabular}{c@{\hspace{4.5cm}}}
			\rnode{DFA1} {
				\psset{nodesep=0.0cm, arrows=->, linewidth=1pt, arrowscale=2, rowsep=1.5cm, colsep=2.5cm, border=0.00cm}
				\fboxrule=1.5pt
				\begin{psmatrix}
					
					\rnode {DFAn1} {\framebox{$\begin{array}{l}
								_1( \ )_1 \dashv \\
								_1( \, _2( \, _3( \, a_4 \\
								_1( \, _2( \, a_6
							\end{array}$}}
					
					&
					
					\rnode {DFAn2} {\framebox{$\begin{array}{l}
								b_5 \\
								)_2 \, )_1 \dashv \\
								)_2 \, _2( \, _3( \, a_4 \\
								)_2 \, _2( \, a_6
							\end{array}$}}
					
					&
					
					\rnode {DFAn3} {\framebox{$\begin{array}{l}
								)_3 \, )_2 \, )_1 \dashv \\
								)_3 \, )_2 \, _2( \, _3( \, a_4 \\
								)_3 \, )_2 \, _2( \, a_6
							\end{array}$}}
					
					\psset{labelsep=0pt}
					
					\nput {180} {DFAn1} {$\to$}
					
					\nput {-90} {DFAn1} {$\downarrow$}
					
					\nput {-90} {DFAn2} {$\downarrow$}
					
					\nput {0} {DFAn3} {$\to$}
					
					\psset{labelsep=3pt}
					
					\nput[labelsep=3pt] {140} {DFAn1} {\large $T_1$}
					
					\nput[labelsep=3pt] {140} {DFAn2} {\large $T_2$}
					
					\nput[labelsep=3pt] {35} {DFAn3} {\large $T_3$}
					
					\ncline {DFAn1} {DFAn2} \naput{$a$}
					
					\nccurve[angleA=70, angleB=110, ncurv=3] {DFAn2} {DFAn2} \nbput[npos=0.75,  labelsep=6pt]{$a$}
					
					\ncarc[arcangle=20] {DFAn2} {DFAn3} \naput{$b$}
					
					\ncarc[arcangle=20] {DFAn3} {DFAn2} \naput{$a$}
			\end{psmatrix}}
			
			\nput[labelsep=2.75] {-4} {DFA1} {\shadowbox{\Large \textsc{DFA $e_2$}}}
			
			\nput[labelsep=1.5] {10} {DFA1} {
				\tabcolsep=0.0cm
				\rnode{minDFA} {
					\begin{tabular}{l}
						\scalebox{0.75}{
							\psframebox[linewidth=1pt, framesep=0.5cm]{
								\psset{arrows=->, arrowscale=2, nodesep=0pt, linewidth=1pt, rowsep=0.375cm}
								\begin{psmatrix}
									\\ \ovalnode[linewidth=1.5pt] 1 {$T_1, \, T_3$} & [colsep=1.5cm] \ovalnode[linewidth=1.5pt] 2 {$T_2$}
									
									\psset{labelsep=0pt}
									
									\nput {180} 1 {$\to$}
									
									\nput {-90} 1 {$\downarrow$}
									
									\nput {0} 2 {$\to$}
									
									\psset{labelsep=3pt, arcangle=20}
									
									\ncarc 1 2 \naput{$a$}
									
									\ncarc 2 1 \naput{$b$}
									
									\nccurve[angleA=60, angleB=120, ncurv=6] 2 2 \nbput[npos=0.25]{$a$}
						\end{psmatrix}}}
				\end{tabular}}
				\nput[labelsep=-0.35cm] {108} {minDFA} {
					minimal DFA
			}}
		\end{tabular}
		\vspace{0.25cm}
	\end{center}
	\caption{The classic DFA, obtained by applying powerset to the parser NFA of Fig.~\ref{fig:NFA-RE2}. It  encodes the language of the linearized syntax trees of RE $e_2$ and behaves as a deterministic parser for $e_2$. The box in the corner shows the minimal DFA.}
	\label{fig:DFA2}
\end{figure}

The NFA that was obtained by the generalized GMY method  is a most simple serial parser for an RE, but, as said, it executes many nondeterministic runs to recognize the input and build the SLPF. If the input is unambiguous, at most one run terminates successfully. Although parallel programming techniques may be applied that allow the serial parser to execute some or all runs in parallel~\cite{DBLP:conf/wia/HolubS09}, yet   extra computation is involved.  It is preferable to reduce to one the number of runs, by determinizing the NFA through the classic powerset construction of an equivalent DFA, e.g., the  DFA of RE $e_2$ in Fig.~\ref{fig:DFA2}. Each DFA state   is associated to a set of NFA states, therefore it includes all and only  the segments of such a set. Therefore, the set of segments reached by many concurrent NFA runs and represented in a SLPF column, coincides with  the set of segments of a single DFA state.  Clearly,  one serial DFA run on the input returns the same SLPF as the multiple NFA runs.
\par
It is important to observe that, for use in a parser, it would be a mistake to minimize the DFA obtained by powerset of the NFA parser.
In fact, if we consider a DFA transition from a source state $s$  to a target state $t$, the segments  in  $t$ have the property of being consecutive, in some LSTs, to those in $s$. But in the minimal DFA, a state  may result form merging two (or more) states $[s, \, t]$, and the latter property would be lost. Consequently,  it would be much more expensive to extract the LSTs from a run of the minimal  DFA.
\paragraph{Multi-entry DFA}
We have now  to consider a text  split into  chunks to be  independently parsed. In order to motivate the choice of a new type of  chunk automaton, we anticipate  two different situations that occur in distinct phases of the parallel parser. If the set of initial segments for each chunk is unknown, then every segment should be taken as initial by the DFA; this occurs in the first phase (\emph{reach})  of the parallel parser (to be described in Sect.~\ref{sec:parallelparser}) and motivates the introduction of the multi-entry DFA. On the other hand, if the set is known,  then the classic DFA can parse the chunk starting from the state that exactly contains those segments; this occurs in the \emph{build} phase.
\par
\begin{figure}[ht]
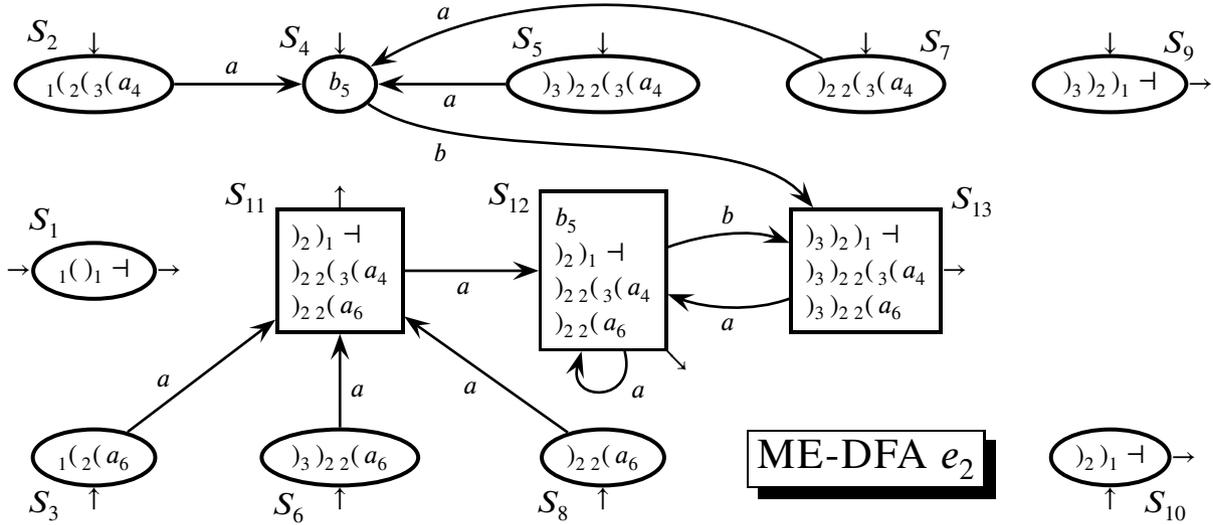

	\begin{center}
		\vspace{0.5cm}
		\rnode {ME-DFA1} {
			\psset{nodesep=0.0cm, arrows=->, linewidth=1.5pt, arrowscale=2, rowsep=1.0cm, colsep=1.125cm, border=0.00cm, nodealign=true}
			\fboxrule=1pt
			\begin{psmatrix}
				\ovalnode {1234} {$_1( \, _2( \, _3( \, a_4$}
				
				&
				
				\ovalnode {5} {$ \ b_5 \ $}
				
				&
				
				\ovalnode {32234} {$)_3 \, )_2 \, _2( \, _3( \, a_4$}
				
				&
				
				\ovalnode {2234} {$)_2 \, _2( \, _3( \, a_4$}
				
				&
				
				\ovalnode {321dashv} {$)_3 \, )_2 \, )_1 \dashv$}
				
				\\
				
				\ovalnode {11dashv} {$_1( \ )_1 \dashv$}
				
				&
				
				\rnode {add} {\framebox{$\begin{array}{l}
							)_2 \, )_1 \dashv \\
							)_2 \, _2( \, _3( \, a_4 \\
							)_2 \, _2( \, a_6
						\end{array}$}}
				
				&
				
				\rnode {DFAn2} {\framebox{$\begin{array}{l}
							b_5 \\
							)_2 \, )_1 \dashv \\
							)_2 \, _2( \, _3( \, a_4 \\
							)_2 \, _2( \, a_6
						\end{array}$}}
				
				&
				
				\rnode {DFAn3} {\framebox{$\begin{array}{l}
							)_3 \, )_2 \, )_1 \dashv \\
							)_3 \, )_2 \, _2( \, _3( \, a_4 \\
							)_3 \, )_2 \, _2( \, a_6
						\end{array}$}}
				
				\\
				
				\ovalnode {126} {$_1( \, _2( \, a_6$}
				
				&
				
				\ovalnode {3226} {$)_3 \, )_2 \, _2( \, a_6$}
				
				&
				
				\ovalnode {226} {$)_2 \, _2( \, a_6$}
				
				&&
				
				\ovalnode {21dashv} {$)_2 \, )_1 \dashv$}
				
				\psset{labelsep=3pt}
				
				\nput {135} {11dashv} {\large $S_1$}
				
				\nput {135} {1234} {\large $S_2$}
				
				\nput {-135} {126} {\large $S_3$}
				
				\nput {135} {5} {\large $S_4$}
				
				\nput[labelsep=4pt] {150} {32234} {\large $S_5$}
				
				\nput {-135} {3226} {\large $S_6$}
				
				\nput {30} {2234} {\large $S_7$}
				
				\nput {-135} {226} {\large $S_8$}
				
				\nput {30} {321dashv} {\large $S_9$}
				
				\nput {-45} {21dashv} {\large $S_{10}$}
				
				\nput {140} {add} {\large $S_{11}$}
				
				\nput {140} {DFAn2} {\large $S_{12}$}
				
				\nput {35} {DFAn3} {\large $S_{13}$}
				
				\psset{labelsep=0pt, linewidth=1pt}
				
				\nput {180} {11dashv} {$\to$}
				
				\nput {90} {321dashv} {$\downarrow$}
				
				\nput {90} {5} {$\downarrow$}
				
				\nput {90} {32234} {$\downarrow$}
				
				\nput {90} {2234} {$\downarrow$}
				
				\nput {90} {1234} {$\downarrow$}
				
				\nput {-90} {126} {$\uparrow$}
				
				\nput {-90} {3226} {$\uparrow$}
				
				\nput {-90} {226} {$\uparrow$}
				
				\nput {-90} {21dashv} {$\uparrow$}
				
				\nput {0} {11dashv} {$\to$}
				
				\nput {0} {321dashv} {$\to$}
				
				\nput {0} {21dashv} {$\to$}
				
				\nput[labelsep=-1pt] {-51} {DFAn2} {\rotatebox{-45}{$\longrightarrow$}}
				
				\nput {0} {DFAn3} {$\to$}
				
				\nput[labelsep=0pt] {90} {add} {$\uparrow$}
				
				\psset{labelsep=3pt, linewidth=1pt}
				
				\ncline {32234} {5} \naput[npos=0.4375]{$a$}
				
				\ncarc[arcangle=-30] {2234} {5} \nbput[npos=0.8125]{$a$}
				
				\nccurve[angleA=-75, angleB=-105, ncurv=2.5] {DFAn2} {DFAn2} \naput[npos=0.25,  labelsep=3pt]{$a$}
				
				\ncarc[arcangle=20] {DFAn2} {DFAn3} \naput{$b$}
				
				\ncarc[arcangle=20] {DFAn3} {DFAn2} \naput{$a$}
				
				\ncline {1234} {5} \naput[npos=0.45]{$a$}
				
				\nccurve[angleA=-35, angleB=130, ncurvA=0.5, ncurvB=0.5] {5} {DFAn3} \nbput[npos=0.2]{$b$}
				
				\ncline {add} {DFAn2} \nbput[npos=0.4375]{$a$}
				
				\ncline {126} {add} \naput[npos=0.325]{$a$}
				
				\ncline {3226} {add} \nbput[npos=0.4]{$a$}
				
				\ncline {226} {add} \naput[npos=0.525]{$a$}
		\end{psmatrix}}
		\nput[labelsep=-0.75cm] {-46} {ME-DFA1} {\shadowbox{\Large \textsc{ME-DFA $e_2$}}}
		\vspace{0.25cm}
	\end{center}
	\caption{Multi-entry DFA (ME-DFA) that encodes the language of the LSTs of RE $e_2$. The oval nodes are singleton sets and are in one-to-one correspondence with those of the parser NFA of $e_2$  (see Fig.~\ref{fig:NFA-RE2}), i.e., $S_j = q_j$ for $1 \leq j \leq 10$. All oval nodes are initial for the ME-DFA, and some of them are also final.}
	\label{fig:MEDFA2}
\end{figure}

Such considerations and a concern for the optimization of the parallel parser motivated the design of a \emph{multi-entry} DFA (ME-DFA), a machine model originally defined in~\cite{DBLP:journals/jcss/GillK74}, which is deterministic except that  instead of one initial state, it has many that are precisely the segments of the numbered RE. The ME-DFA is obtained from the NFA as follows. Let the NFA have $\ell$ states $\set{ \; q_1, \; \ldots, \; q_\ell \; }$ and consider  $\ell$ distinct versions of the NFA, denoted by  $\text{NFA}_j$ with $1 \leq j \leq \ell$, such that  $\text{NFA}_j$ is equal to the NFA except that its unique initial state is $q_j$. Apply the powerset construction separately to each $\text{NFA}_j$, thus obtaining  $\ell$ distinct deterministic automata $\text{DFA}_1$, \ldots, $\text{DFA}_j$, \ldots, $\text{DFA}_\ell$.  For all $\text{DFA}_j$, consider  the states  and merge those that contain exactly the same set of segments, then mark as final the states that contain a final segment. For instance, see the ME-DFA of RE $e_2$ in Fig.~\ref{fig:MEDFA2}.
\par
Notice that the ME-DFA does not necessarily include  all the states of the classic DFA (obtained from the NFA by powerset); e.g., for RE $e_2$ the ME-DFA in Fig.~\ref{fig:MEDFA2} lacks the state $T_1$ of the DFA in Fig.~\ref{fig:DFA2}. In fact, both devices are obtained through powerset, but the starting states are different. We specify the phases where the ME-DFA and DFA are respectively used by the parallel parser:
\begin{description}[leftmargin=1.7cm, style=nextline]
\item[ME-DFA] used  in the first phase, \emph{reach}, by launching for a  chunk each run from every initial state of the ME-DFA, i.e., from every different segment
\item[DFA] used subsequently, in the \emph{build} phase, by launching  for a chunk the  run  from the (unique) DFA state that stores the segments chosen in the join phase
\end{description}
Notice that in both phases every run, once launched, proceeds deterministically. We motivate the use of two different kinds of deterministic machine, which may appear as a  complication. Clearly, the classic DFA could be used in all the phases of a parallel algorithm, by assuming  that \emph{all its states are initial}, thus launching as many runs as there are DFA states. This is the \emph{speculative} approach followed in~\cite{DBLP:conf/wia/HolubS09} and in many other subsequent parallel RE recognizers. It is well-known however, that a DFA equivalent to an NFA may exhibit an exponential growth of the number of states. Such a state explosion may happen   when determinizing our  parser NFA, as illustrated by the parametric family of REs $e \, (k)$ of Ex.~$5$ below.
\paragraph{Example $5$ -- parametric family of REs}
The following  family of REs $e \, (k)$, for  parameter $k \geq 1$, exhibits an exponential growth of the number of states from NFA to DFA. To be valid, a string must have a terminal $a$ in the $(k+1)$-th position from the end:
\begin{center}
\vspace{0.125cm}
	\tabcolsep=0.0cm\def\arraystretch{0}
	\begin{tabular}{c@{\hspace{1.0cm}}l}
        \begin{tabular}{l}
		\large
		\tabcolsep=0.1cm\def\arraystretch{0.75}
		\begin{tabular}{rcl}
			$e \, (k)$ & $=$ & $( \, a \; \mid \; b \, )^\ast \; a \; ( \, a \; \mid \; b \,)^k$ \qquad \normalsize for $k \geq 1$ \\[0.25cm]
			$e_\# \, (k$) & $=$ & $_1( \; _2( \; _3( \; a_4 \; \mid \; b_5 \; )_3 \; {)_2}^\ast \; a_6 \; _7( \; \underbracket[0.75pt]{\, _8( \; a_9 \; \mid \; b_{10} \; )_8 \,}_{\parbox{2.5cm}{\centering \rm \scriptsize repeated $k$ times \par with progressive \par numbering}} \; {)_7}^k \; )_1$
		\end{tabular}
		\end{tabular}
		&
		\tabcolsep=0.0cm\def\arraystretch{0.875}
		\begin{tabular}{c@{\hspace{0.5cm}}l}
			\emph{number} & RE \emph{operator or symbol} \\ \toprule
			$1$ & concatenation (ternary) \\ \midrule
			$2$ & Kleene star \\ \midrule
			$3$ & union \\ \midrule
			$7$ & bounded repetition (App.~\ref{app:otherfeatures}) \\ \midrule
            $8$ & union \\ \midrule
            $4$, $5$, $6$, $9$, $10$ & terminals \\ \bottomrule
		\end{tabular}
\end{tabular}
\vspace{0.125cm}
\end{center}
In Tab.~\ref{tab:statecount} it is shown that, for family $e \, (k)$, both the DFA and the ME-DFA have a \emph{total} state count, with respect to parameter $k$, that grows exponentially with respect to that of the NFA (see also the bottom line of the table); the state count is slightly larger in the ME-DFA than is in the DFA. What matters however, is that the ME-DFA has a number of \emph{initial} states equal to the total state count of the NFA, i.e., the number of segments, which grows only linearly in $k$. Thus, for this  family of REs, a winning strategy  for the reach phase is to use an ME-DFA with a  linear number of initial states, instead of a DFA where exponentially many states are taken as initial. The  performance gain obtainable from using the ME-DFA in the  parallel parser is strongly confirmed by the experimental evaluation on large RE benchmarks reported in Sect.~\ref{sec:experimentation}. \qed
\begin{table}[ht]
	\begin{center}
		\vspace{0.125cm}
		\tabcolsep=0.0cm\def\arraystretch{0.875}
		\begin{tabular}{c@{\hspace{1.75cm}}c@{\hspace{1.5cm}}c@{\hspace{1.25cm}}c@{\hspace{0.5cm}}c@{\hspace{0.75cm}}c}
			\multirow{3}{*}{\Large $k$} & \large \textsc{\bf NFA}  & \large \textsc{\bf DFA} & \multicolumn{3}{c}{\large \textsc{\bf ME-DFA}} \\ \cline{4-6} \\[-0.35cm]
			& total states  & total states  & total states & initial states & $\text{DFA} - \text{ME-DFA ini. states}$ \\[-0.2cm]
			& \footnotesize segments  & \footnotesize segment sets  & \footnotesize segment sets & \footnotesize segments & \\ \toprule
			$1$ & $14$ & $5$    & $20$   & $14$  & $-9$    \\ \midrule
			$2$ & $18$ & $9$    & $30$   & $18$  & $-9$   \\ \midrule
			$3$ & $22$ & $17$   & $44$   & $22$  & $-5$   \\ \midrule
			$4$ & $26$ & $33$   & $66$   & $26$  & $7$   \\ \midrule
			$5$ & $30$ & $65$   & $104$  & $30$  & $35$   \\ \midrule
			$6$ & $34$ & $129$  & $174$  & $34$  & $95$  \\ \midrule
			$7$ & $38$ & $257$  & $308$  & $38$  & $219$  \\ \midrule
			$8$ & $42$ & $513$  & $570$  & $42$  & $471$  \\ \midrule
			$9$ & $46$ & $1025$ & $1088$ & $46$  & $979$ \\ \bottomrule \\[-0.3cm]
			\ldots & \large $4k + 10$ & \large $2^{k + 1} + 1$ & \large $2^{k + 1} + 6k + 10$ & \large $4k + 10$ & \large $2^{k + 1}- 4k - 9$ \\ \bottomrule
		\end{tabular}
		\vspace{0.0cm}
	\end{center}
	\caption{State count of the various FAs (NFA, DFA and ME-DFA) for parsing the parametric RE family $e \, (k)$ of Ex.~$5$.
	The last column shows how many more starting states are in the DFA in comparison with the ME-DFA.} \label{tab:statecount}
\end{table}
\paragraph{Discussion on machine size}
The segment count of an RE has  a non-tight exponential upper bound  with respect to the RE size, see Prop.~\ref{prop:segmentfiniteness} and discussion, but we have found that most often the count is linearly bounded. For instance, assume the size of an RE $e$, to be denoted  $\Vert \, e \, \Vert$, is quantified as the total count of symbols that occur in $e$, i.e., terminals and operators (metasymbols). This amounts to counting how many (numbered) terminals $a_\#$ and parenthesis pairs ``\,$_\#( \ )_\#$\,'' occur in the associated numbered RE $e_\#$. Then the size of the REs $e \, (k)$ in the RE family of Ex.~$5$ is a linear function of their parameter $k \geq 1$, namely $\Vert \, e \, (k) \, \Vert = 3 k + 7$, as in $e_\# \, (k)$ the three symbols $a_9$, $b_{10}$ and ``\,$_8( \ )_8$\,'' are repeated for $k$ times with progressive numbering, and there are seven symbols numbered from $1$ to $7$. Therefore, from Tab.~\ref{tab:statecount}, the segment (NFA state) count of each RE $e \, (k)$ is a linear function of $\Vert \, e \, (k) \, \Vert$, i.e., $\text{segment count} = \frac 4 3 \times \Vert \, e \, (k) \, \Vert + \frac 2 3$. In Sect.~\ref{subsec:results}, Fig.~\ref{fig:segments-vs-REsize}  shows that an approximately linear count of segments with respect to the RE size holds also for a much freer family of REs, which we used as a benchmark in our experimentation campaign.
\par
Similarly to the serial NFA parer, the parallel parser algorithm must also perform a right-to-left analysis, to ensure the cleanness of the SLPF that encodes all the LSTs of the input string. For this, the \emph{reverse} DFA and ME-DFA are needed, i.e., $\widehat\text{DFA}$ and $\widehat\text{ME-DFA}$. These can be obtained by determinizing  the reverse automaton $\widehat\text{NFA}$ (see Eq.~\eqref{eq:reverseNFAdef}), which has the same states, i.e., segments, as the direct NFA, all transitions reversed, and initial and final states switched.
\subsection{Parallel parsing algorithm} \label{subsec:parallelalgorithm}
\begin{figure}[ht]
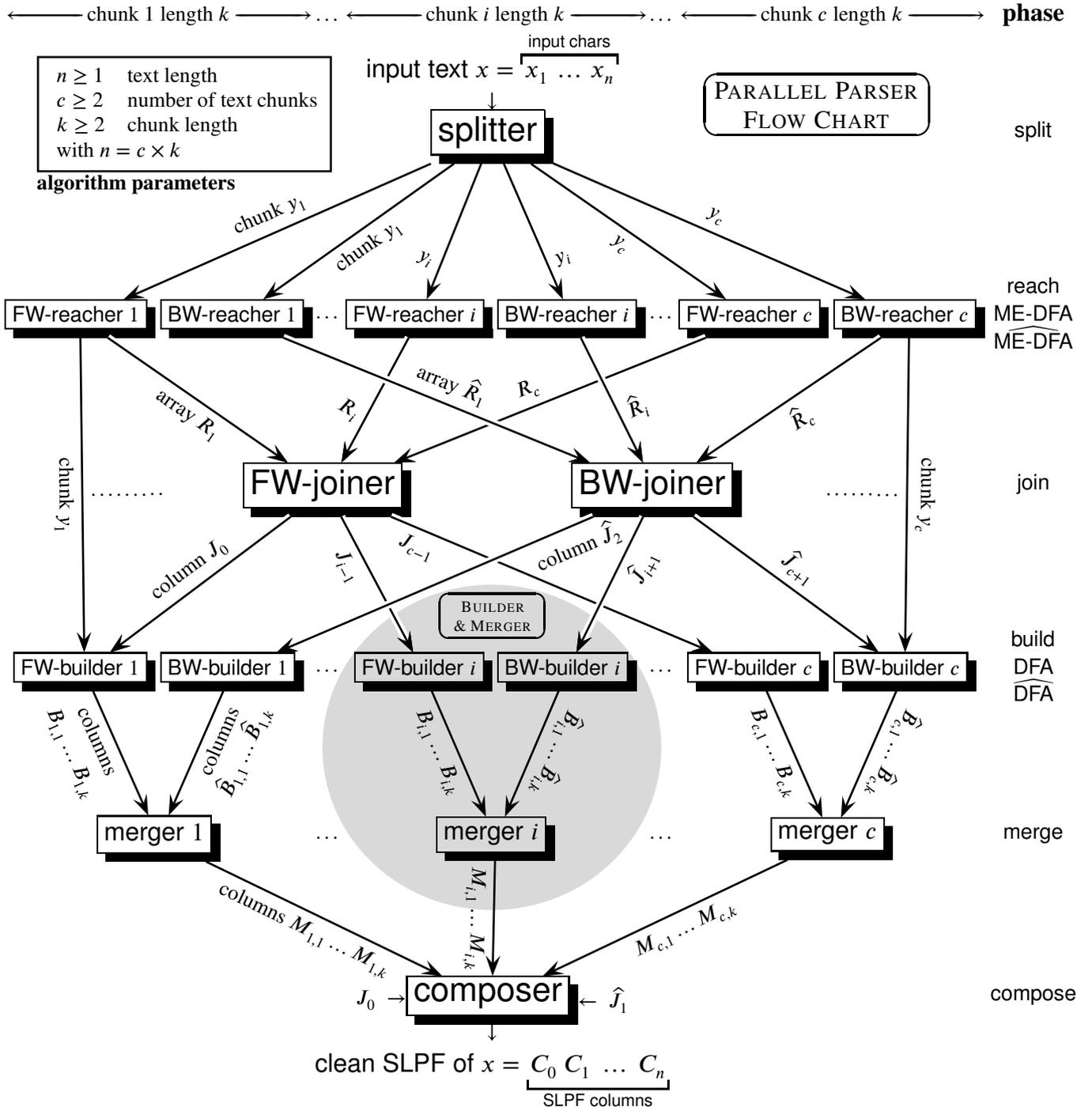

\begin{center}
\vspace{0.25cm}
\psset{linewidth=1pt, arrows=->, arrowscale=2, nodesep=0pt, colsep=0.0625cm, rowsep=1.9cm, border=0.0cm, nodealign=true}
\begin{psmatrix}

$\xleftarrow {\hspace{1.0cm}} \, \text{chunk $1$ length $k$} \, \xrightarrow {\hspace{1.0cm}}$

& \ldots &

$\xleftarrow {\hspace{1.0cm}} \, \text{chunk $i$ length $k$} \, \xrightarrow {\hspace{1.0cm}}$

& \ldots &

$\xleftarrow {\hspace{1.0cm}} \, \text{chunk $c$ length $k$} \, \xrightarrow {\hspace{1.0cm}}$

& \textbf{\large phase} \\[-0.5cm]

\rnode {splitter} {\shadowbox{\Large\sf splitter}} \psspan 5 & \textsf{split} \\

\nput[labelsep=7.0cm] {-90} {splitter} {\ovalnode*[fillcolor=gray!30, fillstyle=solid, linestyle=dashed, linewidth=0.5pt, border=0.00cm] {buildersandmerger} {\tabcolsep=0.0cm\def\arraystretch{0}\begin{tabular}[c]{c} \hspace{3.75cm} \vspace{3.675cm} \end{tabular}}}

\rnode {FWreacher1} {\shadowbox{\sf FW-reacher $1$}} \rnode {BWreacher1} {\shadowbox{\sf BW-reacher $1$}}
& \ldots &
\rnode {FWreacheri} {\shadowbox{\sf FW-reacher $i$}} \rnode {BWreacheri} {\shadowbox{\sf BW-reacher $i$}}
& \ldots &
\rnode {FWreacherc} {\shadowbox{\sf FW-reacher $c$}} \rnode {BWreacherc} {\shadowbox{\sf BW-reacher $c$}} & \parbox[c]{1.5cm}{\centering\sf reach \par ME-DFA \par $\widehat{\text{ME-DFA}}$} \\

\rnode {FWjoiner} {\shadowbox{\Large\sf FW-joiner}} \hspace{2.5cm} \rnode {BWjoiner} {\shadowbox{\Large\sf BW-joiner}} \psspan 5 & \textsf{join} \\

\rnode {FWbuilder1} {\shadowbox{\sf FW-builder $1$}} \rnode {BWbuilder1} {\shadowbox{\sf BW-builder $1$}}
& \ldots &
\rnode {FWbuilderi} {\shadowbox{\sf FW-builder $i$}} \rnode {BWbuilderi} {\shadowbox{\sf BW-builder $i$}}
& \ldots &
\rnode {FWbuilderc} {\shadowbox{\sf FW-builder $c$}} \rnode {BWbuilderc} {\shadowbox{\sf BW-builder $c$}} & \parbox[c]{1.5cm}{\centering\sf build \par DFA \par $\widehat{\text{DFA}}$} \\

\rnode {merger1} {\shadowbox{\large\sf  merger $1$}}
& \ldots &
\rnode {mergeri} {\shadowbox{\large\sf  merger $i$}}
& \ldots &
\rnode {mergerc} {\shadowbox{\large\sf  merger $c$}} & \textsf{merge} \\

\rnode {composer} {\shadowbox{\Large\sf composer}} \psspan 5 & \textsf{compose}

\nput[labelsep=0.0cm] {90} {splitter} {$\arraycolsep=0.0cm\def\arraystretch{0}\begin{array}{c} \textsf{\large input text $x = \overbracket[0.75pt]{ \, x_1 \, \ldots \, x_n \, }^\text{input chars}$} \\[0.25cm] \downarrow \end{array}$}

\nput[labelsep=0.0cm] {-90} {composer} {$\arraycolsep=0.0cm\def\arraystretch{0}\begin{array}{c} \downarrow \\[0.25cm] \textsf{\large clean SLPF of $x = \underbracket[0.75pt]{ \, C_0 \ C_1 \ \ldots \ C_n \, }_\text{SLPF columns}$} \end{array}$}

\nput[labelsep=0.0cm] {180} {composer}  {$J_0 \ \rightarrow$}

\nput[labelsep=0.0cm] {0} {composer}  {$\leftarrow \ \widehat J_1$}

\nput[labelsep=1.65cm] {177.5} {splitter} {
\tabcolsep=0.0cm
\def\arraystretch{0.9}
\begin{tabular}{l}
\psframebox{
\tabcolsep=0.1cm
\begin{tabular}{l@{\hspace{0.375cm}}l}
$n \geq 1$ & text length \\
$c \geq 2$ & number of text chunks \\
$k \geq 2$ & chunk length \\
\multicolumn{2}{l}{with $n = c \times k$}
\end{tabular}} \\
\textbf{algorithm parameters}
\end{tabular}}

\nput[labelsep=2.5cm] {7.5} {splitter} {\Ovalbox{\parbox{3.5cm}{\centering \large \textsc{Parallel Parser} \par \textsc{Flow Chart}}}}

\nput[labelsep=-0.9cm] {90} {buildersandmerger} {\Ovalbox{\parbox{1.5cm}{\footnotesize\centering \textsc{Builder} \par \& \textsc{Merger}}}}

\psset{labelsep=3pt}

\ncline {splitter} {FWreacher1} \nbput[npos=0.50, nrot=:D] {chunk $y_1$} \ncline {splitter} {BWreacher1} \naput[npos=0.50, nrot=:D] {chunk $y_1$}

\ncline {splitter} {FWreacheri} \nbput[npos=0.75, nrot=:L] {$y_i$} \ncline {splitter} {BWreacheri} \naput[npos=0.75, nrot=:L] {$y_i$}

\ncline {splitter} {FWreacherc} \nbput[npos=0.50, nrot=:U] {$y_c$} \ncline {splitter} {BWreacherc} \naput[npos=0.50, nrot=:U] {$y_c$}

\psset{border=0.05cm}

\ncline {FWreacher1} {FWjoiner} \nbput[npos=0.50, nrot=:U] {array $R_1$}

\ncline {FWreacheri} {FWjoiner} \nbput[npos=0.65, nrot=:L] {$R_i$}

\ncline {FWreacherc} {FWjoiner} \nbput[npos=0.550, nrot=:D] {$R_c$}

\ncline {BWreacher1} {BWjoiner} \naput[npos=0.525, nrot=:U] {array $\widehat R_1$}

\ncline {BWreacheri} {BWjoiner} \naput[npos=0.625, nrot=:L] {$\widehat R_i$}

\ncline {BWreacherc} {BWjoiner} \naput[npos=0.50, nrot=:D] {$\widehat R_c$}

\ncline {FWjoiner} {FWbuilder1} \nbput[npos=0.50, nrot=:D] {column $J_0$}

\ncline {FWjoiner} {FWbuilderi} \nbput[npos=0.30, nrot=:U] {$J_{i - 1}$}

\ncline {FWjoiner} {FWbuilderc} \nbput[npos=0.100, nrot=:U] {$J_{c - 1}$}

\ncline {BWjoiner} {BWbuilder1} \naput[npos=0.075, nrot=:D] {column $\widehat J_2$}

\ncline {BWjoiner} {BWbuilderi} \naput[npos=0.30, nrot=:D] {$\widehat J_{i + 1}$}

\ncline {BWjoiner} {BWbuilderc} \naput[npos=0.50, nrot=:U] {$\widehat J_{c + 1}$}

\psset{border=0.00cm}

\ncline {FWbuilder1} {merger1} \nbput[npos=0.35, nrot=:U] {\parbox{1.75cm}{\centering columns \par $B_{1,1} \ldots B_{1,k}$}}

\ncline {BWbuilder1} {merger1} \naput[npos=0.35, nrot=:D] {\parbox{1.75cm}{\centering columns \par $\widehat B_{1,1} \ldots \widehat B_{1,k}$}}

\ncline {FWbuilderi} {mergeri} \nbput[npos=0.425, nrot=:U] {$B_{i,1} \ldots B_{i,k}$}

\ncline {BWbuilderi} {mergeri} \naput[npos=0.425, nrot=:U] {$\widehat B_{i,1} \ldots \widehat B_{i,k}$}

\ncline {FWbuilderc} {mergerc} \nbput[npos=0.425, nrot=:U] {$B_{c,1} \ldots B_{c,k}$}

\ncline {BWbuilderc} {mergerc} \naput[npos=0.425, nrot=:U] {$\widehat B_{c,1} \ldots \widehat B_{c,k}$}

\ncline {merger1} {composer} \nbput[npos=0.45, nrot=:U] {columns $M_{1,1} \ldots M_{1,k}$}

\ncline {mergeri} {composer} \nbput[npos=0.50, nrot=:U] {$M_{i,1} \ldots M_{i,k}$}

\ncline {mergerc} {composer} \naput[npos=0.45, nrot=:D] {$M_{c,1} \ldots M_{c,k}$}

\ncline {FWreacher1} {FWbuilder1} \nbput[nrot=:U] {chunk $y_1$} \naput[nrot=:L] {\ldots\ldots\ldots}

\ncline {BWreacherc} {BWbuilderc} \naput[nrot=:U] {chunk $y_c$} \nbput[nrot=:R] {\ldots\ldots\ldots}
\end{psmatrix}
\vspace{1.0cm}
\end{center}
\caption{Flow chart of the complete parallel parser ($1 \leq i \leq c$), where the shaded area is the unified builder\&merger component to be later discussed in Sect.~\ref{subsec:scheduling} and in Fig.~\ref{fig:builders&merger}.} \label{fig:parallelparser}
\end{figure}

To describe the parallel parser in detail we present different system views of the algorithm, with its components, organization and operations, plus an example, namely:
\begin{itemize}[leftmargin=*, style=standard]
\item  the finite automata that control the parsing phases
\item  the data-structures that store segments and LSTs
\item  a flow chart that displays the algorithmic components and their interfaces
\item  the pseudo-code of the main algorithmic components
\item  and a small yet complete example of parsing
\end{itemize}
As said,  the input text is divided  into chunks to be parsed concurrently. There are two hot-spot parsing phases, called \emph{reach} phase and \emph{build} phase, that are driven by the ME-DFA and DFA, respectively. Given the definition for the serial parser for an RE $e$ (Sect.~\ref{subsec:parserautomaton}), here are the additional definitions  for the parallel parser:
\begin{description}[leftmargin=0.5cm, style=nextline]
\item[DFA] The deterministic FA equivalent to the NFA, obtained through the powerset construction (Sect.~\ref{subsec:deterministicFA}):
\[
\text{DFA} = \big( \, 2^Q, \, \Sigma, \, T_1, \, F_\text{DFA}, \, \delta_\text{DFA} \, \big) \qquad \text{with $\delta_\text{DFA} \colon 2^Q \times \Sigma \to 2^Q$} \hspace{1.0cm} \text{(recall that $Q$ is the NFA state set and $\vert \, Q \, \vert = \ell$)}
\]
Each state of the DFA is a segment set in $2^Q$, denoted $T_j$  ($1 \leq j \leq 2^\ell$). The unique initial state is  $T_1$.
\item[ME-DFA] The multi-entry deterministic FA obtained from the NFA through the (modified) powerset
method (Sect.~\ref{subsec:deterministicFA}):
\[
\text{ME-DFA} = \big( \, 2^Q, \, \Sigma, \, I_\text{ME-DFA}, \, F_\text{ME-DFA}, \, \delta_\text{ME-DFA} \, \big) \qquad \text{with $\delta_\text{ME-DFA} \colon 2^Q \times \Sigma \to 2^Q$}
\]
Each state of the ME-DFA is a segment set, denoted $S_j$ ($1 \leq j \leq 2^\ell$). The set of initial states is:
\[
I_\text{ME-DFA} = \big\{ \; \set{ \, q \, } \; \vert \quad \text{$q$ is a segment of $e$} \; \big\}
\]
By convention the ME-DFA states from $S_1$ to $S_\ell$ denote precisely the singletons, and are numbered as the segments.
Thus, each initial state is a singleton set containing one segment. The transition function $\delta_\text{ME-DFA}$ is extended to input strings in the customary way as $\delta_\text{ME-DFA}^\ast \colon 2^Q \times \Sigma^\ast \to 2^Q$.
\item[text chunk] Decomposition of the input text of length $n$ into $c$ non-empty substrings $y_i$ called chunks, i.e., $x = y_1 \ldots y_i \ldots y_c$ ($2 \leq c \leq n$
and $1 \leq i \leq c$). To simplify the presentation,  assume that the chunks have equal length $\vert \, y_i \, \vert = k \geq 2$, disregarding the possibility of a shorter  last chunk; it must therefore hold $n = c \times k$.
\item[edge segment] For each chunk $y_i$ the \emph{edge} segments are of two types depending on the scan direction: the segments reached by the left-to-right  parsing of string $y_i$, and those reached by the right-to-left  parsing of string $y_{i+1}$. The segment sets $I$ and $F$ of the NFA  store the initial and final edge segments of the whole input text, respectively.
\end{description}
In comparison to the  serial parser of Sect.~\ref{subsec:deterministicFA}, the parallel parser uses more articulated data-structures. We describe  them   and their meaning:
\begin{description}[leftmargin=0.5cm, style=nextline]
\item[array $R$] A series $R_1$, \ldots, $R_c$ of  one-dimensional arrays; each array $R_i$ has $\ell$ elements that are (possibly empty) segment sets. More precisely, each array element $R_{i, \, j}$ is a set of segments, i.e., $R_{i, \, j} \subseteq Q$. The set $R_{i, \, j}$, for $1 \leq j \leq \ell$, stores the edge segments that are reached from segment $q_j$ after scanning the  chunk $y_i$ from left to right.
\item[column $J$] A series of $c + 1$ segment sets, $J_i \subseteq Q$, for $0 \leq i \leq c$, represented as \emph{column} vectors (this also holds for the data-structures \emph{column} $B$, $M$ and $C$ below). Column $J_0$ stores the initial segments and column $J_i$ stores the edge segments of the SLPF reached from any initial segment after scanning the chunk sequence $y_1 \ldots y_i$ from left to right.
\item[column $B$] A series of $n$ segment sets, each set being denoted $B_{i, \, t} \subseteq Q$, for $1 \leq i \leq c$ and $1 \leq t \leq k$. Column $B_{i, \, t}$ stores the segments of the SLPF reached from an initial segment after scanning the text prefix $x_1 \ldots x_r$ with $r = (i - 1) \times k + t$, i.e., between $x_r$ and $x_{r+1}$ (or after $x_n$ if $i = c$ and $t = k$).
\item[backward data-structures] All the above ``forward'' definitions of array series $R$, and of set series $J$ and $B$, have a ``backward'' counterpart, which is computed by scanning the input from right to left, using the automata $\widehat{\text{DFA}}$ or $\widehat{\text{ME-DFA}}$ (Sect.s~\ref{subsubsec:constrParserNFA} and \ref{subsec:deterministicFA}), and is distinguished with a circumflex sign, i.e., $\widehat R$, $\widehat J$ and $\widehat B$. The segments stored in such counterparts are computed analogously to their forward correspondents, but are post-accessible rather than accessible.
\item[column $M$] Same definition and notation as for columns $B$. Column $M_{i, \, t}$ stores the SLPF segments after position
$r = (i - 1) \, \times \, k \, + \, t$ in the text $x$, i.e., between characters $x_r$ and $x_{r + 1}$ (or after $x_n$), that are both accessible and post-accessible.
\item[column $C$] A series of $n + 1$ segment sets, each such set being denoted $C_0, C_r \subseteq Q$, for $1 \leq r \leq n$.
The series $C_0$, $C_1$, \ldots, $C_n$  represents the clean SLPF of the  input text $x$ (see also Sect.~\ref{subsubsec:constrParserNFA} and Fig.~\ref{fig:SLPFe3}).
\end{description}
The flow chart of the parallel parser is shown in Fig.~\ref{fig:parallelparser}. The parser parameters are $n$ (text length), $c$ (number of chunks) and $k$ (chunk length), while $\ell \geq 1$ is the number of segments. The algorithmic components of the parser are the \emph{splitter}, \emph{reacher}, \emph{joiner}, \emph{builder}, \emph{merger} and \emph{composer}, plus the data-structures listed above.
We formally specify the phases of the parallel  algorithm, namely  \emph{split}, \emph{reach}, \emph{join}, \emph{build}, \emph{merge} and \emph{compose}, and their working. Assume $1 \leq i \leq c$ and $1 \leq j \leq \ell$:
\begin{description}[leftmargin=0.5cm, style=nextline]
\item[split phase] The \emph{splitter} divides the text $x$ into a series of $c$ chunks $y_i$, all of size $k$, and the chunks orderly go to both the reach and the build phases.
\item[reach phase] There are $2c$ \emph{reachers}, of which $c$ are of type \emph{forward} (FW) and $c$ of type \emph{backward} (BW):
\begin{equation} \label{eq:reacher}
\textbf{FW} \colon R_{i, \, j} = \bigcup_{S_j \, \in \, I_\text{ME-DFA}} \, \delta_\text{ME-DFA}^\ast \, \left(S_j, \, y_i \right)
\hspace{2.0cm}
\textbf{BW} \colon \widehat R_{i, \, j} = \bigcup_{S_j \, \in \, I_{\widehat{\text{ME-DFA}}} }\, \delta_{\widehat{\text{ME-DFA}}}^\ast \, \left(S_j, \, y_i^R \right)
\end{equation}
\begin{description}[leftmargin=0.75cm, style=multiline]
\item[FW] Reacher $i$ is driven by the ME-DFA, scans chunk $y_i$ forwards and computes the array $R_i$. Thus set $R_{i, \, j}$ stores the edge segments of chunk $y_i$ that are reached starting from any segment $q_j \in Q$.
\item[BW] Reacher $i$ is symmetrical to the corresponding forward one, but it is driven by  $\widehat{\text{ME-DFA}}$, scans chunk $y_i$ backwards, i.e., scans $y_i^R$,
and computes the backward array $\widehat R_i$. Thus set $\widehat R_{i, \, j}$ stores the edge segments of chunk $y_i$ that are reached backwards starting from any segment $q_j \in Q$.
\end{description}
Note that $I_{\widehat{\text{ME-DFA}}} = I_\text{ME-DFA}$, i.e., the reverse ME-DFA and the direct one have the same initial states, which as said can be identified with the segments. Both types of reacher apply their extended transition function $\delta^\ast$ to the whole chunk.
\item[join phase] There are two \emph{joiners},  of type \emph{forward} (FW) and  \emph{backward} (BW) respectively:
\begin{equation} \label{eq:joiner}
\textbf{FW} \colon J_0 = I \quad \text{and} \quad J_i = \bigcup_{q_j \, \in \, J_{i - 1}} R_{i, \, j}
\hspace{3.0cm}
\textbf{BW} \colon \widehat J_{c + 1} = F \quad \text{and} \quad \widehat J_i = \bigcup_{q_j \, \in \, \widehat J_{i + 1}} \widehat R_{i, \, j}
\end{equation}
\begin{description}[leftmargin=0.75cm, style=multiline]
\item[FW] The joiner scans the array series $R$ and orderly computes all the columns $J_0 \, \ldots \, J_c$. Thus column $J_i$ stores the edge segments of chunk $y_i$, or the initial edge segments for $J_0$, that are reached after scanning text $x$ forwards, starting from the initial edge segments.
\item[BW] The joiner scans the backward array series $\widehat R$ and orderly computes all the backward columns $\widehat J_1 \, \ldots \, \widehat J_{c + 1}$, starting from $J_{c+1}$. Thus column $\widehat J_i$ stores the edge segments of chunk $\widehat y_i$, or the final edge segments for $J_{c + 1}$, that are reached after scanning text $x$ backwards, starting from the final edge segments.
\end{description}
Note that columns $J_0$ and $\widehat J_1$ are also  directly accessed by the \emph{compose} phase, to compute column $C_0$.
\item[build phase] There are $2c$ \emph{builders}, half  of type \emph{forward} (FW) and  half \emph{backward} (BW):
\begin{equation} \label{eq:builder}
\def\arraystretch{1.125}
\arraycolsep=0.0cm
\textbf{FW} \ \left\{\begin{array}{r@{\hspace{0.05cm}}c@{\hspace{0.1cm}}l@{\hspace{0.5cm}}l}
B_{i, \, 1} & = & \delta_\text{DFA} \, \left( J_{i - 1}, \, y_{i, \, 1} \right) \\
B_{i, \, t} & = & \delta_\text{DFA} \, \left( B_{i, \, t - 1}, \, y_{i, \, t} \right) & 2 \leq t \leq k
\end{array}\right.
\hspace{1.25cm}
\textbf{BW} \ \left\{\begin{array}{r@{\hspace{0.05cm}}c@{\hspace{0.1cm}}l@{\hspace{0.5cm}}l}
\widehat B_{i, \, k} & = & J_{ i + 1} \\
\widehat B_{i, \, t} & = & \delta_{\widehat{\text{DFA}}} \, \big( \widehat B_{i, \, t + 1}, \, y_{i, \, t + 1} \big) & k - 1 \geq t \geq 1
\end{array}\right.
\end{equation}
\begin{description}[leftmargin=0.75cm, style=multiline]
\item[FW] Builder $i$ is driven by the DFA, scans chunk $y_i$ forwards and computes the columns $B_{i, \, t}$, starting from column $J_{i-1}$. Thus column $B_{i, \, t}$ stores the segments that are obtained \emph{after} character $y_{i, \, t}$ of chunk $y_i$, starting from the initial segments after scanning text $x$. Therefore the builder computes the segments that are accessible for the whole text.
\item[BW] Builder $i$ is symmetrical to the corresponding forward one, but it is driven by the $\widehat{\text{DFA}}$, scans chunk $y_i$ backwards and computes
the backward columns $\widehat B_{i, \, t}$, starting from column $\widehat J_{i + 1}$. Thus column $\widehat B_{i, \, t}$ stores the segments that are obtained \emph{after} character $y_{i, \, t + 1}$ of chunk $y_i$, starting from the final segments after scanning text $x$ backwards. Therefore the builder computes the segments that are post-accessible for the whole text.
\end{description}
Actually column $B_{i, \, k}$ coincides with the column $J_i$ computed by the joiner, thus the FW-builder could skip the last iteration of function $\delta_\text{DFA}$ and just set $B_{i, \, k} = J_i$ (however for long chunks the impact on computation performance is negligible). With this observation, both builder types iterate their transition function $\delta_\text{DFA}$ or $\delta_{\widehat{\text{DFA}}}$ for exactly $k - 1$ times.
\item[merge phase] There are $c$ \emph{mergers}. Merger $i$ uses columns $B_{i, \, t}$ and $\widehat B_{i, \, t}$, and computes columns $M_{i, \, t}$:
\begin{equation} \label{eq:merger}
M_{i, \, t} = B_{i, \, t} \, \cap \, \widehat B_{i, \, t}
\hspace{1.5cm} 1 \leq t \leq k
\end{equation}
From the build phase, it follows that columns $B_{i, \, t}$ and $\widehat B_{i, \, t}$ are produced after scanning character $y_{i, \, t}$ from left to right and after scanning character $y_{i, \, t + 1}$ from right to left, respectively, thus their intersection $M_{i, \, t}$ refers to the same text position between the two characters. Therefore column $M_{i, \, t}$ stores the segments that are both accessible and post-accessible when parsing text $x$, i.e., those of the clean SLPF. In particular it holds $M_{i, \, k} = B_{i, \, k} \, \cap \, \widehat B_{i, \, k} = J_i \, \cap \, \widehat J_{i + 1}$ (see the build phase).
\item[compose phase] The \emph{composer} computes column $C_0 = J_0 \, \cap \, \widehat J_1$ and concatenates the other $n$ columns as $C_r \vert_{\, r \, = \, (i - 1) \, \times \, k \, + \, t} = M_{i, \, t}$, for $1 \leq r \leq n$ and $1 \leq t \leq k$, to output the final clean SLPF.
\end{description}
\begin{table}[ht]
\begin{center}
\definecolor{lightgray}{rgb}{0.875,0.875,0.875}
\vspace{0.125cm}
\tabcolsep=0.0cm
\def\arraystretch{0.5}
\begin{tabular}{p{0.475\textwidth}@{\hspace{0.5cm}}p{0.475\textwidth}} \toprule
\cellcolor{lightgray}
\begin{minipage}[t]{0.475\textwidth}
\begin{algorithm2e}[H]
\setstretch{1.0}
\SetAlgorithmName{component (task)}{}{}
\TitleOfAlgo{\emph{splitter}} \hrule height 0.25pt
\SetKwInOut{Input}{{\i}nput}
\SetKwInOut{Output}{output}
\SetKwInOut{LocVars}{variable~}
\DontPrintSemicolon
\Input{text $x$}
\Output{series $y$ of text chunks}
\LocVars{text positions $r$ and $t$, chunk number $i$} \hrule height 0.25pt
\For (\tcp*[f]{scan whole text}) {\rm $r = 1$ \textbf{to} $n$} {
$i \asgn r \ \text{div} \ c + 1$  \tcp*[r]{compute chunk num}
$t \asgn r \ \text{mod} \ c + 1$  \tcp*[r]{compute text pos}
$y \, [i, \, t] \asgn x \, [r]$ \tcp*[r]{compute chunk}
}
\end{algorithm2e}
\end{minipage}
&
\cellcolor{lightgray}
\begin{minipage}[t]{0.475\textwidth}
\begin{algorithm2e}[H]
\setstretch{1.0}
\SetAlgorithmName{component (task)}{}{}
\TitleOfAlgo{\emph{composer}} \hrule height 0.25pt
\SetKwInOut{Input}{{\i}nput}
\SetKwInOut{Output}{output}
\SetKwInOut{LocVars}{variable~}
\DontPrintSemicolon
\Input{series $M$ of clean SLPF columns, sets $J_0$ and $\widehat J_1$ of edge segments}
\Output{series $C$ of clean SLPF columns}
\LocVars{text positions $r$ and $t$, chunk number $i$} \hrule height 0.25pt
$C \, [0] \asgn J_0 \, \cap \, \widehat J_1$ \tcp*[r]{compute ini column}
\For (\tcp*[f]{scan SLPF columns}) {\rm $r = 1$ \textbf{to} $n$} {
$i \asgn r \ \text{div} \ c + 1$  \tcp*[r]{compute chunk num}
$t \asgn r \ \text{mod} \ c + 1$  \tcp*[r]{compute text pos}
$C \, [r] \asgn M \, [i, \, t]$ \tcp*[r]{compute SLPF col}
}
\end{algorithm2e}
\end{minipage}
\\ \midrule
\cellcolor{lightgray}
\begin{minipage}[t]{0.475\textwidth}
\begin{algorithm2e}[H]
\setstretch{1.0}
\SetAlgorithmName{component (task)}{}{}
\TitleOfAlgo{\emph{FW-reacher} \hfill Eq.~\eqref{eq:reacher}} \hrule height 0.25pt
\SetKwInOut{Input}{{\i}nput}
\SetKwInOut{Output}{output}
\SetKwInOut{LocVars}{variable~}
\DontPrintSemicolon
\Input{set $I_\text{ME-DFA}$ of initial states, text chunk $y_i$, chunk number $i$}
\Output{array $R_i$ of edge segment sets}
\LocVars{segment set $S$, segment $j$, text position $t$} \hrule height 0.25pt
\ForEach (\tcp*[f]{scan segments}) {$\set{ \, j \, } \in I_\textrm{\rm ME-DFA}$} {
$S \asgn \set{ \, j \, }$ \tcp*[r]{initialize $S$}
\For (\tcp*[f]{scan chunk text}) {\rm $t = 1$ \textbf{to} $k$} {
$S \asgn \delta_\textrm{ME-DFA} \, \big( S, \, y \, [i, \, t] \big)$ \tcp*[r]{comp. $\delta^\ast$}
}
$R \, [i, j] \asgn S$ \tcp*[r]{compute $R_{i, \, j}$}
}
\end{algorithm2e}
\end{minipage}
&
\cellcolor{lightgray}
\begin{minipage}[t]{0.475\textwidth}
\begin{algorithm2e}[H]
\setstretch{1.0}
\SetAlgorithmName{component (task)}{}{}
\TitleOfAlgo{\emph{FW-joiner} \hfill Eq.~\eqref{eq:joiner}} \hrule height 0.25pt
\SetKwInOut{Input}{{\i}nput}
\SetKwInOut{Output}{output}
\SetKwInOut{LocVars}{variable~}
\DontPrintSemicolon
\Input{series $R$ of arrays of edge segment sets, set $I$ of initial segments}
\Output{series $J$ of edge segment sets}
\LocVars{chunk number $i$, segment $j$} \hrule height 0.25pt
$J \, [0] \asgn I$ \tcp*[r]{initialize $J_0$}
\For (\tcp*[f]{scan chunk series}) {\rm $i = 1$ \textbf{to} $c$} {
$J \, [i] \asgn \emptyset$ \tcp*[r]{initialize $J_i$}
\ForEach (\tcp*[f]{scan segments}) {$j \in J \, [i-1]$} {
$J \, [i] \asgn J \, [i] \, \cup \, R \, [i, \, j]$ \tcp*[r]{compute $J_i$}
}
}
\end{algorithm2e}
\end{minipage}
\\ \midrule
\cellcolor{lightgray}
\begin{minipage}[t]{0.475\textwidth}
\begin{algorithm2e}[H]
\setstretch{1.0}
\SetAlgorithmName{component (task)}{}{}
\TitleOfAlgo{\emph{FW-builder} \hfill Eq.~\eqref{eq:builder}} \hrule height 0.25pt
\SetKwInOut{Input}{{\i}nput}
\SetKwInOut{Output}{output}
\SetKwInOut{LocVars}{variable~}
\DontPrintSemicolon
\Input{set $J_{i - 1}$ of edge segments, text chunk $y_i$, chunk number $i$}
\Output{subseries $B_i$ of SLPF columns}
\LocVars{text position $t$} \hrule height 0.25pt
$B \, [i, \, 1] \asgn \delta_\textrm{DFA} \, \big( J \, [i-1], \, y \, [i, \, 1] \big)$ \tcp*[r]{comp. $B_{i, \, 1}$}
\For (\tcp*[f]{scan chunk text}) {\rm $t = 2$ \textbf{to} $k$} {
\tcp*[l]{compute $B_{i, \, t}$}
$B \, [i, \, t] \asgn \delta_\textrm{DFA} \, \big( B \, [i, \, t-1], \, y \, [i, \, t] \big)$
}
\end{algorithm2e}
\end{minipage}
&
\cellcolor{lightgray}
\begin{minipage}[t]{0.475\textwidth}
\begin{algorithm2e}[H]
\setstretch{1.0}
\SetAlgorithmName{component (task)}{}{}
\TitleOfAlgo{\emph{merger} \hfill Eq.~\eqref{eq:merger}} \hrule height 0.25pt
\SetKwInOut{Input}{{\i}nput}
\SetKwInOut{Output}{output}
\SetKwInOut{LocVars}{variable~}
\DontPrintSemicolon
\Input{subseries $B_i$ and $\widehat B_i$ of columns, chunk number $i$}
\Output{subseries $M_i$ of clean SLPF columns}
\LocVars{text position $t$} \hrule height 0.25pt
\For (\tcp*[f]{scan chunk text}) {\rm $t = 1$ \textbf{to} $k$} {
\tcp*[l]{compute $M_{i, \, t}$}
$M \, [i, \, t] \asgn B \, [i, \, t] \, \cap \, \widehat B \, [i, \, t]$}
\end{algorithm2e}
\end{minipage} \\ \bottomrule
\end{tabular}
\end{center}
\caption{Pseudo-code of the algorithmic components (tasks) of the parallel parser.
} \label{tab:pseudocode}
\end{table}
The pseudo-code of the algorithmic components of the parallel parser is listed in Tab.~\ref{tab:pseudocode}: splitter, reacher, joiner, builder, merger and composer. Such algorithms compute the formulas in Eq.s~\eqref{eq:reacher}, \eqref{eq:joiner}, \eqref{eq:builder} and \eqref{eq:merger}. The reacher and builder use the transition functions $\delta_\text{ME-DFA}$ and $\delta_\text{DFA}$ of the ME-DFA and DFA, respectively, illustrated in Sect.~\ref{subsec:deterministicFA}. Such functions can be easily implemented as look-up tables, like those shown in  Ex.~$6$ below, and are not detailed. The look-up tables are prepared stored ready for the parallel parser when the segments are computed from the RE, and the ME-DFA and DFA are built. The segment sets are implemented as Boolean vectors, see Sect.~\ref{subsec:serialalgorithmVectorial} and Eq.~\eqref{eq:matrixmul}. Similarly, the arrays of segments sets are implemented as Boolean matrices. In this way all set operations are efficiently executed  as bitwise logical operations. The BW versions of the reacher, joiner and builder are  similar to the FW ones.
\par
To recapitulate, the parallel algorithm first scans the chunks concurrently, finds the edge segments of the SLPF and stores them in the data-structure $J$. Then it scans again the chunks concurrently, starting from the edge segments already identified and stored in the data-structure $J$, finds all the other segments of the SLPF and stores them in the data-structure $B$. Both scans are also carried out backwards and update the respective data-structures. The algorithm is correct, because it actually reproduces all the computations of the NFA (and reverse NFA). Its two crucial phases are reach and build, linked by the join phase. During reach, the ME-DFA reproduces all the NFA computations at the chunk level, thus it identifies the accessible (or post-accessible) edge segments for each chunk separately. The join phase propagates the accessibility property (or post-accessibility in the backward scan) through the chunk series, thus it computes the edge segments of the SLPF. Note that any column, i.e., segment set, produced by join is necessarily a DFA state, because such a set includes only accessible (or post-accessible) segments. For this reason, the segment sets can be passed to the build phase, which is driven by the DFA. Within the builder, the DFA reproduces all the NFA computations at the grain of each input symbol (character), thus it identifies all the accessible (or post-accessible) SLPF segments. Eventually, the merge phase combines the accessibility and post-accessibility properties, by pairwise intersecting the accessible and post-accessible columns in the same position, and so obtains the final clean SLPF. An example clarifies.
\paragraph{Example $6$ -- parallel parser for RE $e_2$}
Consider the RE $e_2$, with segments $q_1 \ldots q_{10}$ (see Tab.~\ref{tab:segments}) simply denoted $1 \ldots 10$. The related NFA, DFA and ME-DFA are shown in Fig.s~\ref{fig:NFA-RE2}, \ref{fig:DFA2} and \ref{fig:MEDFA2}. The NFA states $Q$ are simply the segments $1 \ldots 10$. The states of the DFA and ME-DFA are the segment sets denoted $T$ and $S$, and their transition functions are $\delta_\text{DFA}$ and $\delta_\text{ME-DFA}$, respectively, as follows:
\begin{center}
$\def\arraystretch{1.125}
\arraycolsep=0.0cm
\begin{array}{l@{\hspace{1.5cm}}l@{\hspace{1.5cm}}l}
\underbracket[0.75pt]{
\arraycolsep=0.1cm
\begin{array}{wr{0.5cm}cwl{3.0cm}}
T_1 & = & \set{ \, 1, \, 2, \, 3 \, } = I \\
T_2 & = & \set{ \, 4, \, 7, \, 8, \, 10 \, } = S_{12} \\
T_3 & = & \set{ \, 5, \, 6, \, 9 \, } = S_{13}
\end{array}}_\text{state set of the DFA}
&
\multicolumn{2}{l}{
\arraycolsep=0.1cm
\begin{array}{wr{1.0cm}|ccc}
\delta_\text{DFA} & T_1 & T_2 & T_3 \\ \toprule
a                 & T_2 & T_2 & T_2 \\
b                 & -   & T_3 & -
\end{array}
\hspace{1.25cm}
\underbracket[0.75pt]{Q = \set{1,\ldots,10}}_\text{segments (NFA states)} \qquad \underbracket[0.75pt]{I = \set{1,2,3}}_\text{initial segments} \qquad \underbracket[0.75pt]{F = \set{1,9,10}}_\text{final segments}}
\\ \\
\underbracket[0.75pt]{
\arraycolsep=0.1cm
\begin{array}{wr{0.5cm}cwl{3.0cm}}
S_j    & = & \overbracket[0.75pt]{\set{ \, j \, } \quad \text{for $1 \leq j \leq 10$}}^\text{one state $S_j$ per each segment} \\
S_{11} & = & \set{ \, 7, \, 8, \, 10 \, } \\
S_{12} & = & \set{ \, 4, \, 7, \, 8, \, 10 \, } = T_2 \\
S_{13} & = & \set{ \, 5, \, 6, \, 9 \, } = T_3
\end{array}}_\text{state set of the ME-DFA}
&
\arraycolsep=0.1cm\begin{array}{wr{1.0cm}|cccccccccc|ccc}
& \multicolumn{10}{c}{\text{$I_\text{ME-DFA}$ \ (initial states of the ME-DFA)}} & & & \\
\delta_\text{ME-DFA} & S_1 & S_2 & S_3    & S_4    & S_5 & S_6    & S_7 & S_8    & S_9 & S_{10} & S_{11} & S_{12} & S_{13} \\ \toprule
a                    & -   & S_4 & S_{11} & -      & S_4 & S_{11} & S_4 & S_{11} & -   & -      & S_{12} & S_{12} & S_{12} \\
b                    & -   & -   & -      & S_{13} & -   & -      & -   & -      & -   & -      & -      & S_{13} & -
\end{array}
\end{array}$
\end{center}
For the input text $x = a \, b \, a \, a \, b \, a$ of length $n = 6$, split into $c = 3$ chunks of length $k = 2$, the computation is sketched below:
\begin{center}
$\definecolor{midgray}{rgb}{0.625,0.625,0.625}
\arraycolsep=0.05cm
\def\arraystretch{1.125}
\begin{array}{wl{1.0cm}wc{0.5cm}|wc{0.5cm}|wc{2.25cm}wc{2.25cm}wc{2.25cm}wc{2.25cm}wc{2.25cm}wc{2.25cm}wc{2.25cm}wc{2.25cm}wc{2.25cm}wc{2.25cm}wc{2.25cm}wc{2.25cm}|wc{0.5cm}} \toprule
\textbf{text $x$} &   &   & a \psspan 2 & b \psspan 2 & a \psspan 2 & a \psspan 2 & b \psspan 2 & a \psspan 2 & \\ \toprule
\text{index \ $r$}&   &   & 1 \psspan 2 & 2 \psspan 2 & 3 \psspan 2 & 4 \psspan 2 & 5 \psspan 2 & 6 \psspan 2 & \\ \midrule
\text{index \ $i$}&   & 0 & \cellcolor{midgray} 1 \psspan 4 & \cellcolor{midgray} 2 \psspan 4 & \cellcolor{midgray} 3 \psspan 4 & 4 \\ \midrule
\text{index \ $t$}&   &   & 1 \psspan 2 & 2 \psspan 2 & 1 \psspan 2 & 2 \psspan 2 & 1 \psspan 2 & 2 \psspan 2 & \\ \bottomrule
\textbf{phase}    &   &   & \text{for brevity the functions $\delta_\text{ME-DFA}$ and $\delta_\text{DFA}$ are respectively denoted as $\delta_1$ and $\delta_2$} \psspan{13} \\ \toprule
\text{split}      & y &   & \cellcolor{midgray} y_{1,1} = a \psspan 2 & \cellcolor{midgray} y_{1,2} = b \psspan 2 & \cellcolor{midgray} y_{2,1} = a \psspan 2 & \cellcolor{midgray} y_{2,2} = a \psspan 2 & \cellcolor{midgray} y_{3,1} = b \psspan 2 & \cellcolor{midgray} y_{3,2} = a \psspan 2 & \\ \midrule

\text{reach}      & R &   & \def\arraystretch{1.0} \begin{array}[t]{l} R_{1, \, 2} = \delta_1^\ast \left( S_2, \, a\,b \right) = S_{13} \\ R_{1, \, 5} = \delta_1^\ast \left( S_5, \, a\,b \right) = S_{13} \\ R_{1, \, 7} = \delta_1^\ast \left( S_7, \, a\,b \right) = S_{13} \\ R_{1, \, \text{any other segment}} = \emptyset \end{array} \psspan 4 & \def\arraystretch{1.0} \begin{array}[t]{l} R_{2, \, 3} = \delta_1^\ast \left(S_3, \, a\,a \right) = S_{12} \\ R_{2, \, 6} = \delta_1^\ast \left(S_6, \, a\,a \right) = S_{12} \\ R_{2, \, 8} = \delta_1^\ast \left(S_8, \, a\,a \right) = S_{12} \\ R_{2, \, \text{any other segment}} = \emptyset \end{array} \psspan 4 & \def\arraystretch{1.0} \begin{array}[t]{l} R_{3, \, 4} = \delta_1^\ast \left(S_4, \, b\,a \right) = S_{12} \\ R_{3, \, \text{any other segment}} = \emptyset \end{array} \psspan 4 & \\ \midrule

\text{join}       & J & \rnode{J0A}{I} & \underbracket[0.75pt]{R_{1, \, 1} \cup R_{1, \, 2} \cup  R_{1, \, 3}}_\text{union ranging over set \rnode{J0C}{$I_{\phantom{0}}$}} = \rnode{J1A}{S_{13}} \psspan 4 & \underbracket[0.75pt]{R_{2, \, 5} \cup R_{2, \, 6} \cup  R_{2, \, 9}}_\text{union ranging over set \rnode{J1C}{$S_{13}$}} = \rnode{J2A}{S_{12}} \psspan 4 & \underbracket[0.75pt]{R_{3, \, 4} \cup  R_{3, \, 7} \cup  R_{3, \, 8} \cup  R_{3, \, 10}}_\text{union ranging over set \rnode{J2C}{$S_{12}$}} = \rnode{J3A}{S_{12}} \psspan 4 & \\ \midrule

\text{build}      & B & & \delta_2 \left( \rnode{J0B}{I}, \, a \right) = \rnode{B1A}{T_2} \psspan 2 & \delta_2 \left( \rnode{B1B}{T_2}, \, b \right) = \rnode{B1D}{T_3} \psspan 2 & \delta_2 \left( \rnode{J1B}{S_{13}}, \, a \right) = \rnode{B2A}{T_2} \psspan 2 & \delta_2 \left( \rnode{B2B}{T_2}, \, a \right) = \rnode{B2D}{T_2} \psspan 2 & \delta_2 \left( \rnode{J2B}{S_{12}}, \, b \right) = \rnode{B3A}{T_3} \psspan 2 & \delta_2 \left( \rnode{B3B}{T_3}, \, a \right) = \rnode{B3D}{T_2} \psspan 2 & \\ \bottomrule

\text{reach}^R      & \widehat R &    & \hspace{2pt} \text{omitted} \psspan {12} & \\ \midrule

\text{join}^R       & \widehat J &    & \multicolumn{12}{|c|}{\text{omitted}} & F \\ \midrule

\text{build}^R      & \widehat B &    & \set{4} \psspan 2 & \set{3,6,8} \psspan 2 & \set{2,5,7} \psspan 2 & \set{4} \psspan 2 & \set{3,6,8} \psspan 2 & F \psspan 2 \\ \bottomrule

\text{merge}      & M &         & \rnode{B1C}{T_2} \cap \set{4} \psspan 2 & \rnode{B1E}{T_3} \cap \set{3,6,8} \psspan 2 & \rnode{B2C}{T_2} \cap \set{2,5,7} \psspan 2 & \rnode{B2E}{T_2} \cap \set{4} \psspan 2 & \rnode{B3C}{T_3} \cap \set{3,6,8} \psspan 2 & \rnode{B3E}{T_2} \cap F \psspan 2 &\\ \midrule

\text{comp.}      & C & \set{2} & \set{4} \psspan 2 & \set{6} \psspan 2 & \set{7} \psspan 2 & \set{4} \psspan 2 & \set{6} \psspan 2 & \set{10} \psspan 2 & \\ \bottomrule

\psset{nodesepA=3pt, nodesepB=3pt, linestyle=solid, linewidth=1pt, arrows=->, arrowscale=1.75, border=0.05cm}

\nccurve[angleA=-60, angleB=140] {J0A} {J0B}
\nccurve[angleA=-42.5, angleB=-155] {J0A} {J0C}
\nccurve[angleA=-45, angleB=-135, nodesep=2pt, ncurv=1] {B1A} {B1B}

\nccurve[angleA=-45, angleB=150] {J1A} {J1B}
\nccurve[angleA=-30, angleB=-160] {J1A} {J1C}
\nccurve[angleA=-45, angleB=-135, nodesep=2pt, ncurv=1] {B2A} {B2B}

\nccurve[angleA=-45, angleB=150] {J2A} {J2B}
\nccurve[angleA=-30, angleB=-160] {J2A} {J2C}
\nccurve[angleA=-45, angleB=-135, nodesep=2pt, ncurv=1] {B3A} {B3B}

\nccurve[angleA=-120, angleB=120, nodesep=2pt, ncurv=1] {B1A} {B1C}
\nccurve[angleA=-120, angleB=120, nodesep=2pt, ncurv=1] {B1D} {B1E}

\nccurve[angleA=-120, angleB=120, nodesep=2pt, ncurv=1] {B2A} {B2C}
\nccurve[angleA=-120, angleB=120, nodesep=2pt, ncurv=1] {B2D} {B2E}

\nccurve[angleA=-120, angleB=120, nodesep=2pt, ncurv=1] {B3A} {B3C}
\nccurve[angleA=-120, angleB=120, nodesep=2pt, ncurv=1] {B3D} {B3E}
\end{array}$
\end{center}
The arrows in the simulation indicate the data flow. In the reach phase, the extended transition function $\delta_\text{ME-DFA}^\ast$ is computed stepwise, see the state-transition table of $\delta_\text{ME-DFA}$ above and also the reacher algorithm in Tab.~\ref{tab:pseudocode}. We give here a step-by-step description of the forward computation, starting with notation and initialization:
\begin{description}[leftmargin=1.125cm, style=nextline]
\item[notation and initialization] the ME-DFA initial states (set $I_\text{ME-DFA}$) are the singletons $S_1 = \set { \, 1 \, }$, \ldots,  $S_{10} = \set { \, 10 \, }$, each one containing one segment $1$, \ldots, $10$, see also the ME-DFA in Fig.~\ref{fig:MEDFA2}; the (unique) initial state of the DFA is $T_1 = I = \set{ \, 1, \, 2, \, 3 \, }$, see also the DFA in Fig.~\ref{fig:DFA2}; and the initial states (set $I$) of the NFA are simply the three segments $1$, $2$ and $3$, see also the NFA in Fig.~\ref{fig:NFA-RE2}; for brevity the ME-DFA and DFA transition functions are denoted $\delta_1$ and $\delta_2$, respectively
\end{description}
Then we continue with the sequence of phases of the algorithm:
\begin{description}[leftmargin=1.125cm, style=nextline]
\item[split] the text $x = a\,b\,a\,a\,b\,a$ is split into three chunks $y_1 = a\,b$, $y_2 = a\,a$ and $y_3 = b\,a$ of equal length $k = 2$; the indices $r$ and $t$ enumerate the characters over the whole text and inside each chunk, respectively; index $i$ enumerates the chunks; each text character $x_r$ is also denoted $y_{i, \, t}$ with $r = (i - 1) \, k + t$; e.g., $x_1 = y_{1, \, 1} = a$, $x_2 = y_{1, \, 2} = b$, $x_3 = y_{2, \, 1} = a$, etc.
\item[reach] by using the forward Eq.~\eqref{eq:reacher} (or the reacher algorithm in Tab.~\ref{tab:pseudocode}), the ME-DFA state $R_{i,\,j}$ that contains the edge segments reached by starting from the ME-DFA initial (singleton) state $S_j = \set{ \, j \, }$ and scanning chunk $y_i$, is computed by means of the extended transition function $\delta_1^\ast$; for example:
\begin{itemize}[leftmargin=*, style=standard]
\item for chunk $y_1 = a\,b$ from the initial state $S_2 = \set{ \, 2 \, }$ it is $R_{1, \, 2} = \delta_1^\ast \left( S_2, \, a\,b \right) = \delta_1 \left( \delta_1 \left( S_2, \, a \right), \, b \right) = \delta_1 \left( S_4, \, b \right) = S_{13} = \set{ \, 5, \, 6, \, 9 \, }$, that is, from the edge segment $2$ chunk $1$ reaches the three edge segments $5$, $6$ and $9$
\item while for $y_2 = a\,a$ from $S_3 = \set{ \, 3 \, }$ it is $R_{2, \, 3} = \delta_1^\ast \left(S_3, \, a\,a \right) = S_{12} = \set{ \, 4, \, 7, \, 8, \, 10 \, }$
\end{itemize}
    see alse the NFA in Fig.~\ref{fig:NFA-RE2}, where the transitions between single segments are visible one-by-one; the entire array $R$ of edge segments is computed and shown above, for every chunk and ME-DFA initial state
\item[join] by using the forward Eq.~\eqref{eq:joiner} (or the joiner algorithm in Tab.~\ref{tab:pseudocode}), for each chunk $y_i$ all the segment sets $R_{i, \, j}$ reached starting from any segment $j$ belonging to the previous segment set $J_{i - 1}$ and scanning chunk $y_i$, are united into one new segment set $J_i$; this (recursive) computation starts from uniting the initial states, i.e., segments, of the NFA (or equivalently from the unique initial state of the DFA), i.e., $J_0 = I = \set{ \, 1, \, 2, \, 3 \, } = T_1$; for example:
\begin{itemize}[leftmargin=*, style=standard]
\item for chunk $y_1$ it is $J_1 = \underbracket[0.75pt]{\, R_{1, \, 1} \cup R_{1, \, 2} \cup  R_{1, \, 3} \,}_\text{segments $1$, $2$, $3$ belong to $J_0$} = S_{13} = \set{ \, 5, \, 6, \, 9 \, }$ (note that both $R_{1, \, 1}$ and $R_{1, \, 3}$ are empty), that is, from the initial edge segments $1$, $2$ and $3$ of the NFA, chunk $1$ reaches the edge segments $5$, $6$ and $9$
\item while for $y_2$ it is $J_2 = \underbracket[0.75pt]{\, R_{2, \, 5} \cup R_{2, \, 6} \cup  R_{2, \, 9} \,}_\text{segments $5$, $6$, $9$ belong to $J_1$} = S_{12} = \set{ \, 4, \, 7, \, 8, \, 10 \, }$ (here $R_{2, \, 5} = R_{2, \, 9} = \emptyset$), that is, from the edge segments $5$, $6$ and $9$ previoulsy reached by chunk $1$ (starting from the initial states of the NFA), chunk $2$ reaches the edge segments $4$, $7$, $8$ and $10$
\end{itemize}
all the united edge segments $J$ are computed and shown above, for every chunk; note that by construction the join segment sets $J$ coincide with DFA states (see the state tables above)
\item[build] by using the forward Eq.~\eqref{eq:builder} (or the builder algorithm in Tab.~\ref{tab:pseudocode}), the series of segment sets $B_{i, \, t}$ (columns $B$) reached by wholly scanning the chunks from $y_1$ to $y_{i - 1}$ and the characters from $1$ to $t$ inside chunk $y_i$, i.e, by scanning the whole text $x$ from char $x_1$ to char $x_{r \, = \, (i - 1) \, k \, + \, t}$, is computed by means of the transition function $\delta_2$; for example:
\begin{itemize}[leftmargin=*, style=standard]
\item  for character $y_{1, \, 1} = a$ from the initial segments of the NFA it is $B_{1, \, 1} = \delta_2 \left( J_0 = I = T_1, \, a \right) = T_2 = \set{ \, 4, \, 7, \, 8, \, 10 \, }$, that is, by scanning the first text character $x_{1 \, = \, (1 - 1) \, 2 \, + \, 1} = a$ the segments $4$, $7$, $8$ and $10$ are reached
\item for $y_{2, \, 1} = a$ from the segments $J_1$ reached by the first chunk it is $B_{2, \, 1} = \delta_2 \left( J_1 = S_{13} = T_2, \, a \right) = T_2 = \set{ \, 4, \, 7, \, 8, \, 10 \, }$, that is, by scanning the third text character $x_{3 \, = \, (2 - 1) \, 2 \, + \, 1} = a$ (first of the second chunk) the segments $4$, $7$, $8$ and $10$ are reached (again, see above)
\item while (see above) $B_{1, \, 2} = \set{ \, 5, \, 6, \, 9 \, }$, coincident with the join set $J_1$ (edge segments) as noted above (reach phase)
\end{itemize}
all columns $B$ are computed and shown above, for every text character
\end{description}
Now the phases $\text{reach}^R$, $\text{join}^R$ and $\text{build}^R$ of the backward computation should be run, as of the backward Eq.s~\eqref{eq:reacher}, \eqref{eq:joiner} and \eqref{eq:builder} (or the corresponding backward algorithms in Tab.~\ref{tab:pseudocode}), but for brevity they are omitted and only the result of the backward build is shown, i.e., all columns $\widehat B$. Note that the states of the $\widehat{\text{ME-DFA}}$ and $\widehat{\text{DFA}}$ (reverse ME-DFA and DFA) are also segment sets, but may differ from those of the (direct) ME-DFA and DFA, respectively.  Anyway, the backward columns $\widehat B$ eventually obtained can be verified on the NFA of Fig.~\ref{fig:NFA-RE2}, by reversing the arcs, switching the initial / final segments and scanning backwards each chunk. For instance, to compute the column $\widehat J_1 = \set{ \, 2 ,\, 5, \, 7 \, }$ needed in the compose phase, reverse the NFA of Fig.~\ref{fig:NFA-RE2}, start from column $\widehat J_2$ and scan the reverse chunk $y_1^R = b\,a$ (see also below). We resume and conclude the step-by-step description:
\begin{description}[leftmargin=1.125cm, style=nextline]
\item[merge] by using  Eq.~\eqref{eq:merger} (or the merger algorithm in Tab.~\ref{tab:pseudocode}), the series of segment sets $M_{i, \, t}$ (columns $M$) that are part of the linearized syntax tree(s) (LSTs) of the whole text is computed, by intersecting the corresponding reached states of the DFA and $\widehat{\text{DFA}}$; e.g., $M_{1, \, 1} = T_2 \, \cap \, \set{4} = \set{ \, 4, \, 7, \, 8, \, 10 \, } \, \cap \, \set{ \, 4 \, } = \set{ \, 4 \, }$, $M_{1, \, 2} = T_3 \, \cap \, \set{ \, 3, \, 6, \, 8 \, } = \set{ \, 5, \, 6, \, 9 \, }  \, \cap \, \set{ \, 3, \, 6, \, 8 \, } = \set{ \, 6 \, }$, and so on; all columns $M$ are computed and shown above, for every text character
\item[comp.] the segment sets $C_{r \, = \, (i - 1) \, k \, + \, t} = M_{i, \, t}$ are the SLPF representation, plus set $C_0$; e.g., $C_0 = J_0 \, \cap \, \widehat J_1 = \set{ \, 1, \, 2, \, 3 \, } \, \cap \, \set{ \, 2 ,\, 5, \, 7 \, } = \set{ \, 2 \, }$, $C_1 = M_{1, \, 1} = \set{ \, 4 \, }$, $C_2 = M_{1, \, 2} = \set{ \, 6\, }$, and so on; all columns $C$ are listed above, for every text character, i.e., the whole clean SLPF; see also the forward and backward SLPFs below, and their intersection
\end{description}
The final clean SLPF encodes only one LST, as the text $x$ is not ambiguous. The accessible and post-accessible segments computed in the build phase are as follows, with their transitions as in the NFA and its reverse, i.e., $\widehat{\text{NFA}}$:
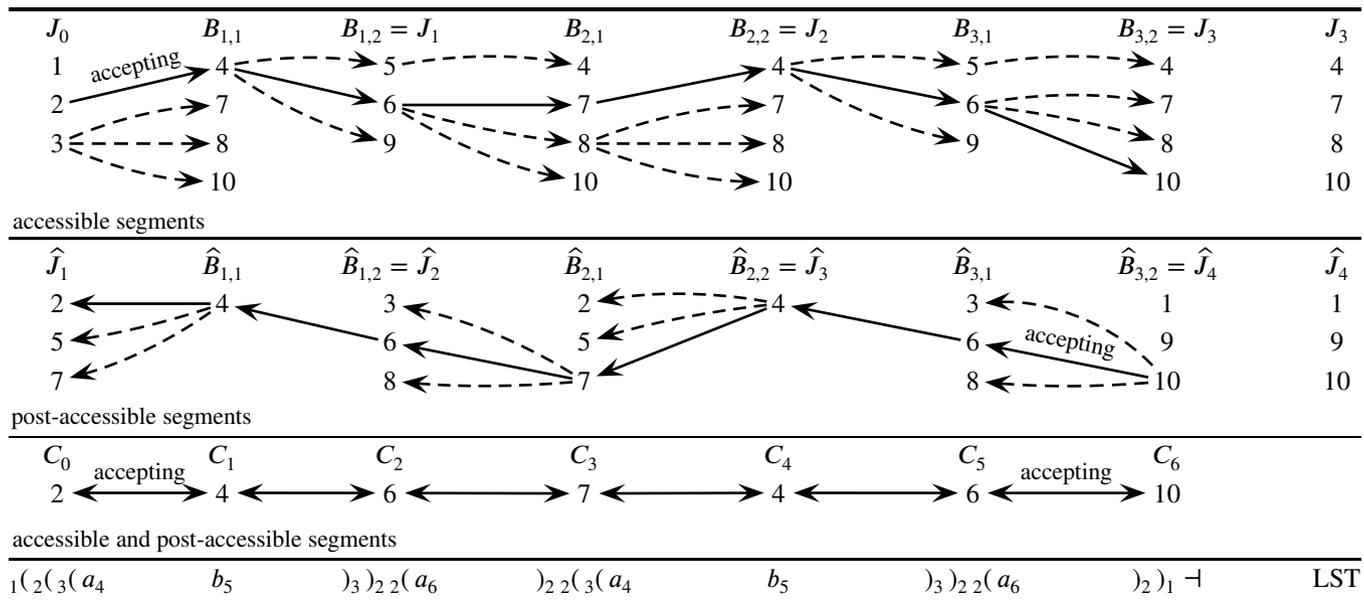
\begin{center}
\vspace{0.25cm}
$\psset{nodesep=2pt, linestyle=dashed, linewidth=1pt, arrows=->, arrowscale=2, colsep=1.25cm, rowsep=0.05cm, border=0.00cm}
\begin{psmatrix} \toprule
J_0      & B_{1,1}      & B_{1,2}=J_1 & B_{2,1}      & B_{2,2}=J_2  & B_{3,1}      & B_{3,2}=J_3  & J_3       \\
1        & 4            & 5           & 4            & 4            & 5            & 4            & 4         \\
2        & 7            & 6           & 7            & 7            & 6            & 7            & 7         \\
3        & 8            & 9           & 8            & 8            & 9            & 8            & 8         \\
         & 10           &             & 10           & 10           &              & 10           & 10        \\[0.5cm] \midrule
\nput[labelsep=5pt] {68} {6,1} {\textrm{\scriptsize accessible segments}}
\ncline[labelsep=3pt, linestyle=solid] {3,1} {2,2} \naput[labelsep=2pt, nrot=:U] {\textrm{\scriptsize accepting}}
\ncarc[arcangle=10] {4,1} {3,2}
\ncline {4,1} {4,2}
\ncarc[arcangle=-10] {4,1} {5,2}
\ncarc[arcangle=10] {2,2} {2,3}
\ncline[linestyle=solid] {2,2} {3,3}
\ncarc[arcangle=-10] {2,2} {4,3}
\ncarc[arcangle=10] {2,3} {2,4}
\ncline[linestyle=solid] {3,3} {3,4}
\ncarc[arcangle=-5] {3,3} {4,4}
\ncarc[arcangle=-10] {3,3} {5,4}
\ncline[linestyle=solid] {3,4} {2,5}
\ncarc[arcangle=10] {4,4} {3,5}
\ncline {4,4} {4,5}
\ncarc[arcangle=-10] {4,4} {5,5}
\ncarc[arcangle=10] {2,5} {2,6}
\ncline[linestyle=solid] {2,5} {3,6}
\ncarc[arcangle=-10] {2,5} {4,6}
\ncarc[arcangle=10] {2,6} {2,7}
\ncarc[arcangle=10] {3,6} {3,7}
\ncarc[arcangle=5] {3,6} {4,7}
\ncline[linestyle=solid] {3,6} {5,7}
\widehat J_1 & \widehat B_{1,1} & \widehat B_{1,2} = \widehat J_2 & \widehat B_{2,1} & \widehat B_{2,2} = \widehat J_3 & \widehat B_{3,1} & \widehat B_{3,2} = \widehat J_4 & \widehat J_4  \\
2        & 4            & 3            & 2            & 4            & 3            & 1            & 1         \\
5        &              & 6            & 5            &              & 6            & 9            & 9         \\
7        &              & 8            & 7            &              & 8            & 10           & 10        \\[0.5cm] \midrule
\nput[labelsep=6pt] {63} {10,1} {\textrm{\scriptsize post-accessible segments}}
\ncarc[arcangle=-25] {9,7} {7,6}
\ncline[linestyle=solid] {9,7} {8,6} \nbput[labelsep=2pt, nrot=:D] {\textrm{\scriptsize accepting}}
\ncarc[arcangle=5] {9,7} {9,6}
\ncline[linestyle=solid] {8,6} {7,5}
\ncarc[arcangle=-10] {7,5} {7,4}
\ncarc[arcangle=-5] {7,5} {8,4}
\ncline[linestyle=solid] {7,5} {9,4}
\ncarc[arcangle=-10] {9,4} {7,3}
\ncline[linestyle=solid] {9,4} {8,3}
\ncarc[arcangle=5] {9,4} {9,3}
\ncline[linestyle=solid] {8,3} {7,2}
\ncline[linestyle=solid] {7,2} {7,1}
\ncarc[arcangle=5] {7,2} {8,1}
\ncarc[arcangle=10] {7,2} {9,1}
C_0      & C_1          & C_2          & C_3          & C_4          & C_5          & C_6          &           \\
\rnode{C0}{2}        & \rnode{C1}{4}            & \rnode{C2}{6}            & \rnode{C3}{7}            & \rnode{C4}{4}            & \rnode{C5}{6}            & \rnode{C6}{10}           &           \\[0.625cm] \midrule
\nput[labelsep=6pt] {56} {12,1} {\textrm{\scriptsize accessible and post-accessible segments}}
_1( \, _2( \, _3( \, a_4 & b_5 & )_3 \, )_2 \, _2( \, a_6 & )_2 \, _2( \, _3( \, a_4 & b_5 & )_3 \, )_2 \, _2( \, a_6 & )_2 \, )_1 \dashv & \text{LST} \\ \bottomrule
\psset{nodesep=3pt, arrows=<->, linestyle=solid}
\ncline {C0} {C1} \naput[labelsep=2pt] {\textrm{\scriptsize accepting}}
\ncline {C1} {C2}
\ncline {C2} {C3}
\ncline {C3} {C4}
\ncline {C4} {C5}
\ncline {C5} {C6} \naput[labelsep=2pt] {\textrm{\scriptsize accepting}}
\end{psmatrix}$
\end{center}
Many segments are canceled after the intersection, thus leaving just one LST with the graphic representation below:
\begin{center}
$\vspace{0.125cm}
\arraycolsep=0.0cm
\begin{array}{c}
\text{LST of $a\,b\,a\,a\,b\,a$} = \, _1( \, _2( \, _3( \, a_4 \, b_5 \, )_3 \, )_2 \, _2( \, a_6 \, )_2 \, _2( \, _3( \, a_4 \, b_5 \, )_3 \, )_2 \, _2( \, a_6 \, )_2 \, )_1 \\[0.75cm]
\tabcolsep=0.0cm\def\arraystretch{0.75}
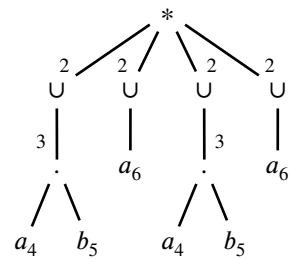
\begin{tabular}{c@{\hspace{0.5cm}}l}
\emph{num} & RE \emph{operator} \\ \toprule
$1$ & Kleene star \\ \midrule
$2$ & union \\ \midrule
$3$ & concatenation \\ \bottomrule
\end{tabular}
\end{array}
\hspace{1.5cm}
\begin{tabular}{c}
\psset{arrows=-, linestyle=solid, linewidth=1pt, levelsep=1.0cm, treesep=0.75cm, nodesep=4pt, labelsep=3pt, treefit=tight, nodealign=true, tpos=0.675}
\pstree{\TR{$\ast$}} {
    \pstree{\TR{$\cup$} \tlput{\footnotesize 2}} {
            \pstree[treesep=0.5cm]{\TR{$.$} \tlput{\footnotesize 3}} {
                \TR{$a_4$}
                \TR{$b_5$}
            }
    }
    \pstree{\TR{$\cup$} \tlput{\footnotesize 2}} {
                \TR{$a_6$}
    }
        \pstree{\TR{$\cup$} \trput{\footnotesize 2}} {
            \pstree[treesep=0.5cm]{\TR{$.$} \trput{\footnotesize 3}} {
                \TR{$a_4$}
                \TR{$b_5$}
            }
    }
        \pstree{\TR{$\cup$} \trput{\footnotesize 2}} {
                \TR{$a_6$}
    }
}
\end{tabular}
\vspace{0.125cm}$
\end{center}
The LST corresponds to the accepting path (consisting of solid edges) shown above, for both the NFA and reverse $\widehat{\text{NFA}}$. \qed
\subsection{Scheduling and optimizations} \label{subsec:scheduling}
Each component of the parallel parsing algorithm (see Tab.~\ref{tab:pseudocode}) can run as a single task. Many tasks can run concurrently, provided the data-dependencies  in  Fig.~\ref{fig:parallelparser} are obeyed. Consequently, granted sufficient computing resources,  the simplest synchronization requirement for the parser to function correctly, is  that the phases work in sequence: each phase is started after the previous one is done, by placing a synchronization barrier between the tasks of the two phases. The splitter task runs first and upon termination, the FW-reacher and BW-reacher tasks run concurrently, as they are independent. After all reachers terminate, the FW-joiner and BW-joiner tasks run concurrently. After both terminate, the FW-builder and BW-builder tasks run concurrently, similarly to  the reach phase and for the same reason. After all builders terminate, the merger tasks run concurrently. Eventually, after all mergers terminate, the composer task runs and outputs the final clean SLPF.
\par
Since the reacher, builder and merger tasks are the most computationally intensive, and  each of them works on a data subset corresponding to one text chunk, we can expect the above concurrent scheduling  to achieve a substantial  speed-up over a serial parser, when the number of processors is large enough.
\paragraph{Scheduling policy}
The best schedule of the tasks of course depends on the actual HW resources, chiefly on  the availability of processors (cores).
We  divide the main tasks into five task pools: FW-reachers, BW-reachers, FW-builders, BW-builders and mergers. For a number $c \geq 2$ of text chunks, each pool consists of $c$ tasks. Furthermore, the reach and build phases are the most time-consuming, and the pool of mergers is scheduled after the pools of reachers and builders.
As a consequence, with  $c$ available processors, their use is optimized by serializing  the pool of FW-reachers and the pool of BW-reachers (say the FW-reachers before the BW-ones), and similarly for the two pools of builders. At best in fact, if each task is exclusively assigned a processor, and all the tasks of the same type process roughly the same amount of data in  the same time, then during the reach, build and merge phases all processors are kept busy. Only the split, join and compose phases underuse the HW resources, but their computational load is low. Of course, if the number of processors is limited ($< c$), a few concurrent tasks in a pool will be scheduled on the same processor and the time performance of the whole parser may degrade.
\par
 We assume that the memory is global and shared across all tasks. As the tasks in a pool do not perform any concurrent read or write operations on the same data-structure elements, this memory model does not require any synchronization between the tasks, other than the scheduling of the phases described above (see Fig.~\ref{fig:parallelparser}). In the experimentation, we keep to the above described scheduling and architectural model. A detailed performance analysis for various benchmarks is carried out in Sect.~\ref{sec:experimentation}, assuming the number of processors is $\geq c$ (number of chunks).
\par
In case the number of processors is larger, another more parallel scheduling may be preferable.
Assuming the number of processors is $2c$, all the reacher tasks can be concurrently executed, i.e., the FW and BW pools together; and subsequently do similarly for all the builder tasks. This improves the time performance of the computationally heavy reach and build phases, although it causes some resource underuse in the lighter merge phase. Furthermore,  with $2c$ processors available and all the FW-builder and BW-builder tasks running concurrently, a merger task can be scheduled at the earliest time when both its  corresponding FW and BW builders meet in the midpoint of the chunk; then  it can proceed in opposite directions towards the chunk  endpoints. Unfortunately, this may require  fine-tuned synchronization between corresponding builders and merger, and  it may be preferable to integrate  the corresponding FW and BW builders and  merger into a single task, to reduce  synchronization overhead. The resulting organization is shown in Fig.~\ref{fig:builders&merger}. We no longer insist on such performance improvements based on the reorganization of the tasks, as the structure of the parallel algorithm does not change substantially.
\paragraph{Memory optimization}
The memory consumption of the parallel parser is the sum of the memory footprints of  its data-structures. Their sizes, including those of the DFA and ME-DFA, depend of course on the number $\ell$ of segments (states of the parser NFA), which in turn depends on the RE. The most memory-consuming data-structures are $B$ (and its BW counterpart $\widehat B$), $M$ and $C$, which are similar and have comparable sizes. In fact, each of them is an array of $n + 1$ (text length) segment sets, and each segment set takes one or more memory words as a Boolean vector of size $\ell$  (see also Sect.~\ref{subsec:serialalgorithmVectorial}). We discuss possible  memory optimizations, starting from a very profitable one. Instead of allocating   separate memory areas to the data-structures $B$, $\widehat B$ and $M$, the latter $M$ can used in all phases. Structure $M$ is first (temporarily) assigned the result of the forward build phase, and later  (definitely) overwritten by the backward build and merge phases united (analogously to what was hinted for the serial parser in  Fig.~\ref{fig:builders&merger}). Similarly, memory for the data-structure $C$ is not really necessary and the data-structure $M$ can be directly output. By such optimizations, the only large data-structure left is $M$, which has a size proportional to the text length $n$.
\begin{figure*}[ht]
\begin{center}
\definecolor{lightgray}{rgb}{0.875,0.875,0.875}
\tabcolsep=0.0cm
\def\arraystretch{1.0}
\begin{tabular}{c} \toprule
\cellcolor{lightgray}
\begin{minipage}[c]{0.95\textwidth}
\begin{algorithm2e}[H]
\setstretch{1.0}
\SetAlgorithmName{component (task)}{}{}
\TitleOfAlgo{\emph{builder\&merger} \hfill unification of Eq.s~\eqref{eq:builder} and \eqref{eq:merger} -- see also Tab.~\ref{tab:pseudocode} and Fig.~\ref{fig:parallelparser}} \hrule height 0.25pt
\SetKwInOut{Input}{{\i}nput}
\SetKwInOut{Output}{output}
\SetKwInOut{LocVars}{variable~}
\DontPrintSemicolon
\Input{sets $J_{i-1}$ and $\widehat J_{i+1}$ of edge segments, text chunk $y_i$, chunk number $i$}
\Output{subseries $M_i$ of clean SLPF columns}
\LocVars{temporary column $\emph{TMP}$, text position $t$ \hfill (separate memory for subseries $B_i$ and $\widehat B_i$ is not allocated)} \hrule height 0.25pt
\Begin (\tcp*[f]{\textbf{forward build} by \textbf{DFA} and \textbf{temporarily} store in $M$}) {
$M \, [i, \, 1] \asgn \delta_\textrm{DFA} \, \big( J \, [i - 1], \, y \, [i, \, 1] \big)$ \tcp*[r]{build $B_{i, \, 1}$ and store in $M_{i, \, 1}$}
\For (\tcp*[f]{scan text chunk forwards}) {\rm $t = 2$ \textbf{up to} $k$} {
$M \, [i, \, t] \asgn \delta_\textrm{DFA} \, \big( M \, [i, \, t-1], \, y \, [i, \, t] \big)$ \tcp*[r]{build $B_{i, \, t}$ and store in $M_{i, \, t}$}
}
}
\Begin (\tcp*[f]{\textbf{backward build} by \textbf{reverse DFA} + \textbf{merge} and \textbf{definitely} store in $M$}) {
$\emph{TMP} \asgn \widehat J \, [i + 1]$ \tcp*[r]{build $\widehat B_{i, \, k}$ and store in $\mathit{TMP}$}
$M \, [i, \, k] \asgn M \, [i, \, k] \, \cap \, \emph{TMP}$ \tcp*[r]{merge $B_{i, \, k} \, \cap \, \widehat B_{i, \, k}$ and store in $M_{i, \, k}$}
\For (\tcp*[f]{scan text chunk backwards}) {\rm $t = k - 1$ \textbf{down to} $1$} {
$\emph{TMP} \asgn \delta_\text{reverse DFA} \, \big( \emph{TMP}, \, y \, [i, \, t + 1] \big)$ \tcp*[r]{build $\widehat B_{i, \, t}$ and store in $\mathit{TMP}$}
$M \, [i, \, t] \asgn M \, [i, \, t] \, \cap \, \emph{TMP}$ \tcp*[r]{merge $B_{i, \, t} \, \cap \, \widehat B_{i, \, t}$ and store in $M_{i, \, t}$}
}
}
\end{algorithm2e}
\end{minipage} \\ \bottomrule
\end{tabular}
\end{center}
\caption{Unification of the FW / BW builders and the merger, to save memory and directly produce the clean SLPF.} \label{fig:builders&merger}
\end{figure*}
\par
Fig.~\ref{fig:builders&merger} shows how to unify the FW / BW builder and merger tasks for a chunk, so as to have a single task that allocates only one memory array. A pool of $c$ (number of chunks) such \emph{builder\&merger} tasks can run concurrently in one phase, which executes the aggregate work previously carried out by the three consecutive phases of FW-build, BW-build and merge.
\par
In contrast, the other data-structures, i.e., $R$ and $J$ (and their BW counterparts $\widehat R$ and $\widehat J$), which are also arrays of segment sets, have a size proportional to the number of chunks $c$. Thus their size is much smaller than that of structure $M$, since it is advisable to keep $c \ll n$ to ensure that the chunks are long enough. Our tool eventually parses the text and optimizes the memory by using a single series of $n + 1$ columns, as detailed above and continued in Sect.~\ref{sec:tool}.
Memory could be further optimized by representing the SLPF in a different way and by compressing it, see  App.~\ref{app:SLPFoptimization} for more details.
\paragraph{Automaton optimization}
As an optimization, the DFA  can be merged into the ME-DFA. With this straightfoward extension, implemented in the tool but for brevity not described here, the ME-DFA becomes the single final device that drives both phases (build and reach) of our parallel parser, the only difference between DFA and ME-DFA being  the choice of the state from which to start a run.
\section{Tool design} \label{sec:tool}
Our parallel parser is implemented by a SW tool, written in the Java language and  available on the GitHub repository \RePar. Its purpose is twofold: as a demo of our method and as a tool -- though not a professional one -- for parsing large benchmarks. We describe the structure and functions of the tool and we discuss some optimization techniques. In App.~\ref{app:otherfeatures} we describe some additional RE features skipped in earlier  sections, but  supported or easy to incorporate in our tool.
\subsection{Tool structure} \label{subsec:toolstructure}
The tool realizes the parallel parsing algorithm described in Sect.~\ref{subsec:parallelalgorithm}, and consists of two main parts that operate on user-supplied data: ($\mathit{i}$) generation of the parser for the given RE $e$, and ($\mathit{ii}$) parallel parsing of the given  text $x$. Its output is  the shared linearized parse forest (SLPF) of text $x$ according to RE $e$. To generate the parser (part $\mathit{i}$), the tool analyzes RE $e$ and creates the associated numbered RE $e_\#$ as of Sect.~\ref{subsec:numberedRE}; then it computes all the segments and determines their followers as of Sect.~\ref{subsubsec:segment-computation}, by the algorithm in Fig.~\ref{fig:segmentalgo} and formula Eq.~\eqref{eq:folsegdef}; at last,  it builds the NFA, and subsequently the DFA and ME-DFA (both direct and reverse), as of Sect.s~\ref{subsubsec:constrParserNFA} and \ref{subsec:deterministicFA}.
To parse the text (part $\mathit{ii}$), the tool applies the parallel parsing algorithm described in Sect.~\ref{subsec:parallelalgorithm}, driven by the DFA and ME-DFA, and parameterized with the number of threads, i.e., chunks, specified by the user; see Tab.~\ref{tab:pseudocode} and Fig.~\ref{fig:builders&merger} for the various component tasks of the parallel algorithm. The tool can also run the same algorithm with just one chunk, i.e., the whole text, in which case it operates  serially. In other words, the tool can either perform parallel matching and parsing with a user-defined number of threads, or just serial matching and parsing.  Furthermore, the tool can be downgraded to merely function as a recognizer.
\paragraph{Parallel parser}
The implementation language is Java (release $19$). Concurrency is achieved by means of the Java Thread model, which
is built on the native thread model of the underlying OS (Linux); thus each task of the algorithm, see Tab.~\ref{tab:pseudocode} and Fig.~\ref{fig:builders&merger}, is realized as a Java Thread. The scheduling policy was described in Sect.~\ref{subsec:scheduling}, with a number of threads equal to the number $c$ of text chunks, for the pools of concurrent threads. The tool creates a thread pool and runs it by means of an \emph{ExecutorService} method, which
allows  to wait for the termination of all the threads in a pool, and to collect their results. This method serializes the algorithm phases, which is the only synchronization requirement as earlier explained. The tool  provides a few other methods needed for RE matching (mentioned in Sect.~\ref{sec:introduction}), to extract from the SLPF the substrings that correspond to user-selected sub-REs.
\par
For efficiency of the Java Runtime Environment, the five thread pools described in Sect.~\ref{subsec:scheduling} (FW-reachers, BW-reachers, FW-builders, BW-builders, and mergers), each consisting of $c$ concurrent threads, are partially fused. Each pair of FW and BW reachers that work on the same text chunk is unified into a single thread (computing, say, first FW then BW), thus resulting in one pool of $c$ FW-BW \emph{reacher} threads. Upon termination of the reacher threads, the two joiners (FW and BW) are executed by a single thread, without performance loss since the join phase  weights for less than one percent of the total parse time. As anticipated in Sect.~\ref{subsec:scheduling}, for efficiency  each pair of FW and BW builders and the merger that work on the same chunk are combined into a single thread; see the algorithm in Fig.~\ref{fig:builders&merger}. Thus, the build and merge phases are fused, and there is one pool of $c$ \emph{{\rm FW-BW} builder\&merger} threads. In summary, the tool  has two pools of $c$ threads, one pool for reach and the other for build\&merge, plus a single thread for join. It works in three serialized phases: reach, join, and build\&merge. The main program encapsulates everything and performs I/O.
\par
As said, memory is global and shared across all threads. More precisely,  the tool  uses for both the build and merge operations a single SLPF data-structure, i.e., an array of segment sets,  the size of which equals the text length (plus one); the same array also stores the final clean SLPF, as of Sect.~\ref{subsec:scheduling}. Thus in our experimentation, the split and compose phases are unnecessary.
\par
As a remark, when the number of processors is very large, the computer architecture may not support the global shared-memory model. The solution is to partition text and SLPF into blocks that are moved across global and local memory. In such a case the splitter and composer can be used to move the data blocks (a sketch is in Tab.~\ref{tab:pseudocode}).
\paragraph{Serial parser}
Clearly, the tool performs serial parsing of the  undivided text, when the number of  chunks in the parallel parsing algorithm is  set to $c = 1$. The serial algorithm is thus divided into exactly three FW-BW threads for reach, join and build\&merge, respectively, which are run in sequence on a single processor. Such a serial one-chunk parser mainly serves for performance comparison, to measure the speed-up of the parallel parser as a function of the number of concurrent threads.
\par
Incidentally, one may consider two other versions of serial parser that could be taken as reference for performance comparison: ($\mathit{i}$) the one illustrated in Sect.~\ref{subsec:serialalgorithmVectorial}, which is driven by the NFA and is based on  matrix computation, and another ($\mathit{ii}$) driven by the DFA of Sect.~\ref{subsec:deterministicFA}, based on a transition function stored as a look-up table (not described in detail). Essentially, version ($\mathit{ii}$) is the determinization of ($\mathit{i}$). We discuss how such versions would score as reference.  The NFA parser ($\mathit{i}$) is  too slow, being driven by a nondeterministic automaton. The DFA parser  ($\mathit{ii}$)  could  be used as  reference, due to its similarity with the parallel parser; it might even look preferable to the one-chunk parser, which pays for the  reach and join phases, useless for undivided text. Yet we do not use it in our experimentation in Sect.~\ref{subsec:results}, as its time speed-up  over the one-chunk parser is close to one (measurements are reported in Fig.~\ref{fig:bigdata-performance}), so that in practice it does not matter which serial parser is chosen.
\subsection{Other functions of the tool} \label{subsec:toolfunctions}
We outline the support for RE matching and recognition offered by the tool, in addition to the main function as a parser for REs.
\paragraph{Match extraction}
Two methods are supported for extracting matches from an SLPF. Indeed the SLPF structure, though primarily designed for
encoding syntax trees, easily allows the user to search subtrees or just strings. Such methods produce lists of index pairs (offsets) in the text,
giving the positions of the matches:
\begin{description}[leftmargin=2.5cm, style=multiline]
\item[\emph{getMatches} $( \ )$]  extracts the matches of the strings that match a group (parenthesized subexpression), or a group inside another
\item[\emph{getChildren} $( \ )$]   extracts the matches of the child strings of a match
\end{description}
Method \emph{getMatches} is typically called to get the matches of a desired RE group, and  performs a visit of the  forest  encoded in the SLPF. Then it is called again to get the matches of any inner groups, wherever they occur in the text.
Method \emph{getChildren} serves  to access the SLPF as a forest, but without using groups.
Matches can be extracted efficiently since the SLPF is clean, hence all the occurrences on a path of the subtrees that represent the target match are guaranteed to be part of some parse tree.
\par
For instance, such methods can be used to support matching in semi-structured data languages, e.g., JSON, which have
a list structure that allows the user to easily  describe the desired matches  by an RE. The methods can also be used to access the contents of the SLPF by means of the DOM (Document Object Module) specification, similarly to what is supported by the XML and HTML (meta-)languages.
\paragraph{Matching user queries by means of parsing}
The matchers for REs, such as \emph{grep}, are typically used to find all the occurrences of certain user-specified patterns within a text. For such a job our parser is more accurate than an RE matcher, since it allows the user to specify the context of the occurrence, thus improving the selectivity of the search. We illustrate by means of an example, i.e., the case of searching for certain fields in HTML documents.
\paragraph{Example $7$ -- Searching for fields in an HTML document}
We want to search for certain values of the fields (attributes) \verb!id! and \verb!nam! inside the HTML headers of type \verb!h3!, namely \verb!<h3 id = ... nam = ... >!, where the ellipsis can be strings or numbers in quotes. To specify the positions to match, the user will edit the RE by inserting some \emph{extra parenthesis pairs} around the positions wanted (see App.~\ref{app:otherfeatures}). For instance, here is an RE for searching such a header in an HTML document:
\[
_1( \, \langle \, \mathit{h3} \; \mathit{id} = \, _2( \, \text{``\,$\Sigma^+$\,''} \, )_2 \; \mathit{nam} \, = \, _3( \, \text{``\,$\Sigma^+$\,''} \, )_3 \, \rangle \, )_1 \hspace{1.5cm} \text{where $\Sigma$ stands for any char other than the quotes}
\]
An ordinary RE matcher will independently match the fields \verb!id! and \verb!nam!, thus returning also some false positive. \qed
\paragraph{Mere recognizer}
The tool can  act as a parallel recognizer, by avoiding all the parsing operations that are unnecessary for recognition. Thus it only executes the forward reach and join, then it  notifies whether the text is accepted or not, and stops. It skips the forward build, all the backward operations (reach, join and build) and the merge, and of course it does not produce the SLPF. A version of the tool that \emph{only} performs recognition is also presented in~\cite{DBLP:conf/ppopp/BorsottiBMC25,borsotti2024minimizingspeculationoverheadparallel}, with some further optimization.
\subsection{Tool optimization} \label{subsec:tooloptimization}
The SW tool closely follows the parallel parsing algorithm of Sect.~\ref{sec:parallelparser} and the guidelines of Sect.~\ref{subsec:toolstructure}. In addition, it  implements a few optimizations for mitigating the consumption of HW resources (processors and memory) that the parallel algorithm may impose. Mitigation concerns load balancing, speculation heuristics for the ME-DFA, reduction of the memory footprint of the SLPF, and also the matching function offered in addition to parsing.
\paragraph{Load balancing}
In both the reach and build phases, two text chunks (of identical length) may need rather different processing times, thus
the threads that finish more quickly will remain idle and cause resource underuse. To keep all  threads working, we apply load balancing separately
to the pool of reachers and to the pool of builder\&merger components. Each chunk is split into a small
number of pieces, here called fragments, that are placed in a FIFO queue. Each thread in a pool processes one
fragment at a time, instead of a whole chunk. The threads repeatedly extract from the queue the head fragment and elaborate it,
until the queue gets empty. We have found that four fragments per chunk suffice to ensure a good balance of the thread lifetimes
in each pool.
\paragraph{Techniques for mitigating speculation cost}
In the reach phase each text chunk is speculatively parsed starting from  several  states, i.e., segments.
To reduce the number of starting states, we use two techniques: ($\mathit{i}$) the reduction of the number of states from which a reacher starts, by using an ME-DFA as explained in Sect.~\ref{subsec:deterministicFA}, and ($\mathit{ii}$) a fine-grained adjustment of the chunk edges by shifting to a nearby position so as to find an advantageous position having fewer starting states. In particular, technique ($\mathit{i}$) is novel and quite effective.
\par
Furthermore, the computation of a reacher, which executes over its  chunk one ME-DFA run per starting state to the chunk end (unless it dies before), could be organized in two different ways. The first possibility is to compute  all the runs for the same chunk  in lockstep, by executing one ME-DFA transition per run, under the current character, before moving on to the next character. The second possibility, which turns out to be faster,  executes all the transitions of a single run, by scanning the whole chunk or going as far as possible before moving on to another run.
\par
Our experience  also confirms the several past findings (see Sect.~\ref{sec:relatedwork} -- \emph{Related work}) that  the number of active states decreases very rapidly after scanning the first few characters of a chunk. Therefore the \emph{reach+join} phases pay lesser speculation overhead.
\paragraph{SLPF optimization}
Since the SLPF memory footprint may be very high for long and ambiguous texts, we have developed some techniques to encode and compress the SLPF; see App.~\ref{app:SLPFoptimization} for  details.
\section{Experimentation} \label{sec:experimentation}
We report in Sect.~\ref{subsec:results} the performance measurements for our parsing algorithm on a parallel computing platform. The measurements were collected in an experimentation campaign with the benchmarks described in Sect.~\ref{subsec:benchmarks}. The parser tool was applied to several benchmarks of REs and texts, thus obtaining comparative performance  evaluations for time and speedup. The computing platform is a shared-memory multi-core Dell PowerEdge R$7425$ server featuring two AMD EPYC $7551$ $64$-bit CPUs (see \AMDEPYC), each with $32$ cores (ISA x$64$). In total the platform has $64$ identical cores running at a $2.0$ GHz clock frequency (such cores are called processors when discussing scheduling policy and results). Each core has an L$1$ cache memory of $96$ Kbyte ($64$K for instructions and $32$K for data) and an L$2$ cache memory of $512$ KByte. Each CPU (with $32$ cores) has an L$3$ cache memory of $64$ MByte (shared across the $32$ cores). The two CPUs, i.e., all $64$ cores, share a main memory of $512$ GByte. The operating system is  Debian GNU/Linux $6$.$1$.$0$-$13$-amd$64$, configured with an SMP $6$.$1$.$55$-$1$-x$86$-$64$ kernel and the PREEMPT$\_$DYNAMIC option enabled. Furthermore, the same level of compilation optimization was consistently used, and during the experimentation campaign the platform ran exclusively our tool, so the measurements of computational time and memory consumption are not affected by external factors.
\subsection{Benchmarks} \label{subsec:benchmarks}
As benchmarks we chose both real-life and synthetic examples, with a large corpus of texts. All the benchmarks and their text corpora are listed in Tab.~\ref{tab:benchmarks}. The real-life benchmarks include:
\begin{itemize}[leftmargin=*]
\item  structured files, such as biometric sequences (FASTA) and  traffic data logs (TRAFFIC)
\item unstructured files (BIBLE), where the interesting parts are buried in a large text that is not relevant, such as HTML
\end{itemize}
The synthetic benchmarks (BIGDATA and REGEN) were generated by means of our previous tool REgen~\cite{DBLP:conf/wia/BorsottiBCM19}, which produces random REs and random valid texts to be parsed. In particular, benchmark REGEN has been (randomly) extracted from a large collection of REs and texts of increasing size and length (for a total of $1,000$ REs and $100,000$ texts), which was generated by REgen and is fully available on \REgen. For the first four benchmarks in Tab.~\ref{tab:benchmarks}, we report the results for a single RE and several texts, while for REGEN we report the results for  a set of $60$ REs\footnote{Of the public benchmarks available, very few include both REs and corresponding collections of texts. For instance, for FASTA just the text is available, thus we had to write a few queries of biological interest. To circumvent this problem and also to experiment with larger REs, we turned to synthetic REs and random texts generated  by REgen, see \REgen.}.
\par
More precisely, for all the benchmarks listed in Tab.~\ref{tab:benchmarks}, each RE comes with six valid texts of increasing length; see Fig.s \ref{fig:bigdata-speedup}, \ref{fig:bigdata-performance}, \ref{fig:parsing} and \ref{fig:regen-speedupvsRE&text} for the text lengths. The length of the text is the number of characters (bytes). The size of the RE is quantified as the total count of symbols, i.e., terminals and operators (metasymbols), that occur in the original RE $e$, and is denoted by $\Vert \, e \, \Vert$ or simply $\Vert \, \text{RE} \, \Vert$; see also Ex.~$5$ in Sect.~\ref{subsec:deterministicFA} for a sample computation of the RE size for a parametric family of REs.
\par
\begin{table}[ht]
\begin{center}
\tabcolsep=0.0cm
\def\arraystretch{1.125}
\begin{tabular}{l@{\hspace{1.0cm}}r@{\hspace{1.0cm}}r@{\hspace{1.0cm}}l@{\hspace{1.0cm}}r}
\textbf{name} & \textbf{RE num} & \textbf{RE size} (symbols) & \textbf{text description} & \textbf{max text length} (bytes) \\ \toprule
BIGDATA & single & $9$ & random text produced by REgen~\cite{DBLP:conf/wia/BorsottiBCM19} & $13$ M \\ \midrule
BIBLE & single & $31$ & book in HTML (The Bible) & $4$ M \\ \midrule
FASTA & single & $102$ & biometric data (DNA sequences)~\cite{SUBRAMANIAN2019100269} & $765$ K \\ \midrule
TRAFFIC & single & $123$ & system log of network traffic & $11$ M \\ \midrule
REGEN & $60$ & from $40$ to $99$ & random text produced by REgen~\cite{DBLP:conf/wia/BorsottiBCM19} & $160$ M  \\ \bottomrule
\end{tabular}
\end{center}
\caption{Benchmarks of REs and text corpora used in the experimentation campaign.  For all benchmarks the alphabet is ASCII with $7$ bits, i.e., $128$ characters.} \label{tab:benchmarks}
\end{table}
\noindent
Here are a description of each benchmark and its source:
\begin{description}[leftmargin=1.75cm, style=multiline]
\item[BIGDATA] One small random RE extracted from the large collection of REs generated by means of our tool REgen~\cite{DBLP:conf/wia/BorsottiBCM19}. The random text is also
extracted from the same collection. Source: \REgen
\item[BIBLE] One middle-sized RE applied to an HTML file that contains the entire Bible.
We have described the titles of the h$3$ subsections through the RE.
This allows us to model the file as a text where, buried in parts
that are not of  interest, there are instances of that RE. Source: \bible
\item[FASTA] One large RE applied to a text file that contains DNA sequences~\cite{SUBRAMANIAN2019100269} in the FASTA format.
We have described the FASTA format through an RE that represents any such file.
\par\noindent Source: \fasta
\item[TRAFFIC] One large RE applied to a syslog file of network traffic, which consists of sequences of
traffic records. We have described the file through an RE that represents any such syslog traces.
\par\noindent Source: \traffic
\item[REGEN] From the large collection of REs generated by means of REgen~\cite{DBLP:conf/wia/BorsottiBCM19}, we have extracted at random $60$  REs of increasing size
and for each of them we have produced valid random text of increasing length.
\par\noindent Source: \REgen
\end{description}
 All the data needed to reproduce the measurements are available on public sites (see the source links listed above).
\par
Unfortunately, there is no guarantee that corpora for different benchmarks exercise
the RE constructs to a similar degree of coverage. This has the annoying consequence that the performance measured for two different
cases may diverge not only because of the intrinsic RE difference, but also because the
corpora may be more or less representative. The reader is thus warned not to draw hasty
conclusions about the comparative performance of the parser applied to different
REs.
\par
Our measurements include only the parsing time and exclude the I/O time. In order to do so, the texts are first loaded into memory and then are parsed, thus eventually producing an SLPF in the main memory.
\subsection{Results} \label{subsec:results}
We report the results of our experimentation. For the first four benchmarks listed in Tab.~\ref{tab:benchmarks}, we measured the total \emph{parsing time} of the parallel parser as a function of two independent variables: \emph{number of threads} and \emph{text length}. For the synthetic benchmark REGEN, the measurements are a function of the same two variables plus a third independent variable: RE \emph{size}, i.e., number of symbols. The total time comprises all the phases of the parallel algorithm. However, only reach and build\&merge are relevant, while join takes less than onepercent of the total. The reach phase takes the longest time, because it performs operations similar to those of the build\&merge phase, but speculatively on all segments.
\par
We start by showing the \emph{absolute computation time} of the parser, for most benchmarks. Then the parser performance is more extensively displayed and discussed, for all benchmarks, in the form of \emph{speed-up}, that is, the reciprocal of the \emph{ratio} between the total parsing time of the parallel parser with $c$ (number of chunks) threads and the time with a single thread, i.e., the time of serial parsing (see Sect.~\ref{subsec:toolstructure} -- \emph{serial parser}). For completeness, we also measured in one case (benchmark BIGDATA) the recognition time (see Sect.~\ref{subsec:toolfunctions} -- \emph{mere recognizer}) and we obtained the recognition speed-up, in the same way as for parsing.
\par
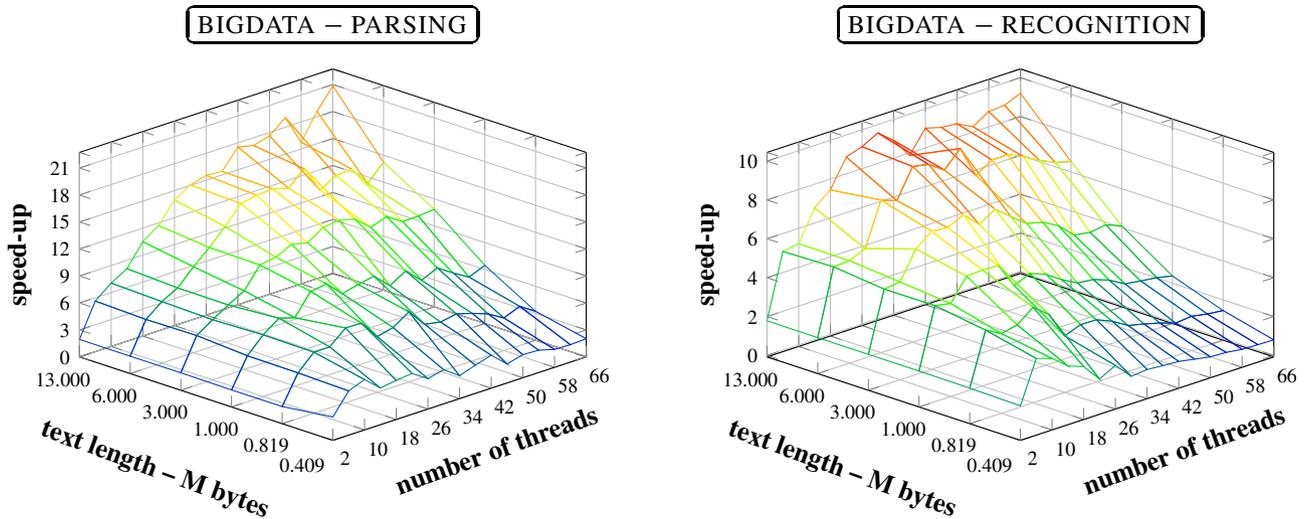
\begin{figure}[ht]
\tikzexternaldisable
\begin{center}
\begin{subfigure}[t][][c]{0.4625\textwidth}
\begin{tikzpicture}
\begin{axis} [
    title={\Ovalbox{\large \textsc{bigdata -- parsing}}},
    width=1.0\textwidth,
    height=0.275\textheight,
    view={315}{30},
    xtick={2,10,...,66},
    ztick={0,3,...,22},
    ytick={1,2,...,6},
    yticklabels={0.409 $\quad$, 0.819, 1.000, 3.000, 6.000, 13.000},
    unbounded coords = jump,
    grid,
    every tick label/.append style={font=\footnotesize},
    every axis y label/.style={at={(ticklabel cs:0.5)}, rotate=90, anchor=near ticklabel},
    every axis x label/.style={at={(ticklabel cs:0.5)}, anchor=near ticklabel},
    xlabel style={rotate=20},
    xlabel={\normalsize\bf number of threads},
    ylabel style={rotate=-110},
    ylabel={\normalsize\bf text length -- M bytes},
    zlabel={\normalsize\bf speed-up},
    colormap={pos}{color(0)=(blue) color(0.4)=(green) color(0.6)=(yellow) color(0.8)=(orange) color(1.0)=(red)}
]
\addplot3[mesh, mark=.] table [x=x, y=y, z=z] {bigdata-parsing-grinder64.dat};
\end{axis}
\end{tikzpicture}
\end{subfigure}
\qquad
\begin{subfigure}[t][][c]{0.4625\textwidth}
\begin{tikzpicture}
\begin{axis} [
    title={\Ovalbox{\large \textsc{bigdata -- recognition}}},
    width=1.0\textwidth,
    height=0.275\textheight,
    view={315}{30},
    xtick={2,10,...,66},
    ztick={0,2,...,22},
    ytick={1,2,...,6},
    yticklabels={0.409 $\quad$, 0.819, 1.000, 3.000, 6.000, 13.000},
    unbounded coords = jump,
    grid,
    every tick label/.append style={font=\footnotesize},
    every axis y label/.style={at={(ticklabel cs:0.5)}, rotate=90, anchor=near ticklabel},
    every axis x label/.style={at={(ticklabel cs:0.5)}, anchor=near ticklabel},
    xlabel style={rotate=20},
    xlabel={\normalsize\bf number of threads},
    ylabel style={rotate=-110},
    ylabel={\normalsize\bf text length -- M bytes},
    zlabel={\normalsize\bf speed-up},
    colormap={pos}{color(0)=(blue) color(0.4)=(green) color(0.6)=(yellow) color(0.8)=(orange) color(1.0)=(red)}
]
\addplot3[mesh, mark=.] table [x=x, y=y, z=z] {bigdata-recognition-grinder64.dat};
\end{axis}
\end{tikzpicture}
\end{subfigure}
\end{center}
\caption{Left: plot of the speed-up of multiple vs single thread (one-chunk serial parser) parallel parsing for benchmark BIGDATA. Right: the same plot type as on the left, but for recognition.} \label{fig:bigdata-speedup}
\tikzexternalenable
\end{figure}

The scheduling policy was explained in Sect.~\ref{subsec:scheduling} and refined in Sect.~\ref{subsec:toolstructure}. It assumes that each one of the two pools of concurrent tasks, i.e., the pool of FW-BW reachers and the pool of FW-BW builder\&merger components, has a size of $c$ threads, and that there are at least $c$ processors (cores) available. Thus we can assume that each concurrent thread is scheduled on one processor. On the computing platform used,  the number of available processors is constant and set to $64$.
\paragraph{Measure of absolute parsing time}
We include some information about the absolute running time, which may interest the readers acquainted with other RE text-processing applications. However, it is important to keep in mind that parsing is a more thorough operation than recognition and matching.
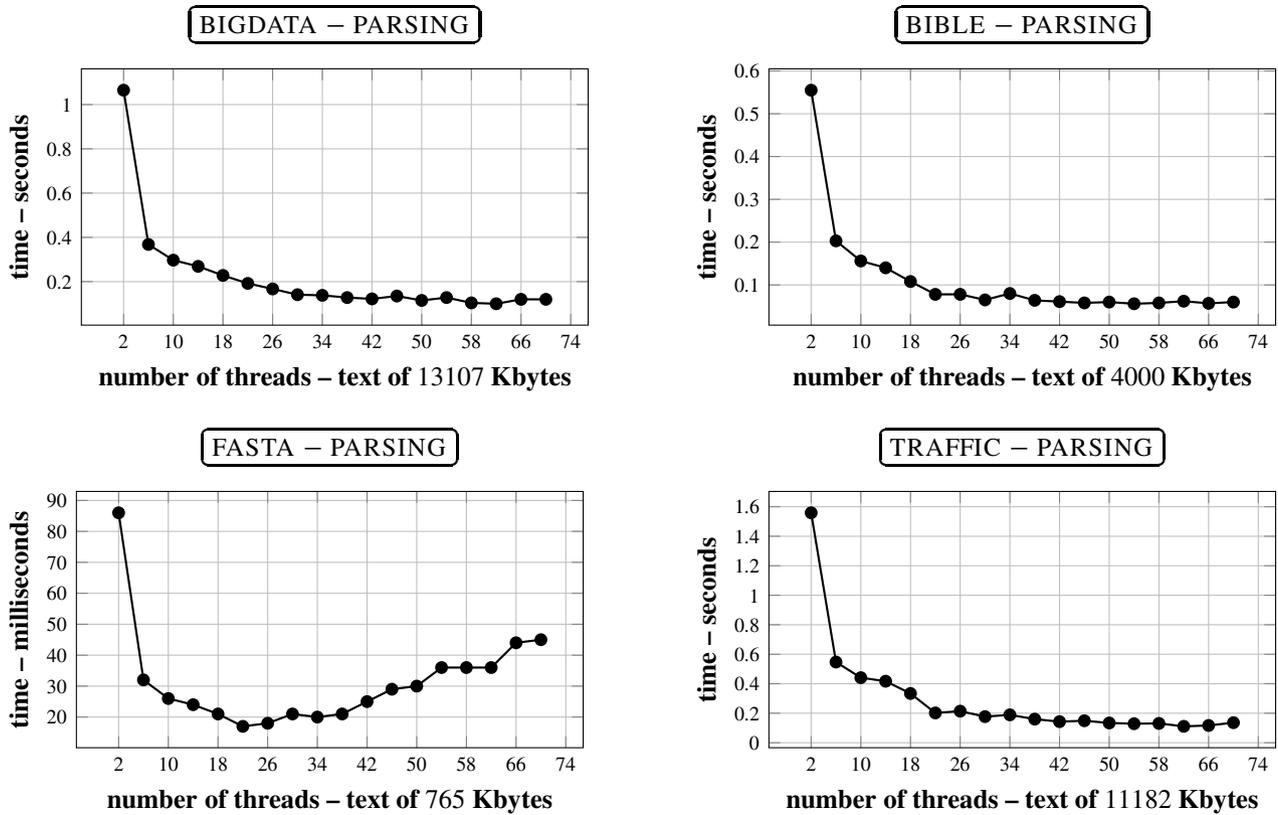
\begin{figure}[ht]
\tikzexternaldisable
\begin{center}
\begin{subfigure}[c][][c]{0.4625\textwidth}
\begin{tikzpicture}
\begin{axis} [
    title={\Ovalbox{\large \textsc{bigdata -- parsing}}},
    width=1.0\textwidth,
    height=0.6\textwidth,
    xtick={2,10,...,70},
    ytick={0,200,...,1100},
    scaled y ticks=base 10:-3,
    ytick scale label code/.code={},
    unbounded coords = jump,
    grid,
    every tick label/.append style={font=\footnotesize},
    xlabel={\normalsize\bf number of threads -- text of $13107$ Kbytes},
    ylabel={\normalsize\bf time -- seconds},
]
\addplot[mesh, mark=*, thick, color=black] table [x=thread, y=time] {bigdata-time-maxlength.dat};
\end{axis}
\end{tikzpicture}
\end{subfigure}
\qquad
\begin{subfigure}[c][][c]{0.4625\textwidth}
\begin{tikzpicture}
\begin{axis} [
    title={\Ovalbox{\large \textsc{bible -- parsing}}},
    width=1.0\textwidth,
    height=0.6\textwidth,
    xtick={2,10,...,70},
    ytick={0,100,...,700},
    scaled y ticks=base 10:-3,
    ytick scale label code/.code={},
    unbounded coords = jump,
    grid,
    every tick label/.append style={font=\footnotesize},
    xlabel={\normalsize\bf number of threads -- text of $4000$ Kbytes},
    ylabel={\normalsize\bf time -- seconds},
]
\addplot[mesh, mark=*, thick, color=black] table [x=thread, y=time] {bible-time-maxlength.dat};
\end{axis}
\end{tikzpicture}
\end{subfigure}
\par
\vspace{0.25cm}
\begin{subfigure}[c][][c]{0.4625\textwidth}
\begin{tikzpicture}
\begin{axis} [
    title={\Ovalbox{\large \textsc{fasta -- parsing}}},
    width=1.0\textwidth,
    height=0.6\textwidth,
    xtick={2,10,...,70},
    ytick={0,10,...,100},
    unbounded coords = jump,
    grid,
    every tick label/.append style={font=\footnotesize},
    xlabel={\normalsize\bf number of threads -- text of $765$ Kbytes},
    ylabel={\normalsize\bf time -- milliseconds},
]
\addplot[mesh, mark=*, thick, color=black] table [x=thread, y=time] {fasta-time-maxlength.dat};
\end{axis}
\end{tikzpicture}
\end{subfigure}
\qquad
\begin{subfigure}[c][][c]{0.4625\textwidth}
\begin{tikzpicture}
\begin{axis} [
    title={\Ovalbox{\large \textsc{traffic -- parsing}}},
    width=1.0\textwidth,
    height=0.6\textwidth,
    xtick={2,10,...,70},
    ytick={0,200,...,1600},
    scaled y ticks=base 10:-3,
    ytick scale label code/.code={},
    unbounded coords = jump,
    grid,
    every tick label/.append style={font=\footnotesize},
    xlabel={\normalsize\bf number of threads -- text of $11182$ Kbytes},
    ylabel={\normalsize\bf time -- seconds},
]
\addplot[mesh, mark=*, thick, color=black] table [x=thread, y=time] {traffic-time-maxlength.dat};
\end{axis}
\end{tikzpicture}
\end{subfigure}
\end{center}
\caption{Absolute times of parallel parsing for four different benchmarks of Tab.~\ref{tab:benchmarks}. \label{fig:timing}}
\tikzexternalenable
\end{figure}
In Fig.~\ref{fig:timing} we plot the  times for parsing the longest respective texts of benchmarks BIGDATA, BIBLE, FASTA and TRAFFIC, by using a variable number of threads, on the Dell PowerEdge R$7425$ platform described at the beginning of Sect.~\ref{sec:experimentation}. Three benchmarks generally run under one second, while FASTA takes a few tens of milliseconds (its text is shorter).
\paragraph{Speed-up of  parsing and of just  recognition (first benchmark)}
The 3D diagrams in Fig.~\ref{fig:bigdata-speedup} plot the speed-up of the parallel parsing and recognition algorithms as a function of the number of threads and the text length, for benchmark BIGDATA. The speed-up of the parallel algorithms is substantial, especially for parsing, up to $21$ times the serial one-chunk execution. For a fixed text length, the parser speed-up initially grows with the number of threads, then it  saturates and at last slowly decays; this happens especially for short texts, because the chunks are shorter. On the other hand, for a fixed number of threads, the parser speed-up grows approximately linearly with the text length (with a lower growth for small numbers of threads), because the chunks become longer. In fact, with short chunks the work of the reach, join and build\&merge tasks is so small that it does not compensate for the  overhead due to multitasking. The recognizer speed-up behaves similarly to the parser one, but it saturates earlier due to the minor complexity of the algorithm.
\par
\begin{figure}[ht]
\tikzexternaldisable
\begin{center}
\begin{subfigure}[t][][c]{0.4625\textwidth}
\begin{tikzpicture}
\begin{axis} [
    title={\Ovalbox{\large \textsc{bigdata -- parsing}}},
    width=0.875\textwidth,
    height=0.175\textheight,
    ytick={0.85,0.90,0.95,1.00,1.05,1.10,1.15},
    xtick={1,2,...,6},
    xticklabels={409\,K\quad,819\,K,1\,M,3\,M,6\,M,13\,M},
    unbounded coords = jump,
    grid,
    every tick label/.append style={font=\footnotesize},
    xlabel={\normalsize\bf text length -- bytes},
    ylabel={\normalsize\bf time ratio},
]
\addplot[mesh, mark=*, thick, color=black] table [x=x, y=y] {bigdata-parsing-comparison-grinder64.dat};
\end{axis}
\end{tikzpicture}
\end{subfigure}
\qquad
\begin{subfigure}[t][][c]{0.4625\textwidth}
\begin{tikzpicture}
\begin{axis} [
    title={\Ovalbox{\large \textsc{bigdata -- recognition}}},
    width=0.875\textwidth,
    height=0.175\textheight,
    ytick={0.30,0.35,0.40,0.45,0.50,0.55,0.60,0.65,0.70},
    xtick={1,2,...,6},
    xticklabels={409\,K\quad,819\,K,1\,M,3\,M,6\,M,13\,M},
    unbounded coords = jump,
    grid,
    every tick label/.append style={font=\footnotesize},
    xlabel={\normalsize\bf text length -- bytes},
    ylabel={\normalsize\bf time ratio},
]
\addplot[mesh, mark=*, thick, color=black] table [x=x, y=y] {bigdata-recognition-comparison-grinder64.dat};
\end{axis}
\end{tikzpicture}
\end{subfigure}
\end{center}
\caption{Left: plot of the ratio of serial parsing times, for the one-chunk parser and the DFA-based serial parser outlined in Sect.~\ref{subsec:toolstructure} -- \emph{serial parser}. Right: the same plot type as on the left, but for recognition.} \label{fig:bigdata-performance}
\tikzexternalenable
\end{figure}
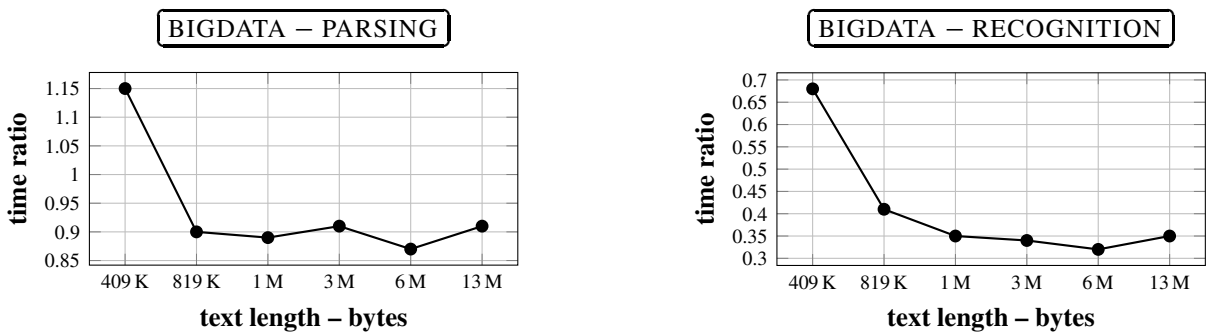
\begin{figure}[ht]
\tikzexternaldisable
\begin{center}
\begin{subfigure}[c][][c]{0.4625\textwidth}
\begin{tikzpicture}
\begin{axis} [
    title={\Ovalbox{\large \textsc{bigdata -- parsing}}},
    width=1.0\textwidth,
    height=0.8\textwidth,
    view={315}{30},
    xtick={2,10,...,66},
    ztick={0,3,...,24},
    ytick={1,2,...,6},
    yticklabels={0.409$\quad$,0.819,1.000,3.000,6.000,13.000},
    unbounded coords = jump,
    grid,
    every tick label/.append style={font=\footnotesize},
    every axis y label/.style={at={(ticklabel cs:0.5)}, rotate=90, anchor=near ticklabel},
    every axis x label/.style={at={(ticklabel cs:0.5)}, anchor=near ticklabel},
    xlabel style={xshift=0pt, rotate=20},
    xlabel={\normalsize\bf number of threads},
    ylabel style={yshift=0pt, rotate=-110},
    ylabel={\normalsize\bf text length -- M bytes},
    zlabel={\normalsize\bf speed-up},
    colormap={pos}{color(0)=(blue) color(0.4)=(green) color(0.6)=(yellow) color(0.8)=(orange) color(1.0)=(red)}
]
\addplot3[mesh, mark=.] table [x=x, y=y, z=z] {bigdata-parsing-grinder64.dat};
\end{axis}
\end{tikzpicture}
\end{subfigure}
\qquad
\begin{subfigure}[c][][c]{0.4625\textwidth}
\begin{tikzpicture}
\begin{axis} [
    title={\Ovalbox{\large \textsc{bible -- parsing}}},
    width=1.0\textwidth,
    height=0.8\textwidth,
    view={315}{30},
    xtick={2,10,...,66},
    ztick={0,2,...,24},
    ytick={1,2,...,6},
    yticklabels={1.5$\quad$,2.0,2.5,3.0,3.5,4.0},
    unbounded coords = jump,
    grid,
    every tick label/.append style={font=\footnotesize},
    every axis y label/.style={at={(ticklabel cs:0.5)}, rotate=90, anchor=near ticklabel},
    every axis x label/.style={at={(ticklabel cs:0.5)}, anchor=near ticklabel},
    xlabel style={xshift=0pt, rotate=20},
    xlabel={\normalsize\bf number of threads},
    ylabel style={yshift=0pt, rotate=-110},
    ylabel={\normalsize\bf text length -- M bytes},
    zlabel={\normalsize\bf speed-up},
    colormap={pos}{color(0)=(blue) color(0.4)=(green) color(0.6)=(yellow) color(0.8)=(orange) color(1.0)=(red)}
]
\addplot3[mesh, mark=.] table [x=x, y=y, z=z] {bible-parsing-grinder64.dat};
\end{axis}
\end{tikzpicture}
\end{subfigure}
\par
\vspace{0.25cm}
\begin{subfigure}[c][][c]{0.4625\textwidth}
\begin{tikzpicture}
\begin{axis} [
    title={\Ovalbox{\large \textsc{fasta -- parsing}}},
    width=1.0\textwidth,
    height=0.8\textwidth,
    view={315}{30},
    xtick={2,10,...,66},
    ztick={0,1,...,24},
    ytick={1,2,...,6},
    yticklabels={658$\quad$,679,701,722,744,765},
    unbounded coords = jump,
    grid,
    every tick label/.append style={font=\footnotesize},
    every axis y label/.style={at={(ticklabel cs:0.5)}, rotate=90, anchor=near ticklabel},
    every axis x label/.style={at={(ticklabel cs:0.5)}, anchor=near ticklabel},
    xlabel style={xshift=0pt, rotate=20},
    xlabel={\normalsize\bf number of threads},
    ylabel style={yshift=0pt, rotate=-110},
    ylabel={\normalsize\bf text length -- K bytes},
    zlabel={\normalsize\bf speed-up},
    colormap={pos}{color(0)=(blue) color(0.4)=(green) color(0.6)=(yellow) color(0.8)=(orange) color(1.0)=(red)}
]
\addplot3[mesh, mark=.] table [x=x, y=y, z=z] {fasta-parsing-grinder64.dat};
\end{axis}
\end{tikzpicture}
\end{subfigure}
\qquad
\begin{subfigure}[c][][c]{0.4625\textwidth}
\begin{tikzpicture}
\begin{axis} [
    title={\Ovalbox{\large \textsc{traffic -- parsing}}},
    width=1.0\textwidth,
    height=0.8\textwidth,
    view={315}{30},
    xtick={2,10,...,66},
    ztick={0,3,...,24},
    ytick={1,2,...,6},
    yticklabels={1.572$\quad$,3.166,4.762,7.971,9.578,11.182},
    unbounded coords = jump,
    grid,
    every tick label/.append style={font=\footnotesize},
    every axis y label/.style={at={(ticklabel cs:0.5)}, rotate=90, anchor=near ticklabel},
    every axis x label/.style={at={(ticklabel cs:0.5)}, anchor=near ticklabel},
    xlabel style={xshift=0pt, rotate=20},
    xlabel={\normalsize\bf number of threads},
    ylabel style={yshift=0pt, rotate=-110},
    ylabel={\normalsize\bf text length -- M bytes},
    zlabel={\normalsize\bf speed-up},
    colormap={pos}{color(0)=(blue) color(0.4)=(green) color(0.6)=(yellow) color(0.8)=(orange) color(1.0)=(red)}
]
\addplot3[mesh, mark=.] table [x=x, y=y, z=z] {traffic-parsing-grinder64.dat};
\end{axis}
\end{tikzpicture}
\end{subfigure}
\end{center}
\caption{Speed-up of multiple vs single thread parallel parsing for four different benchmarks (see Tab.~\ref{tab:benchmarks}). \label{fig:parsing}}
\tikzexternalenable
\end{figure}
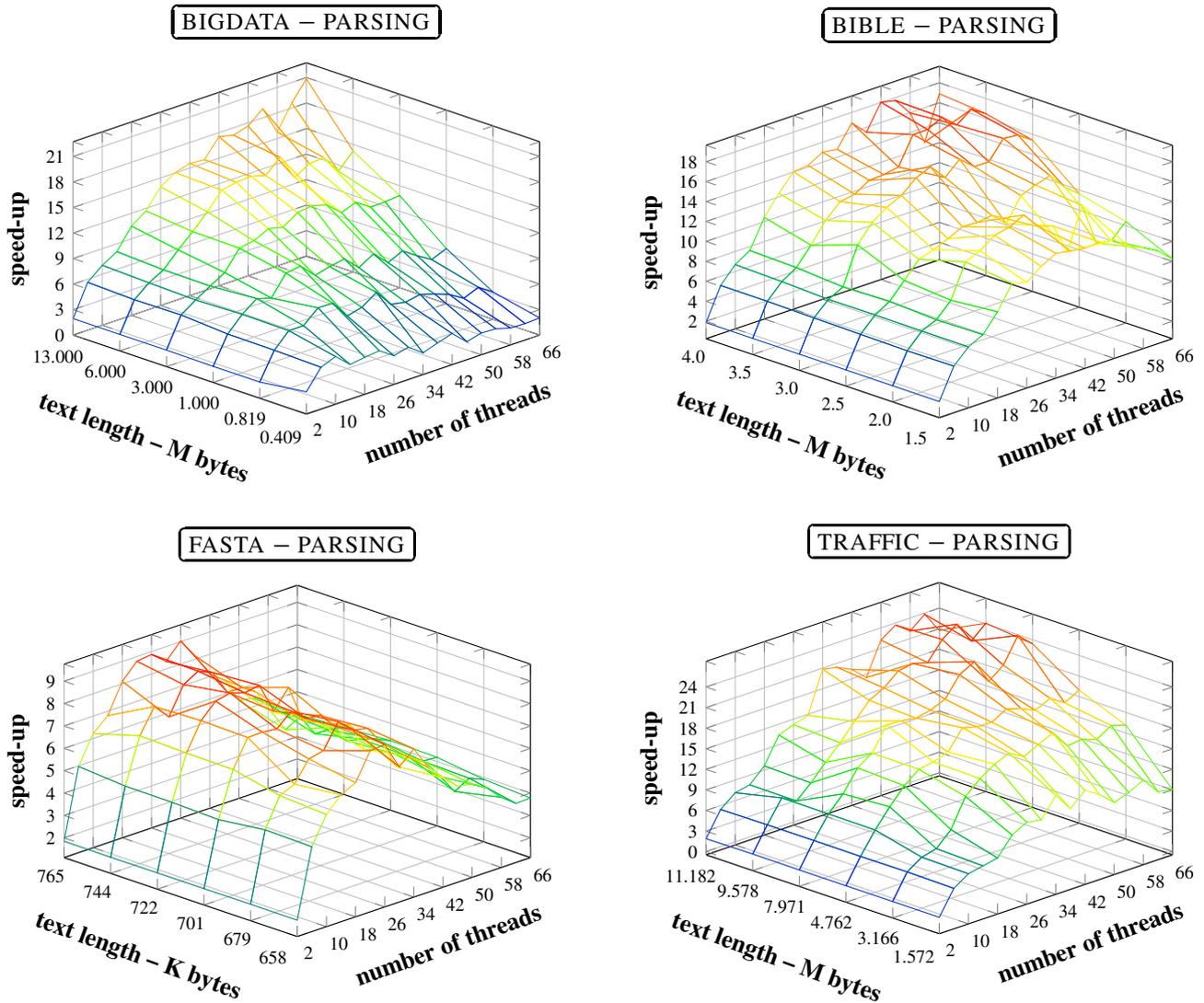

The number of available processors is decoupled from the number of text chunks $c$, i.e., concurrent threads, which ranges from $2$ to $66$. Thus for $c = 64$ chunks the scheduling policy optimizes the utilization of HW resources, as said in Sect.~\ref{subsec:scheduling}. Coherently, parsing achieves maximum speed-up when the number of threads is close to $64$ and the texts are long. Recognition behaves similarly, though its maximum speed-up (about $9$) is lower than for parsing, again due to the simplicity of the algorithm.
\paragraph{Remark on the choice of the serial parser as term of comparison}
We have verified that the serial one-chunk parser used in measuring does not pay an overhead due to its having the same two-level program structure as the parallel parser. The 2D diagram in Fig.~\ref{fig:bigdata-performance} (left) plots the ratio of the time spent by the parallel parsing algorithm when using a single chunk, over the time of the serial DFA-based parser obtained by determinizing the serial parser NFA (Sect.~\ref{subsec:deterministicFA} and Sect.~\ref{subsec:toolstructure} -- \emph{serial parser}); the abscissa shows the text length. After a transient for short texts, the ratio stabilizes and approaches one. This behaviour confirms that the choice of  the serial parser used as comparison term has a negligible impact (just a linear rescaling) on the speed-up obtained. The behaviour is similar for recognition (Fig.~\ref{fig:bigdata-performance}, right).
\paragraph{Speed-up for parsing of real-life benchmarks}
The 3D diagrams in Fig.~\ref{fig:parsing} plot the speed-up of the parallel parsing algorithm as a function of the number of threads and the text length, for the first four benchmarks in Tab.~\ref{tab:benchmarks}, three of which come from real-life applications. Each of these benchmarks uses a single RE, specific for the related application. However, their REs have different sizes spanning one order of magnitude, so that the RE size is also somewhat probed (the parser speed-up of benchmark BIGDATA plotted in Fig.~\ref{fig:bigdata-speedup} is repeated in Fig.~\ref{fig:parsing}, for comparison with the other three benchmarks). The slopes of the parser speed-up plots agree with what already observed and commented for benchmark BIGDATA. The FASTA benchmark exhibits an early saturation of  speed-up, followed by  decay, with respect to the number of threads, because the text length is significantly smaller than  for the other benchmarks. This means (again) that as the number of  chunks grows, their length becomes too small to give a  performance gain that outbalances  the system overhead due to the larger number of threads. A similar trend to saturation is also  observed for benchmark BIBLE, which has text lengths between those of FASTA and those of BIGDATA and TRAFFIC. The maximum speed-up ranges between $9$ and $24$ depending on the benchmark. Their (relatively limited) differences in the plots are justifiable as dimensional and structural diversities of the involved REs and texts to parse.
\par
\begin{figure}[ht]
\tikzexternaldisable
\begin{center}
\rnode {toppart} {
\begin{subfigure}[t][][c]{1.0\textwidth}
\rnode {figA} {
\begin{subfigure}[t][][c]{0.4625\textwidth}
\begin{tikzpicture}
\begin{axis} [
    title={\Ovalbox{\large \textsc{regen -- parsing -- text length}}},
    width=1.0\textwidth,
    height=0.275\textheight,
    view={315}{30},
    xtick={2,10,...,66},
    ztick={0,2,...,18},
    ytick={1,2,...,6},
    yticklabels={1.2$\quad$,1.5,1.8,2.1,2.4,2.7},
    unbounded coords = jump,
    grid,
    every tick label/.append style={font=\footnotesize},
    every axis y label/.style={at={(ticklabel cs:0.5)}, rotate=90, anchor=near ticklabel},
    every axis x label/.style={at={(ticklabel cs:0.5)}, anchor=near ticklabel},
    xlabel style={rotate=20},
    xlabel={\normalsize\bf number of threads},
    ylabel style={rotate=-110},
    ylabel={\normalsize\bf text length -- M bytes},
    zlabel={\normalsize\bf speed-up},
    colormap={pos}{color(0)=(blue) color(0.4)=(green) color(0.6)=(yellow) color(0.8)=(orange) color(1.0)=(red)}
]
\addplot3[mesh, mark=.] table [x=x, y=y, z=z] {regen-parsing-textlength-tripled-grinder64.dat};
\end{axis}
\end{tikzpicture}
\end{subfigure}}
\qquad
\rnode {figB} {
\begin{subfigure}[t][][c]{0.4625\textwidth}
\begin{tikzpicture}
\begin{axis} [
    title={\Ovalbox{\large \textsc{regen -- parsing -- RE size}}},
    width=1.0\textwidth,
    height=0.275\textheight,
    view={315}{30},
    xtick={2,10,...,66},
    ztick={0,2,...,18},
    ytick={1,2,...,6},
    yticklabels={40$\quad$,50,60,70,80,90},
    unbounded coords = jump,
    grid,
    every tick label/.append style={font=\footnotesize},
    every axis y label/.style={at={(ticklabel cs:0.5)}, rotate=90, anchor=near ticklabel},
    every axis x label/.style={at={(ticklabel cs:0.5)}, anchor=near ticklabel},
    xlabel style={rotate=20},
    xlabel={\normalsize\bf number of threads},
    ylabel style={rotate=-110},
    ylabel={\normalsize\bf RE size -- symbols},
    zlabel={\normalsize\bf speed-up},
    colormap={pos}{color(0)=(blue) color(0.4)=(green) color(0.6)=(yellow) color(0.8)=(orange) color(1.0)=(red)}
]
\addplot3[mesh, mark=.] table [x=x, y=y, z=z] {regen-parsing-RElength-tripled-grinder64.dat};
\end{axis}
\end{tikzpicture}
\end{subfigure}}
\end{subfigure}}
\end{center}
\begin{center}
\rnode {bottompart} {
\begin{subfigure}[t][][c]{1.0\textwidth}
\rnode {figC} {
\begin{subfigure}[t][][c]{0.4625\textwidth}
\begin{tikzpicture}
\begin{axis} [
    width=1.0\textwidth,
    height=0.275\textheight,
    view={315}{30},
    xtick={2,10,...,66},
    ztick={0,4,...,32},
    ytick={1,2,...,6},
    yticklabels={64$\quad$,80,96,112,128,144},
    unbounded coords = jump,
    grid,
    every tick label/.append style={font=\footnotesize},
    every axis y label/.style={at={(ticklabel cs:0.5)}, rotate=90, anchor=near ticklabel},
    every axis x label/.style={at={(ticklabel cs:0.5)}, anchor=near ticklabel},
    xlabel style={rotate=20},
    xlabel={\normalsize\bf number of threads},
    ylabel style={rotate=-110},
    ylabel={\normalsize\bf text length -- M bytes},
    zlabel={\normalsize\bf speed-up},
    colormap={pos}{color(0)=(blue) color(0.4)=(green) color(0.6)=(yellow) color(0.8)=(orange) color(1.0)=(red)}
]
\addplot3[mesh, mark=.] table [x=x, y=y, z=z] {regen-parsing-textlength-onehundredsixtytimes-grinder64.dat};
\end{axis}
\end{tikzpicture}
\end{subfigure}}
\qquad
\rnode {figD} {
\begin{subfigure}[t][][c]{0.4625\textwidth}
\begin{tikzpicture}
\begin{axis} [
    width=1.0\textwidth,
    height=0.275\textheight,
    view={315}{30},
    xtick={2,10,...,66},
    ztick={0,4,...,32},
    ytick={1,2,...,6},
    yticklabels={40$\quad$,50,60,70,80,90},
    unbounded coords = jump,
    grid,
    every tick label/.append style={font=\footnotesize},
    every axis y label/.style={at={(ticklabel cs:0.5)}, rotate=90, anchor=near ticklabel},
    every axis x label/.style={at={(ticklabel cs:0.5)}, anchor=near ticklabel},
    xlabel style={rotate=20},
    xlabel={\normalsize\bf number of threads},
    ylabel style={rotate=-110},
    ylabel={\normalsize\bf RE size -- symbols},
    zlabel={\normalsize\bf speed-up},
    colormap={pos}{color(0)=(blue) color(0.4)=(green) color(0.6)=(yellow) color(0.8)=(orange) color(1.0)=(red)}
]
\addplot3[mesh, mark=.] table [x=x, y=y, z=z] {regen-parsing-RElength-onehundredsixtytimes-grinder64.dat};
\end{axis}
\end{tikzpicture}
\end{subfigure}}
\end{subfigure}}
\end{center}
\nput[labelsep=-1.5cm] {90} {toppart} {\Ovalbox{\large \textsc{short texts}}}
\nput[labelsep=-0.5cm] {90} {bottompart} {\Ovalbox{\large \textsc{long texts}}}
\nput[labelsep=-1.5cm] {140} {figA} {\Ovalbox{\large \textsc{A}}}
\nput[labelsep=-1.5cm] {40} {figB} {\Ovalbox{\large \textsc{B}}}
\nput[labelsep=-0.6125cm] {135} {figC} {\Ovalbox{\large \textsc{C}}}
\nput[labelsep=-0.6125cm] {45} {figD} {\Ovalbox{\large \textsc{D}}}
\caption{Speed-up of multiple vs single thread parallel parsing as a function of text-length / RE-size for benchmark REGEN. \label{fig:regen-speedupvsRE&text}}
\tikzexternalenable
\end{figure}
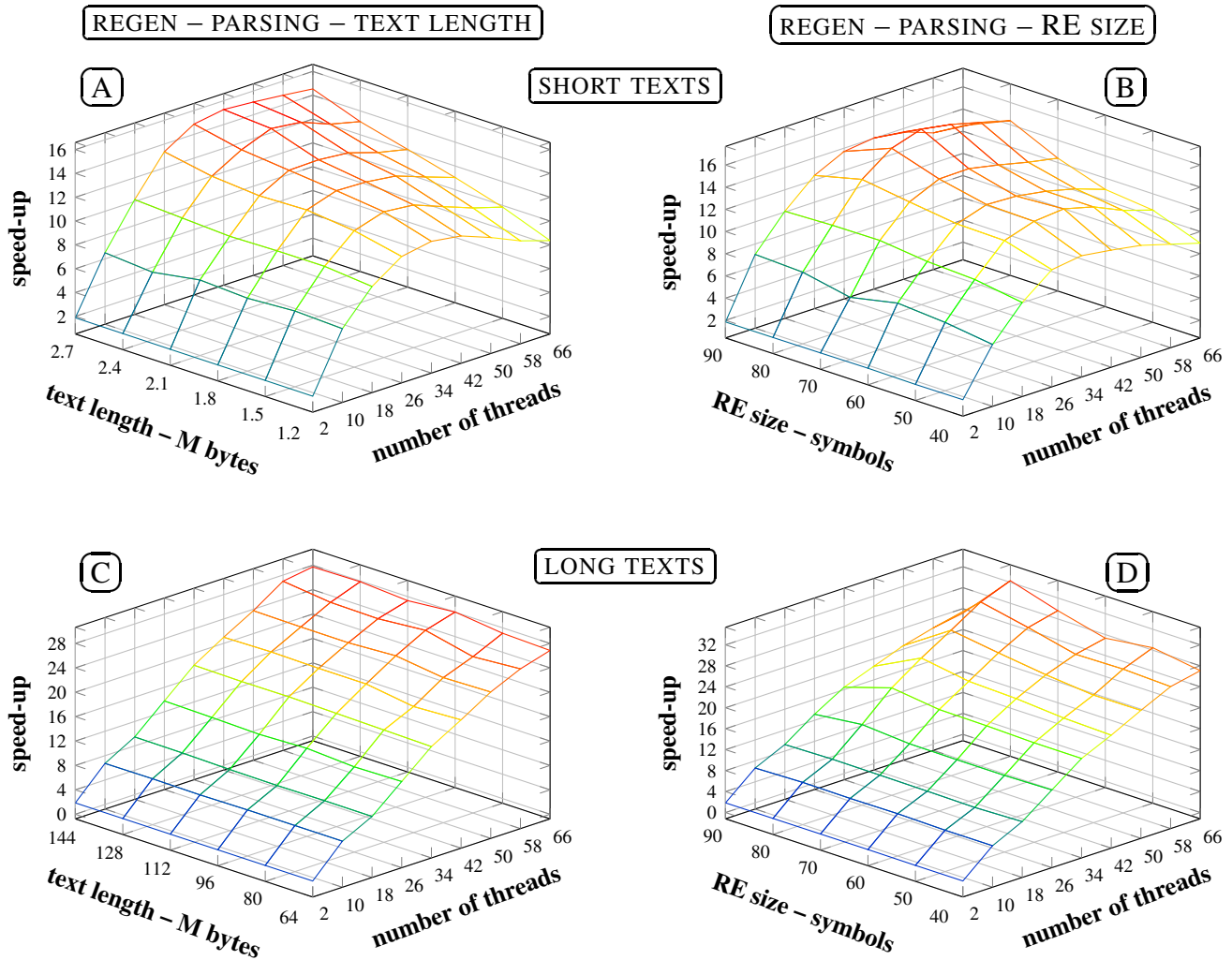

We also experimentally evaluated on these four benchmarks the benefit of using the ME-DFA, instead of the DFA, in the reach phase as explained in Sect.~\ref{subsec:deterministicFA}, to mitigate the overhead deriving from the speculative approach. The gain in performance depends on the analyzed RE and can reach $73 \, \%$ of the entire parsing time.
\paragraph{Using a synthetic benchmark to probe a larger space of REs}
We have seen that for the first four benchmarks of Tab.~\ref{tab:benchmarks}, the  speed-up consistently grows with the number of threads and the text length, except for some cases of early saturation caused by shorter texts. Thus, we have not yet seen evidence of the inherent upper bound of the parallel parsing algorithm itself. Clearly, the speed-up may not trespass the  value $64$, i.e., the number of available processors, thus sooner or later an effective upper bound $\leq 64$ will be reached. To exert the parallel parsing algorithm to its actual limit, we have used the fifth benchmark (REGEN) in Tab.~\ref{tab:benchmarks}, which has much longer texts and several random REs of different sizes. The size of the RE is an additional variable, the impact of which on performance can be analyzed by means of REGEN test data. For benchmark REGEN, we report the speed-up as a function of  text length and number of threads, or as a function of  RE size and number of threads. The results are plotted in 3D in Fig.~\ref{fig:regen-speedupvsRE&text} for two series of texts: short (top part) and long (bottom part), their length ratio being about $50$. Since in a 3D plot just two independent variables can be visualized, one of them is the number of threads, and the other one is respectively the text length in Fig.~\ref{fig:regen-speedupvsRE&text} A, C
and the RE size in Fig.~\ref{fig:regen-speedupvsRE&text} B, D. The missing variable in A and C is the RE size, which is set to the average over the variable domain.
Similarly, for plots B and D the text length is set to the average of text lengths.
\par
\noindent
We comment each plot separately:
\begin{description}[leftmargin=*, style=nextline]
\item[A -- short texts -- text length] the speed-up grows slowly as the text length increases for a fixed number of threads;  it grows linearly with the number of threads,
but saturates before using all available processors
\item[B -- short texts -- RE size] the speed-up grows slowly with the RE size for a fixed number of threads, until it saturates and tends to slowly decay;
 it grows linearly with the number of threads, but again saturates before using all available processors
\item[C -- long texts -- text length] the speed-up is constant for a fixed number of threads, independently of the text length, and it grows linearly with the number of threads
without saturating
\item[D -- long texts -- RE size] the speed-up grows slowly with the RE size for a fixed number of threads, until it saturates and tends to slowly decay; it grows linearly with the number of threads, without saturating
\end{description}
In all  cases, the speed-up grows linearly with the number of threads. Presumably, the saturation with short texts is caused by the parallelization overhead, i.e., the system time spent in managing threads. For long texts, with about twenty threads, i.e., text chunks, the parallel parser is already one order of magnitude faster than the serial one. Maximum speed-up is achieved when the number of threads and  available processors are equal (which as said is $64$), and its value is about half such a number. We have verified that it does not grow any further with longer texts. Therefore, since benchmark REGEN contains random REs, such a limited maximum speed-up can be taken as a structural limit of the parallel algorithm scheme itself (see Fig.~\ref{fig:parallelparser}), irrespective of the specific FAs that drive the various phases (primarily reach and build\&merge) of the algorithm.
\par
\begin{figure}[ht]
\tikzexternaldisable
\begin{center}
\begin{tikzpicture}
\begin{axis} [
    title={\Ovalbox{\large \textsc{segment (NFA state) scattering vs RE size -- REGEN collection}}},
    width=0.90\textwidth,
    height=0.325\textheight,
    clip mode=individual,
    xmin=8, xmax=101,
    ymin=7, ymax=1450,
    xtick={0,10,...,100},
    ytick={0,100,...,1500},
    grid,
    every tick label/.append style={font=\footnotesize},
    xlabel={\normalsize\bf size of the RE -- symbols},
    ylabel={\normalsize\bf number of segments -- NFA states},
    mark size=1.75pt,
    colormap={pos}{color(0)=(blue) color(0.25)=(green) color(0.50)=(yellow) color(0.75)=(orange) color(1)=(red)},
    colorbar,
    colorbar style={ ytick={0,100,...,1500} }
]
\addplot [
    only marks,
    mark=*,
    domain=9:100,
    scatter,
    scatter src=y
] table [x=n1, y=n2] {fa-complete.prn};
\addplot [no markers, ultra thick, magenta] table [x=n1, y={create col/linear regression={y=n2}}] {fa-complete.prn}
node [
    above,
    sloped,
    pos=0.30,
    magenta
] {\parbox{9.5cm}{\large \centering $\text{regression line equation}$ \\[0.15cm] $\text{number of segments} \approx \pgfmathprintnumber[precision=1] {\pgfplotstableregressiona} \times \Vert \, \text{RE} \, \Vert \pgfmathprintnumber[precision=1] {\pgfplotstableregressionb}$
\\[0.15cm] $\text{linear (Pearson) correlation coeff.} \approx 0.52$ \\ \vspace{3.25cm}}};
\end{axis}
\end{tikzpicture}
\end{center}
\caption{Scatter plot of the number of segments as function of RE size for the complete REGEN collection. \label{fig:segments-vs-REsize}}
\tikzexternalenable
\end{figure}
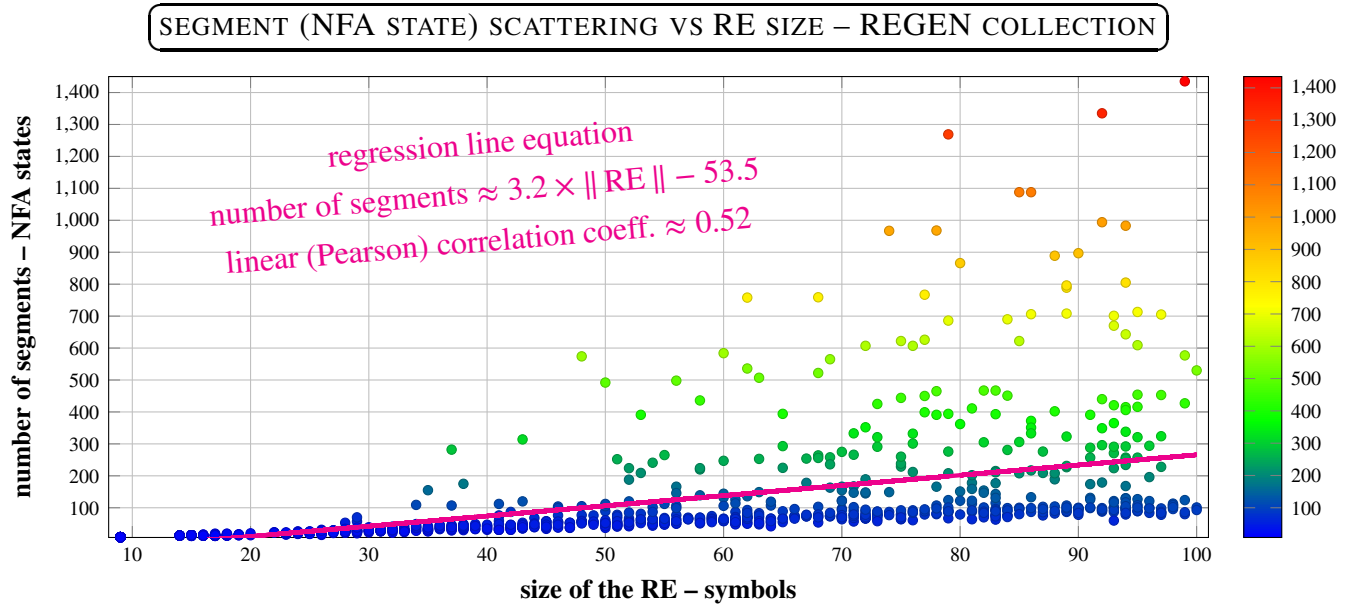

In fact, the chunk FAs that drive the algorithm process one input symbol per move (operating as real-time automata). For  long enough texts, the speed-up is intuitively expected to be independent of the text length, to grow linearly with respect to the number of threads, and to saturate when the number of threads comes close to the number of available processors. From Fig.~\ref{fig:regen-speedupvsRE&text} the speed-up is approximately  constant when the text length changes, especially for long texts. Therefore, the expectation is confirmed by the  plots in Fig.~\ref{fig:regen-speedupvsRE&text}. How the speed-up depends on the RE size is  more difficult to intuitively justify, yet it is approximately constant (with some limited irregularity) up to a certain size, then it tends to decay, possibly due to the increasing structural complexity of the RE itself, which may be higher for the larger REs.
\paragraph{Remark on the modeling and measure of the RE complexity}
The total number of symbols in an RE is a simple  parameter to rank RE complexity  from the user perspective. On the other hand, the number of RE segments, i.e., NFA states, would be a better complexity measure that is  related to the structure of the parallel parsing algorithm, since the FAs driving the parsers are constructed from the segments.   For this reason we have studied the relationship between segment number and RE size, i.e., symbol count.
A scatter plot of the segment count with respect to the RE size is shown in Fig.~\ref{fig:segments-vs-REsize} for the complete REGEN collection, which contains $1,000$ random REs of size from $9$ to $100$ symbols. Each plotted mark corresponds to one or more REs with that combination of size and segment count. The numbers of segments cover the range $[8 - 1,435]$ and the average number is about $107$ segments.
\par
The scatter plot shows a visible trend for the segment count to be directly proportional to the RE size $\Vert \, \text{RE} \, \Vert$. In fact, most marks are well aligned with the regression line of the entire diagram, with few exceptions that have much more numerous segments. Overall, for sufficiently large REs the segment count is expected to be about $3.2$ times the RE size. The linear (Pearson) correlation coefficient is about $0.52$, that is, distant from weak ($< 0.3$) and tending to strong ($> 0.7$). Thus the scatter plot demonstrates that, even if the speed-up values of the REGEN benchmark in Fig.~\ref{fig:regen-speedupvsRE&text} were experimentally measured as a function of the segment count, instead of the RE size, their shape would be roughly linearly scaled, and the qualitative analysis of the speed-up would not change. Therefore, we may conclude that the observed speed-up behaviour is a structural property of the parallel parsing algorithm, and not a feature of the benchmarks or their parameters.
\par
As said, in most cases an SLPF column is encoded in one $64$-bit memory word. Thus the overall memory footprint is proportional to the text length, plus a (small) constant amount of memory that depends only on the RE. For the longest texts of the REGEN benchmark of Tab.~\ref{tab:benchmarks}, the SLPF size ranges between $1$ and $2$ GBytes, compatible with the memory capacity of a laptop computer.
\paragraph{Discussion of speed-up upper bound}
The speed-up upper limit of about half the number of processors deserves  closer analysis. As already observed, the hot-spot phases of the parallel parsing algorithm  are reach (driven by the ME-DFA) and build\&merge (by the DFA), which are serialized and have  similar duration. In other words, the parallel parser is a two-stage algorithm. When the numbers of threads and available processors get closer, we can figure that each processor elaborates exactly one text chunk: it  first executes the reach phase, then the build\&merge one for about the same duration (independently of load balancing). The serial parser we compare with, is  the one-chunk parallel one, thus  it behaves as a two-stage algorithm as well, with a single thread. In the discussion of Fig.~\ref{fig:bigdata-performance},  we argued that the one-chunk parser behaviour resembles the DFA-based serial parser, which just  executes  the build\&merge  on the undivided text. Such a performance similarity of the two serial parsers can be explained by the much lighter weight of reach executed by the parallel one-chunk parser on the undivided text; therefore the time taken by reach is negligible with respect to the time of build\&merge. In fact, the ME-DFA   parses the undivided text and needs to start exclusively from the initial segments of the RE, thus saving the overhead to start from \emph{all} the segments of the RE, with a reduction of speculation degree. All things considered, the parallel parser is expected to use one processor per chunk for twice as long as the serial parser does, thus achieving a maximum speed-up of about half the number of available processors. Real-life benchmarks come more or less close to this speed-up limit depending on their peculiarities, while the REGEN benchmark, which consists of several random REs and numerous texts, comes very close. Our experimentation campaign convincingly supports the evidence of such an effective speed-up bound.
\paragraph{Memory space consumption}
The memory footprint of the parallel parser is mainly due to the SLPF forest, the size of which linearly depends on the text length.
The size of the hash table of the SLPF (see App.~\ref{app:SLPFoptimization}) is negligible  as it does not depend on the text length.
The other data-structures, which comprise some arrays (see Sect.~\ref{subsec:parallelalgorithm}) as well as the ME-DFA and DFA state-transition tables, are also negligible and independent of the text length.
\paragraph{Time to generate the parallel parser}
So far we have not considered the time needed by our tool to generate the parallel parser for a given RE. However, in real life, there are situations where also the parser generation time has to be accounted for, since it delays the delivery of answers to user queries. Then, one may wonder whether the greater complexity of generating a parser instead of a simpler RE matcher could penalize the users. Fortunately, this is not the case: the parallel-parser generation time is very short. We report measurements of the parser generation time for the REs of the benchmarks on a commodity desktop computer at $2.20$ GHz: for BIGDATA $5$ ms (milliseconds), for BIBLE $6$ ms, for FASTA $16$ ms and for TRAFFIC $29$ ms.  As expected, the longer times are for the more complex REs. In  practice, users avoid very long REs, which are error-prone and poorly understandable. Therefore, it is safe to say that the time cost of parser generation does not reduce practical usability.
\paragraph{Summary of results}
The various experimental measurements carried out, with different real-life and synthetic benchmarks (see Fig.s~\ref{fig:parsing} and \ref{fig:regen-speedupvsRE&text}), demonstrate a speed-up of the parallel parsing algorithm compared to the serial one of at least one order of magnitude on fewer than twenty processors, up to a limit of about half the number of available processors. Furthermore, for long texts the speed-up is independent of the text length and scales linearly with the number of threads, until it saturates  at
the above mentioned limit. The memory consumption scales only linearly with the text length. Therefore, we can safely conclude that the parallel parser is effective for practical applications, especially where speed is a relevant requirement and the amount of text to be parsed is large.
The time to construct the parser from a given RE is  suitable to practical use.
\section{Related work} \label{sec:relatedwork}
Fast  processing of regular languages has motivated   much   research for  many years, even before parallel computer architectures became a commodity. In this short survey we focus on some works  on parallel RE recognition / matching, which bear relation to our own work.
As said in the introduction, we are not aware of any preceding work on parallel  (RE) parsing, for which we  just mention the main existing serial algorithms at the end of this section.
\par
The works  considered use a deterministic FA (DFA) obtained  by means of a classic RE-to-FA converter, which in most cases is based on the Gluskhov / McNaughton-Yamada (GMY) algorithm, followed by determinization. For some real-life RE patterns, the latter transformation may cause a quadratic and even exponential growth of the DFA state complexity, as analyzed in the influential paper~\cite{YuChenZhifengDiao2006}.
\paragraph{Foundations on parallel FA computation}
Long before parallel computation technology reached maturity,
Gill~\cite{DBLP:journals/jcss/GillK74}  realized that for some regular languages, recognition can be effected much more economically with a parallel network of identical DFAs, each one starting in a different initial state. Such a machine model, called multi-entry DFA, is a precursor of almost all the subsequent works on parallel FA computation, including ours. Our machine model ME-DFA, see Sect.~\ref{subsec:deterministicFA}, is  a novel  multi-entry automaton optimized to reduce the number of active states, i.e., segments, within a chunk worker.
\par
The very influential paper~\cite{DBLP:journals/jacm/LadnerF80} in the area of the  algorithmic complexity has shown how to parallelize any computational problem specified by a finite-state transducer, e.g., lexical analysis.
An early parallel  implementation for such  problems is reported in~\cite{DBLP:journals/cacm/HillisS86}  as a significant case of data-parallel  programming.
\par
As said in previous sections, the influential paper~\cite{DBLP:conf/wia/HolubS09}  sums up  such early efforts,  and presents and analyzes the  data-parallel algorithm used in almost any subsequent work on parallel RE matching.  For brevity, we call such an approach the ``standard (parallelization) model''.
The standard approach operates in three  phases, which are: (\emph{i}) splitting the input in chunks, (\emph{ii}) speculative parallel scanning of each chunk by means of a DFA to obtain the exit state reached for each entry state, and (\emph{iii}) orderly joining  the  entry-exit state pairs to obtain the successful computations of the input.
\paragraph{Parallel recognition and matching}
In recent years the need to  access huge data-collections by means of RE queries, as well as  the security need for real-time inspection of transmitted  packets, have motivated much practical work on parallelization for various computer architectures.   We  just mention a few works that are closer to ours in their conceptual approach, and we leave aside  the papers  that focus  on  specific parallel architectures, such as FPGAs or GPUs.
\par
Recent works on the standard model primarily  address its major weakness, namely the overhead caused by the need to start each chunk automaton in all DFA  states. Complex REs imply  larger DFAs, with the consequence that the overhead may outweight any gain brought by parallelization.
Different optimizations have been proposed, the effectiveness of which depends not just on the RE size, but also on the RE structure and particularly on the presence of ambiguities.
\par
A frequently voiced  observation is that the overhead due to speculation is related to the number of states that remain active during scanning, after some transitions. Experimental measurements, e.g., in~\cite{DBLP:conf/asplos/MytkowiczMS14,DBLP:journals/ijpp/KoJHB14,FuZheLi2017}, performed on the number of active states after one, two, etc, transitions, agree that the process converges to a constant in a majority of cases after a few steps. For instance, for a benchmark considered in~\cite{DBLP:conf/asplos/MytkowiczMS14} ``\textsl{around $90 \, \%$ of DFAs converge to $16$ active states or less after a mere $10$ steps}''. Although most authors agree on the rapid convergence to a small set of active states,  the reported figures are quite different and depend on the choice of  REs and input texts. Such figures have been exploited in~\cite{DBLP:conf/asplos/MytkowiczMS14,DBLP:journals/ijpp/KoJHB14} to take advantage of  various sources of data parallelism available within one processor, including vector (SIMD) instructions and memory-level parallelism. In our project we do not explicitly consider  instruction-level parallelism. Another common trick, called \emph{state aggregation} in~\cite{FuZheLi2017}, consists of performing two (or more) transitions in a computation step, to reduce the number of steps for reaching  convergence.
At last, the simple local optimization  proposed in the PaREM  recognizer / matcher~\cite{DBLP:conf/cse/MemetiP14} is worth mentioning.  To reduce the number of starting states, the last terminal, say $b$,  of each chunk $y_i$ is used, ahead of scanning chunk $y_{i+1}$ to discard from the starting set any  state that is not the target of a $b$-labeled transition in the parser automaton. For certain families of REs such a simple  optimization has been found effective;it would be interesting to experiment it in combination with our  RE parser. 
\par
We have recently adapted the ME-DFA approach to reduce speculation in a parallel RE recognizer, thus achieving a significant speedup compared to parallel recognizers based on the standard approach. Experimental results are reported in~\cite{borsotti2024minimizingspeculationoverheadparallel}, obtained via the related tool available on \parRErectool.
\par
A secondary concern is to balance the work on chunks to minimize idle waiting by the processors. For instance, if all the chunks have the same length, the first processor would finish much earlier since the first chunk task  starts in just one state and thus avoids speculative state-transitions. The matchers in~\cite{DBLP:journals/ijpp/KoJHB14,FuZheLi2017} use certain RE/DFA-dependent heuristics
to roughly estimate in advance the optimal length of each chunk.
\par
We operate differently to  balance the worker activity independently of the RE, by resorting to  a classical parallel programming method.
Consider the reacher tasks in Fig.~\ref{fig:parallelparser}. The input is initially split into a number of equal chunks that is larger than the number of processors. Each reacher takes an unprocessed chunk from the pool and, as soon as it finishes processing the chunk, it takes another one, if any is left. At the end, all the results of the reachers are combined by the joiners. We have found that having four times more chunks than there are processors achieves a good balancing without penalizing the join overhead (see Sect.~\ref{subsec:tooloptimization} -- \emph{load balancing}).
\paragraph{RE parsing}
Parsing  of RE has been  less investigated than  recognition / matching in the serial case, and this is the first attempt to parallelize. Notice that parallel parsing  has  been investigated in the more general setting of  context-free languages, e.g., in~\cite{Zhao+AlParParserHTML13,DBLP:journals/scp/BarenghiCMPP15}, and also for specific  mark-up languages such as JSON. Such parsers are also based on the data-parallel approach, but they   use a pushdown automaton instead of an FA. For  regular languages, we do not expect such parsers  to be competitive with RE parsers, but experimental measurements remain to be done.
Yet, some ideas introduced in recent parallel  parsers for semi-structured data, e.g., JSON and XML, might be also useful for REs. An example from~\cite{Jiang+AlScalableParallelProcessingSemistructData2019} is the idea of compiling the queries and the JSON syntax that defines the input,  into a single (pushdown) automaton.
\par
We recall that RE parsers come in  two categories. The \emph{partial} parsers  return just one syntax tree, either chosen at random or, more typically, the prior tree according to the POSIX criterion, e.g., in~\cite{DBLP:conf/wia/OkuiS10,DBLP:conf/flops/SulzmannL14,DBLP:journals/spe/BorsottiT21,DBLP:journals/corr/abs-2206-01398}, or according to the greedy criterion, e.g., in~\cite{DBLP:conf/icalp/FrischC04}.
The \emph{total} parsers  return all the syntax trees, either by sequentially enumerating them~\cite{DBLP:journals/ijfcs/SulzmannL17}, or in a packed forest representation as we did in  our precedent Berry-Sethi Parser BSP~\cite{DBLP:journals/acta/BorsottiBCM21},  which has been the baseline for our parallel development.
Clearly, partial parsers are not intended for RE matching.
\par
The SLPF structure we use to encode the set of all the parses of a given string, is similar in function to the \emph{shared packed parse forest} (SPPF) used in grammar parsing~\cite{Tomita86,Johnson1991} for encoding all the parse trees of ambiguous sentences. As we argue in App.~\ref{app:compareSLPFandSPPF}, the memory footprint of the SLPF, the size of which grows linearly with the string length, is better than that of the SPPF, which is non-linear polynomial~\cite{DBLP:journals/scp/ScottJ18} or in some cases even non-polynomial~\cite{DBLP:journals/vlc/BorsottiBCM23} with respect to the string length.
\par
Our parallel total parser,  described in the preceding sections, is a novel contribution and its structure (see Fig.~\ref{fig:parallelparser}) is more articulated than in the recognizers / matchers. In Fig.~\ref{fig:parallelparser} the reachers essentially perform recognition and pass to the builders the information needed to avoid the very expensive creation of  linearized syntax trees that cover some chunks but not the whole input.
The  two cascaded phases of reach and build, each one in turn parallel, are a major design innovation.
\par
Since the parser NFA and the derived DFA may be quite larger than the corresponding automata used by  RE recognizers, we effectively reduce  the number of initial states by means of the multi-entry DFA technique, see the ME-DFA in Sect.~\ref{subsec:deterministicFA}, also a novel contribution.
\section{Conclusion} \label{sec:conclusion}
The present parallel parser is innovative in two senses, namely functionality and design. To our knowledge, it is the first parallel algorithm that parses a text defined by an RE  and returns one or more syntax trees in a compact forest representation;  the forest precisely represents how the different input parts are matched by RE subexpressions.
The parser organization is much more sophisticated than  in the many proposed parallel RE recognizers and matchers  that operate  multiple speculative parsing threads for each  chunk of the input text. RE parsing is conceptually equivalent  to RE recognition by means of a nondeterministic finite-state machine NFA, the successful runs of which encode the syntax trees of the input. However the NFA is  too slow to be practical, and is replaced by a novel machine, the multi-entry DFA, which combines the state reduction of a nondeterministic machine with the speed of deterministic state transitions. This, jointly with other optimizations such as  two  scanning phases at different granularity,  effectively controls  the cost of  speculative execution. The parser performance on commodity multi-core computers is good and achieves a significant and scalable   speed-up  over serial execution, for a collection of real-life and synthetic benchmarks.
\par
Concerning future  developments,  we  consider to extend the tool in various directions, in particular:  least editing distance to correct input texts, and   complete support for the partial syntax modeling feature (see App.~\ref{app:otherfeatures}). Another practically relevant development  is to extend the parallel  parser from REs, i.e, regular languages,  towards semi-structured data representations such as JSON. This development would involve a careful combination of finite-state methods,  general parsing methods~\cite{DBLP:journals/toplas/ScottJ06,DBLP:journals/vlc/BorsottiBCM23}, and parallel parsing for locally parsable languages, e.g., in~\cite{DBLP:journals/scp/BarenghiCMPP15,DBLP:conf/hpcasia/LiT23}.
\section*{Acknowledgments}
To Filippo Carloni and Stefano Zanero for referring us to several benchmarks of regular expressions. To Alessandro Barenghi for helping us in the experimentation campaign on the computing platform.
\bibliography{automatabib}
\appendix
\section{-- Other features of REs} \label{app:otherfeatures}
\begin{figure}[ht]
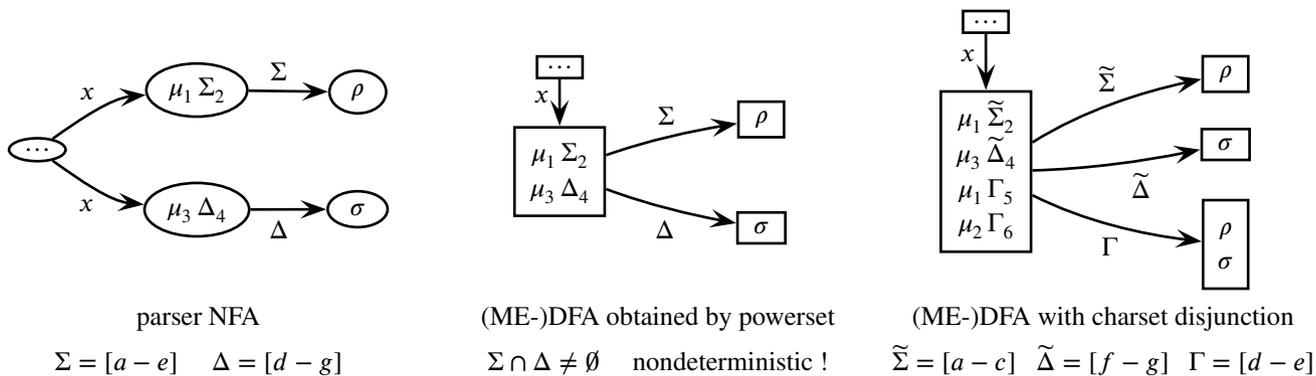

\begin{center}
\psset{labelsep=3pt, nodesep=0pt, linewidth=1pt, arrows=->, arrowscale=2, colsep=1.0cm, rowsep=0.5cm, border=0.0cm}
\begin{tabular}{c@{\hspace{1.0cm}}c@{\hspace{0.75cm}}c}
\begin{tabular}[c]{c}
\psset{rowsep=0.25cm}
\begin{psmatrix}
& \ovalnode {sNFA1} {$\mu_1 \, \Sigma_2$} & \ovalnode {tNFA1} {$ \ \rho \ $} \\

\ovalnode{sourceNFA} {$\ldots$} \\

& \ovalnode {sNFA2} {$\mu_3 \, \Delta_4$} & \ovalnode {tNFA2} {$ \ \sigma \ $} \\

\ncline {sNFA1} {tNFA1} \naput[npos=0.375]{$\Sigma$}

\ncline {sNFA2} {tNFA2} \nbput[npos=0.375]{$\Delta$}

\ncarc[arcangle=15] {sourceNFA} {sNFA1} \naput{$x$}

\ncarc[arcangle=-15] {sourceNFA} {sNFA2} \nbput{$x$}
\end{psmatrix}
\end{tabular}

&

\begin{tabular}[c]{c}
\psset{colsep=1.5cm, rowsep=0.25cm, labelsep=3pt}
\begin{psmatrix}

\rnode {sourceDFAndet} {\psframebox{$\ldots$}} \\

\rnode {sDFAndet} {\psframebox{$\begin{array}{c} \mu_1 \, \Sigma_2 \\ \mu_3 \, \Delta_4 \end{array}$}} & \begin{tabular}[c]{c} \rnode {tDFA1ndet} {\psframebox{$ \ \rho \ $}} \\[1.0cm] \rnode {tDFA2ndet} {\psframebox{$ \ \sigma \ $}} \end{tabular} \\

\ncline {sourceDFAndet} {sDFAndet} \nbput[npos=0.375]{$x$}

\ncarc[arcangle=5] {sDFAndet} {tDFA1ndet} \naput{$\Sigma$}

\ncarc[arcangle=-5] {sDFAndet} {tDFA2ndet} \nbput{$\Delta$}
\end{psmatrix}
\end{tabular}

&

\begin{tabular}[c]{c}
\psset{colsep=2.0cm, rowsep=0.25cm, labelsep=3pt}
\begin{psmatrix}

\rnode {sourceDFA} {\psframebox{$\ldots$}} \\

\rnode {sDFA} {\psframebox{$\begin{array}{c} \mu_1 \, \widetilde \Sigma_2 \\ \mu_3 \, \widetilde \Delta_4 \\ \mu_1 \, \Gamma_5 \\ \mu_2 \, \Gamma_6 \end{array}$}} & \begin{tabular}[c]{c} \rnode {tDFA1} {\psframebox{$ \ \rho \ $}} \\[0.5cm] \rnode {tDFA2} {\psframebox{$ \ \sigma \ $}} \\[0.5cm]\rnode {tDFA3} {\psframebox{$\begin{array}{c} \rho \\ \sigma \end{array}$}} \end{tabular} \\

\ncline {sourceDFA} {sDFA} \nbput[npos=0.375]{$x$}

\ncarc[arcangle=10] {sDFA} {tDFA1} \naput{$\widetilde \Sigma$}

\ncarc[arcangle=-5] {sDFA} {tDFA2} \nbput[npos=0.625]{$\widetilde \Delta$}

\ncarc[arcangle=-10] {sDFA} {tDFA3} \nbput{$\Gamma$}
\end{psmatrix}
\end{tabular}

\\[-0.25cm]

parser NFA & (ME-)DFA obtained by powerset & (ME-)DFA with charset disjunction

\\[0.125cm]

$\Sigma = [a - e]$ \quad $\Delta = [d - g]$

&

$\Sigma \cap \Delta \not = \emptyset$ \quad nondeterministic !

&

$\widetilde \Sigma = [a - c]$ \ \ $\widetilde \Delta = [f - g]$ \ \ $\Gamma = [d - e]$
\end{tabular}
\end{center}
\caption{Generalized segments for modeling character sets and how to obtain the deterministic parser.} \label{fig:charset}
\end{figure}

The basic RE features supported were introduced in Sect.~\ref{subsec:regexp}. The \emph{atomic} expressions are the \emph{alphabetical} symbols; the \emph{basic} (or standard) operators are \emph{concatenation}, \emph{union}, and Kleene \emph{star} or \emph{cross}; all these features can be freely combined, except that infinite ambiguity was excluded. Here are a few other RE features our tool supports or is designed to support by customization. First we briefly consider \emph{escaped} symbols, \emph{wildcards} and \emph{sets} or \emph{intervals} of symbols (in Unicode format), \emph{bounded repetition} operators, \emph{empty string} symbols, \emph{infinite ambiguity}, and \emph{extra parenthesis}, which are all supported by our tool. Eventually we consider \emph{partial syntax} modeling, which is currently unsupported but could be similarly added to the tool. All such additional features can be realized by simply updating the notion of segment of Sect.~\ref{subsubsec:foundationsRE2NFA} and the segment computation algorithm of Fig.~\ref{fig:segmentalgo}, depending on the specific feature wanted. The rest of the parsing method is unchanged, that is:
\begin{itemize}[leftmargin=*, style=standard]
\item the classic followers and the segment followers are defined as of Eq.~\eqref{eq:classicfollowers} and Eq.~\eqref{eq:folsegdef}, respectively
\item the parser NFA is defined as in Sect.~\ref{subsec:parserautomaton}, and its computations represent the LSTs and whole SLPF
\end{itemize}
Thus both the serial and parallel parsing algorithms still function with no changes, as they are derived from the parser NFA.
\paragraph{Character features}
In order to perform measurements in real cases, the RE features below are necessary for our tool, as they occur frequently:
\begin{description}[leftmargin=1.75cm, style=multiline]
\item[escape] a slash ``\,$\setminus$\,'' forces any symbol to be a terminal, e.g., ``\,$\setminus \, ($\,'' turns an isolated open parenthesis into a terminal
\item[wildcard] a dot ``\,.\,'' matches any terminal character (not newline), e.g., ``\,$a\,.\,b$\,'' is any string of length $3$ with endpoints $a$ and $b$
\item[char set and interval] a pair of square brackets $[ \, \ldots \, ]$ indicates a subset of terminal characters that consists of individual
characters or intervals thereof, e.g., $[a - z, \, 0]$ for the subset of all the Latin-$1$ lowercase letters plus digit $0$; set complement is also supported, e.g., $[\widehat{\;\;}\, 0 - 9]$ denotes all the characters except for the decimal digits
\item[Unicode] any Unicode character is a valid terminal symbol (according to the Unicode version supported by Java)
\end{description}
For efficiency, character sets should not be represented by explicitly listing all the set members. For instance, reducing the interval $[a-z]$ to a union $a \, \mid \, b \, \mid \, \ldots \, \mid \, z$ would result into many segments (at least one per each character $a$, $b$, \ldots, $z$) with an identical behaviour, i.e., the same followers. It suffices to define a single generalized segment $\mu \, \Sigma_\#$, where the end-letter is a character set $\Sigma$ (or a character interval or a wildcard), numbered in a unique way, which can match any input symbol in the set, e.g., $\Sigma = [a-z]$. Such a generalized segment becomes a state of the parser NFA, just like an ordinary segment, and the arcs from the state are simply labeled with the set name (unnumbered), e.g., $\Sigma$. A state of the ME-DFA or DFA however, might contain two or more generalized segments. For the (ME-)DFA to be deterministic, such segments must not match the same input symbols, lest the state has two or more outgoing arcs (directed to different target states) that match the same input. This requires to partition the character sets into disjoint subsets, at the cost of a limited number of additional segments, as exemplified in Fig.~\ref{fig:charset}.
Our tool supports all these features, which in the benchmarks turned out to be very effective for reducing the number of segments, i.e., states and transitions of the parser (see Sect.~\ref{subsec:results}). The support of Unicode blends with the one of character sets. The parser builds a map from characters (all Java Unicode) to integers representing
character subsets. This map is an array of blocks (planes), where identical blocks are shared.
\paragraph{Bounded repetition}
The tool  supports some operators of \emph{bounded repetition}, in the variants (in POSIX notation) $e \set{h}$, $e \set{h, \, k}$ and $e \set{h,}$ (with $0 \leq h < k$), and of \emph{optionality} $e? = e \set{0, \, 1}$. A bounded repetition cannot be directly expanded into basic RE operators, e.g., $a^3$ into $_1( \, a_2 \cdot a_2 \cdot a_2 \, )_1$ because a symbol would occur multiple times with the same numbering in the numbered RE. But for the parser NFA to be constructed through the GMY algorithm, each symbol in the RE  must be uniquely identified.
\par
\begin{figure}[ht]
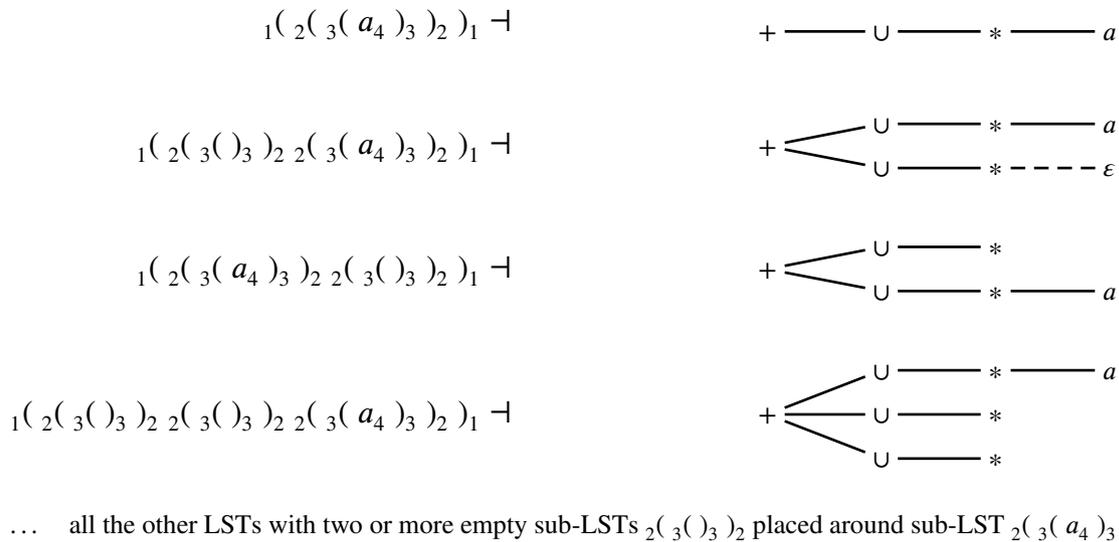

\begin{center}
\tabcolsep=0.0cm\begin{tabular}{r@{\hspace{3.25cm}}l}
\large $_1( \; _2( \; _3( \; a_4 \; )_3 \; )_2 \; )_1 \dashv$ &
\tabcolsep=0.0cm\begin{tabular}{c}
\psset{treemode=R, nodesep=3pt, linewidth=1pt, arrows=-, levelsep=1.5cm, treefit=tight, treesep=0.375cm}
\pstree{\TR{$+$}} {
    \pstree{\TR{$\cup$}} {
        \pstree{\TR{$\ast$}} {
            \TR{$a$}
        }
    }
}
\end{tabular} \\[1.0cm]
\large $_1( \; _2( \; _3( \; )_3 \; )_2 \; _2( \; _3( \; a_4 \; )_3 \; )_2 \; )_1 \dashv$ &
\tabcolsep=0.0cm\begin{tabular}{c}
\psset{treemode=R, nodesep=3pt, linewidth=1pt, arrows=-, levelsep=1.5cm, treefit=tight, treesep=0.375cm}
\pstree{\TR{$+$}} {
    \pstree{\TR{$\cup$}} {
        \pstree{\TR{$\ast$}} {
            \TR{$a$}
        }
    }
    \pstree{\TR{$\cup$}} {
        \pstree[linestyle=dashed]{\TR{$\ast$}} {
            \TR{$\varepsilon$}
        }
    }
}
\end{tabular} \\[1.0cm]
\large $_1( \; _2( \; _3( \; a_4 \; )_3 \; )_2 \; _2( \; _3( \; )_3 \; )_2 \; )_1 \dashv$ &
\tabcolsep=0.0cm\begin{tabular}{c}
\psset{treemode=R, nodesep=3pt, linewidth=1pt, arrows=-, levelsep=1.5cm, treefit=tight, treesep=0.375cm}
\pstree{\TR{$+$}} {
    \pstree{\TR{$\cup$}} {
        \TR{$\ast$}
    }
    \pstree{\TR{$\cup$}} {
        \pstree{\TR{$\ast$}} {
            \TR{$a$}
        }
    }
}
\end{tabular} \\[1.0cm]
\large $_1( \; _2( \; _3( \; )_3 \; )_2 \; _2( \; _3( \; )_3 \; )_2 \; _2( \; _3( \; a_4 \; )_3 \; )_2 \; )_1 \dashv$ &
\tabcolsep=0.0cm\begin{tabular}{c}
\psset{treemode=R, nodesep=3pt, linewidth=1pt, arrows=-, levelsep=1.5cm, treefit=tight, treesep=0.375cm}
\pstree{\TR{$+$}} {
    \pstree{\TR{$\cup$}} {
        \pstree{\TR{$\ast$}} {
            \TR{$a$}
        }
    }
    \pstree{\TR{$\cup$}} {
        \TR{$\ast$}
    }
    \pstree{\TR{$\cup$}} {
        \TR{$\ast$}
    }
}
\end{tabular} \\[1.0cm]
\multicolumn{2}{l}{\ldots \quad all the other LSTs with two or more empty sub-LSTs $_2( \; _3( \; )_3 \; )_2$ placed around sub-LST $_2( \; _3( \; a_4 \; )_3 \; )_2$}
\end{tabular}
\end{center}
\caption{The first shortest LSTs of the infinitely ambiguous string $a$ of RE $e_5$ above (left), and their graphic trees (right).} \label{fig:infamb}
\end{figure}

To circumvent this problem, the argument of a bounded operator is assigned an iteration number, e.g., the $3$-rd power $a^3$ becomes $_1( \, a_{2, 1} \cdot a_{2, 2} \cdot a_{2, 3} \, )_1$ and results in three distinct segments with different end-letters $a_{2, 1}$, $a_{2, 2}$ and $a_{2, 3}$. If the argument is a subexpression, the iteration number is applied recursively to all the numbered symbols in the subexpression. This allows the user to uniquely identify each symbol, and to extract specific matches.
\paragraph{Empty (or null) string}
An RE may contain empty string symbols $\varepsilon$, which are necessary for modeling nullable subexpressions. We exemplify this feature on  RE $e_4 = ( \, a \; \mid \; \varepsilon \, ) \, \cdot \, b$:
\begin{itemize}[leftmargin=*, style=standard]
\item a syntax tree may have empty leaf nodes, e.g., the tree for string $b$: \
\begingroup
\tabcolsep=0.0cm \def\arraystretch{0}
\begin{tabular}[c]{c}
\psset{treemode=R, nodesep=3pt, linewidth=1pt, arrows=-, levelsep=1.5cm, treefit=tight, treesep=0.375cm}
\pstree{\TR{$\cdot$}} {
    \TR{$b$}
    \pstree{\TR{$\cup$}} {
        \TR{$\varepsilon$}
    }
}
\end{tabular}
\endgroup
\item also the empty string $\varepsilon$ is numbered, e.g., $e_{4\#} = \, _1( \, _2( \, a_3 \; \mid \; \varepsilon_4 \, )_2 \; b_5 \, )_1$ with
\begingroup
\tabcolsep=0.0cm\def\arraystretch{0.5}
\begin{tabular}[c]{c@{\hspace{0.25cm}}l}
$\#$ & \emph{operator} \\ \toprule
$1$ & concatenation \\ \midrule
$2$ & union \\ \bottomrule
\end{tabular}
\endgroup
\item a meta-prefix may contain numbered symbols $\varepsilon_\#$, e.g., segment ``\,$\underbracket[0.75pt]{_1( \, _2( \, \varepsilon_4 \, )_2}_\text{meta-prefix} \; b_5$\,'' of RE $e_4$
\item the segments are computed by extending their meta-prefix also with numbered symbols $\varepsilon_\#$
\end{itemize}
For instance, for string $b$ the parser NFA of RE $e_4$ has the accepting path \begingroup \tabcolsep=0.35cm \def\arraystretch{0} \begin{tabular}[c]{c} \psset{nodesep=0pt, linewidth=1pt, arrows=->, arrowscale=2, nodealign=true, colsep=0.75cm, rowsep=0.0cm} \begin{psmatrix} \ovalnode {N1} {$_1( \, _2( \, \varepsilon_4 \, )_2 \; b_5$} & \ovalnode {N2} {$)_1 \dashv$} \nput[labelsep=0pt] {180} {N1} {$\to$} \nput[labelsep=0pt] {0} {N2} {$\to$} \ncline {N1} {N2} \naput[npos=0.375] {$b$} \end{psmatrix} \end{tabular} \endgroup and computes the LST $_1( \, _2( \, \varepsilon_4 \, )_2 \; b_5 \, )_1$, which corresponds to the graphic tree above.
\paragraph{Infinite ambiguity}
An RE that generates at least one string with an infinite number of LSTs, is said to be \emph{infinitely ambiguous}. For generality our parser does not reject such REs, which in the literature are often blamed as  problematic. Instead, it returns a finite sample of the LSTs, which is still representative of the generality. An example is RE $e_5$ in Eq.~\eqref{eq:infamb}, with two-level iteration:
\begin{equation}
\vspace{0.125cm}
\label{eq:infamb}
\begin{array}{rcl}
e_5 & = & \big( \, a^\ast \; \mid \; a \; b \, \big)^+ \\[0.5cm]
e_{5\#} & = & _1( \; _2( \; _3( \; a_4 \; )_3^\ast \; \mid \; _5( \; a_6 \; b_7 \; )_5 \; )_2 \; {)_1}^+
\end{array}
\hspace{3.5cm}
\tabcolsep=0.0cm\def\arraystretch{0.75}
\begin{tabular}{c@{\hspace{0.5cm}}l}
$\#$ & \emph{operator} \\ \toprule
$1$ & cross \\ \midrule
$2$ & union \\ \midrule
$3$ & star \\ \midrule
$5$ & concatenation \\ \bottomrule
\end{tabular}
\vspace{0.125cm}
\end{equation}
The language generated by  RE $e_5$ contains the string $a$, which is infinitely ambiguous (actually all the strings of $L \, (e_5)$ are so). The first four of an infinite series of LSTs for $a$, along with the corresponding graphic tree, are shown in Fig.~\ref{fig:infamb}; whether or not to append the empty string $\varepsilon$ to a star node that iterates zero times is irrelevant (we generally omit it).
\par
Our parser recognizes the presence of infinite ambiguity in an RE and returns just the first few LSTs. When extending a segment meta-prefix, it suffices to allow numbered parentheses ``\,$_\#($\,'' and ``\,$)_\#$\,'', and possibly also empty strings $\varepsilon_\#$, to occur in the meta-prefix two or more times up to a fixed limit, so that the number of segments remains finite. The segment computation algorithm simply has to use counters for the numbered symbols. This approach is  similar to that discussed in~\cite{DBLP:journals/acta/BorsottiBCM21}, where more detail can be found.
\paragraph{Extra parenthesis}
We say that, in an RE, any parenthesis pair that is not needed to specify the scope of an operator or a change of the standard operator precedence, is an \emph{extra parenthesis}. In the positive sense, an extra parenthesis can be placed to specify how to associate the operands of an otherwise associative operator, e.g., $a \, ( \, b \, c \, )$ is an extra parenthesis since it is equivalent to $a\,b\,c$, or to outline a subexpression that for some reason deserves attention. For instance, the RE $a \; \mid \; ( \, a \, b \, c \, )$ uses an extra parenthesis since it is equivalent to $a \; \mid \; a \, b \, c$, but it allows the user to find all the occurrences of the term $a \,b\, c$ in the text.
\par
Placing extra parenthesis pairs is a subjective choice, which some RE matchers offer under the name of \emph{grouping}. Anyway, our parsing method directly supports it. The segment construction algorithm applies as-is also to the extra parenthesis, e.g., RE $a \, ( \, b \, c \, )$ is numbered as $_1( \, a \, _2( \, b \, c \, )_2 \, )_1$ and the segments may contain also the numbered extra parenthesis ``\,$_2($\,'' and ``\,$)_2$\,''.
\paragraph{Partial syntax modeling}
So far, all the operators that occur in an RE  are represented as numbered parenthesis pairs in the corresponding LST. On the other hand, a user may prefer to mask certain uninteresting syntactic aspects, to obtain  neater  syntax trees. Our parser has been designed to be customizable with respect to such user requirement. A simple example illustrates how to do.
\par
Suppose the user wants to mask the union operator (numbered $2$) in the  RE $e_2$ reported below (see also Ex.~$2$). To achieve this, the numbered RE $e_{2\#}$ is simplified as shown in the RE $\widetilde e_{2\#}$ in Eq.~\eqref{eq:masking}.
\begin{equation}
\vspace{0.125cm}
\label{eq:masking}
e_2 = \big( \, a \; b \; \mid \; a \, \big)^\ast \hspace{1.0cm}
\begin{array}{rcl@{\hspace{0.5cm}}l}
e_{2\#} & = & \, _1( \; _2( \; _3( \; a_4 \; b_5 \; )_3 \; \mid \; a_6 \; )_2 \; {)_1}^\ast & \text{with numbered union} \\[0.2cm]
\widetilde e_{2\#} & = & \, _1( \; _3( \; a_4 \; b_5 \; )_3 \; \mid \; a_6 \; {)_1}^\ast & \text{without numbered union}
\end{array}
\hspace{1.0cm}
\tabcolsep=0.0cm\def\arraystretch{0.75}
\begin{tabular}{c@{\hspace{0.5cm}}l}
$\#$ & \emph{operator} \\ \toprule
$1$ & star \\ \midrule
$2$ & union \\ \midrule
$3$ & concatenation \\ \bottomrule
\end{tabular}
\vspace{0.125cm}
\end{equation}
The numbered parenthesis pair $_2( \ )_2$ is canceled (along with the square brackets) and the original numbering, though now discontinuous, is retained for comparison. Now the syntax tree of the  string $a \, a \, b \; a$ is condensed as shown in Fig.~\ref{fig:masking}.
\begin{figure}[ht]
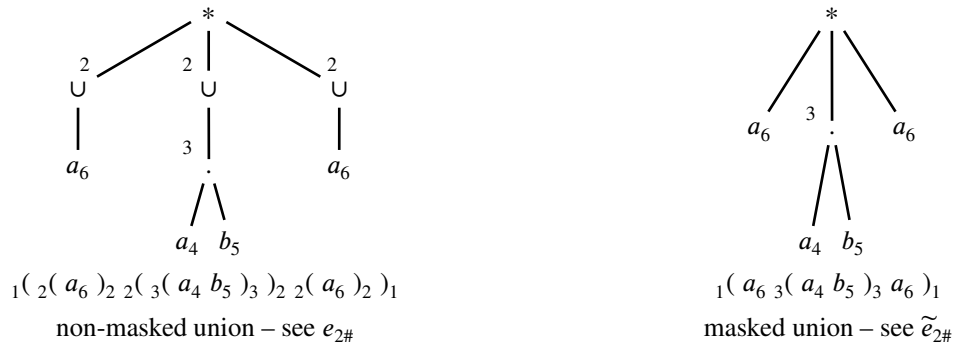

\begin{center}
\psset{arrows=-, linestyle=solid, linewidth=1pt, nodesep=4pt}
\tabcolsep=0.0cm\def\arraystretch{0.5}
\begin{tabular}{c@{\hspace{4.0cm}}c}
\tabcolsep=0.0cm\def\arraystretch{0}
\begin{tabular}[c]{c}
\pstree[treesep=1.5cm, levelsep=1.0cm, tpos=0.675]{\TR{$\ast$}}{
    \pstree{\TR{$\cup$} \tlput{\footnotesize 2}}{
        \TR{$a_6$}
    }
    \pstree[tpos=0.65]{\TR{$\cup$} \tlput{\footnotesize 2}}{
        \pstree[treesep=0.25cm]{\TR{$.$} \tlput[tpos=0.75]{\footnotesize 3}}{
            \TR{$a_4$}
            \TR{$b_5$}
        }
    }
    \pstree{\TR{$\cup$} \trput{\footnotesize 2}}{
        \TR{$a_6$}
    }
}
\end{tabular}

&
\tabcolsep=0.0cm\def\arraystretch{0}
\begin{tabular}[c]{c}
\pstree[treesep=0.75cm, levelsep=1.5cm, tpos=0.675]{\TR{$\ast$}}{
    \TR{$a_6$}
    \pstree[treesep=0.25cm]{\TR{$.$} \tlput[tpos=0.875]{\footnotesize 3}}{
        \TR{$a_4$}
        \TR{$b_5$}
    }
    \TR{$a_6$}
}
\end{tabular}

\\ \\

$_1( \; _2( \; a_6 \; )_2 \; _2( \; _3( \; a_4 \; b_5 \; )_3 \; )_2 \; _2( \; a_6 \; )_2 \; )_1$

&

$_1( \; a_6 \; _3( \; a_4 \; b_5 \; )_3 \; a_6 \; )_1$ \\ \\

non-masked union -- see $e_{2\#}$ & masked union -- see $\widetilde e_{2\#}$
\end{tabular}
\end{center}
\caption{Partial syntax modeling with masking of the union operator of the RE $e_2$ above.} \label{fig:masking}
\end{figure}
The tree still shows that the star iterates three times. Furthermore, if also the concatenation operator is masked in the RE, the syntax tree becomes simply $_1( \; a_6 \; a_4 \; b_5 \; a_6 \; )_1$.
\par
To support partial syntax modeling, it suffices to simplify the numbering policy for the RE, whereas the parsing method is unchanged. Several systematic or application-specific masking policies can be devised that are not discussed here. It is important to note that any numbered RE, whatever simplification is applied, remains linear, i.e., each terminal occurs only once, thus allowing the same parser construction method to be applied. Partial modeling may increase parsing speed, since the parser automaton has fewer states and  ambiguities may be reduced or  eliminated.
\section{-- Comparison of SLPF and SPPF} \label{app:compareSLPFandSPPF}
\begin{figure}[ht]
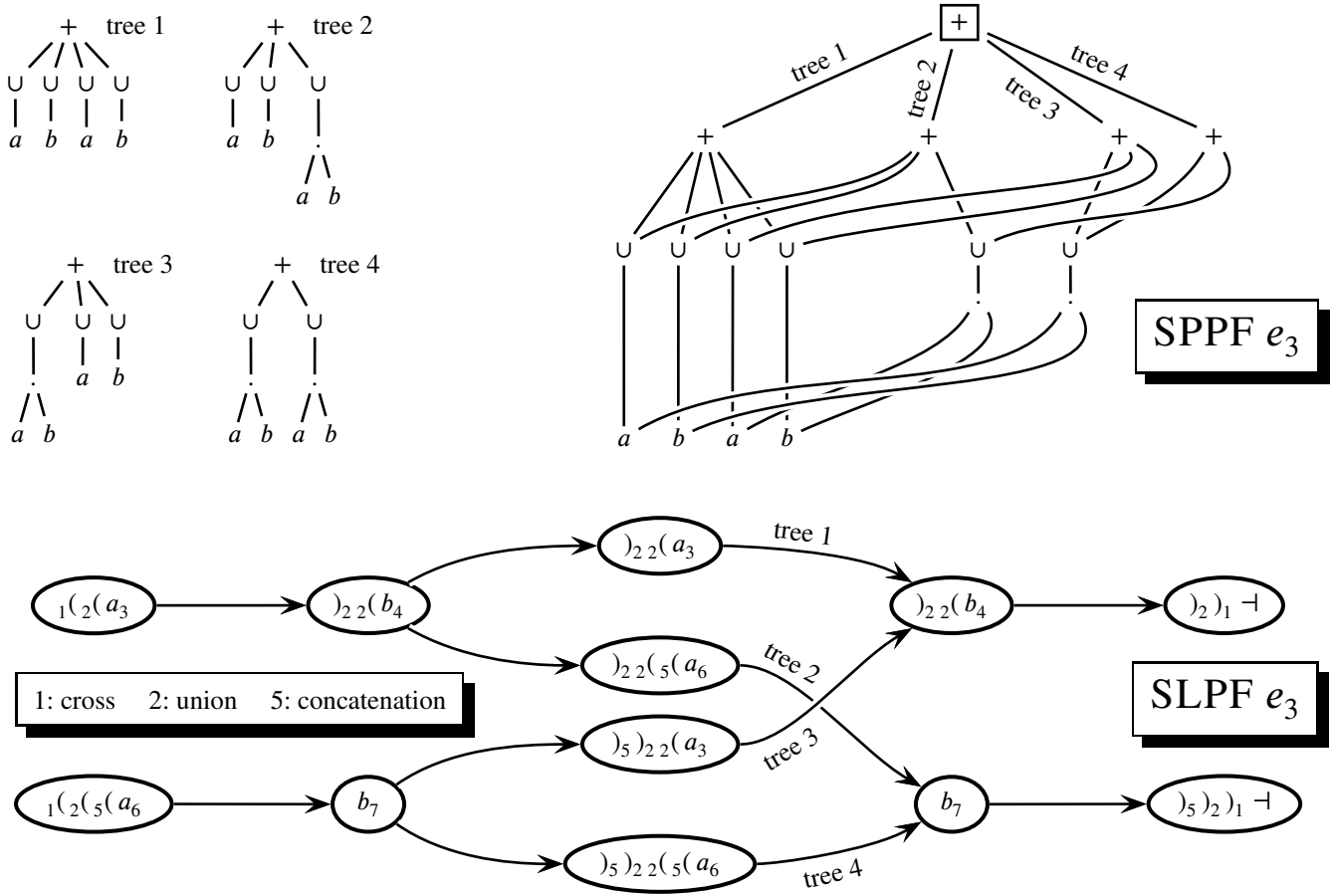

	\begin{center}
		\tabcolsep=0.0cm
		\begin{tabular}{c@{\hspace{2.5cm}}c}
			\tabcolsep=0.0cm
			\begin{tabular}{c@{\hspace{1.25cm}}c}
				\psset{nodesep=3pt, arrows=-, linewidth=1pt, levelsep=0.75cm, treesep=0.25cm, treefit=loose}
				\pstree{\TR[name=path1]{$+$}} {
					\pstree{\TR{$\cup$}} {
						\TR{$a$}
					}
					\pstree{\TR{$\cup$}} {
						\TR{$b$}
					}
					\pstree{\TR{$\cup$}} {
						\TR{$a$}
					}
					\pstree{\TR{$\cup$}} {
						\TR{$b$}
					}
				}
				
				&
				
				\psset{nodesep=3pt, arrows=-, linewidth=1pt, levelsep=0.75cm, treesep=0.25cm, treefit=loose}
				\pstree{\TR[name=path2]{$+$}} {
					\pstree{\TR{$\cup$}} {
						\TR{$a$}
					}
					\pstree{\TR{$\cup$}} {
						\TR{$b$}
					}
					\pstree{\TR{$\cup$}} {
						\pstree{\TR{$.$}} {
							\TR{$a$}
							\TR{$b$}
						}
					}
				}
				
				\\ \\
				
				\psset{nodesep=3pt, arrows=-, linewidth=1pt, levelsep=0.75cm, treesep=0.25cm, treefit=loose}
				\pstree{\TR[name=path3]{$+$}} {
					\pstree{\TR{$\cup$}} {
						\pstree{\TR{$.$}} {
							\TR{$a$}
							\TR{$b$}
						}
					}
					\pstree{\TR{$\cup$}} {
						\TR{$a$}
					}
					\pstree{\TR{$\cup$}} {
						\TR{$b$}
					}
				}
				
				&
				
				\psset{nodesep=3pt, arrows=-, linewidth=1pt, levelsep=0.75cm, treesep=0.25cm, treefit=loose}
				\pstree{\TR[name=path4]{$+$}} {
					\pstree{\TR{$\cup$}} {
						\pstree{\TR{$.$}} {
							\TR{$a$}
							\TR{$b$}
						}
					}
					\pstree{\TR{$\cup$}} {
						\pstree{\TR{$.$}} {
							\TR{$a$}
							\TR{$b$}
						}
					}
				}
			\end{tabular}
			
			\psset{labelsep=0.375cm}
			
			\nput 0 {path1} {tree $1$}
			
			\nput 0 {path2} {tree $2$}
			
			\nput 0 {path3} {tree $3$}
			
			\nput 0 {path4} {tree $4$}
			
			&
			
			\rnode {SPPF} {
				\tabcolsep=0.0cm
				\begin{tabular}{c}
					\psset{nodesep=3pt, arrows=-, linewidth=1pt, levelsep=1.5cm, treesep=1.0cm}
					\pstree{\TR{\psframebox{$+$}}} {
						\pstree[treesep=0.5cm]{\TR[name=K1]{$+$} \nbput[labelsep=3pt, nrot=:D] {tree $1$}} {
							
							\pstree[levelsep=2.5cm]{\TR[name=U1]{$\cup$}} {
								\TR[name=a1]{$a$}
							}
							\pstree[levelsep=2.5cm]{\TR[name=U2]{$\cup$}} {
								\TR[name=b2]{$b$}
							}
							\pstree[levelsep=2.5cm]{\TR[name=U3]{$\cup$}} {
								\TR[name=a3]{$a$}
							}
							\pstree[levelsep=2.5cm]{\TR[name=U4]{$\cup$}} {
								\TR[name=b4]{$b$}
							}
						}
						\pstree{\TR[name=K2]{$+$} \nbput[labelsep=3pt, nrot=:D, npos=0.625] {tree $2$}} {
							
							\phantom{
								\pstree{\TR{$\cup$}} {
									\TR{$.$}
								}
							}
							\pstree[levelsep=0.625cm]{\TR[name=U34]{$\cup$}} {
								\TR[name=C34]{$.$}
							}
						}
						\pstree{\TR[name=K3]{$+$} \nbput[labelsep=3pt, nrot=:U] {tree $3$}} {
							
							\pstree[levelsep=0.625cm]{\TR[name=U12]{$\cup$}} {
								\TR[name=C12]{$.$}
							}
							\phantom{
								\pstree{\TR{$\cup$}} {
									\TR{$.$}
								}
							}
						}
						\pstree{\TR[name=K4]{$+$} \naput[labelsep=3pt, nrot=:U] {tree $4$}} {
							
						}
					}
					\psset{border=0.05cm}
					
					\nccurve[angleA=-135, angleB=30, ncurvA=0.5] {K2} {U1}
					
					\nccurve[angleA=-120, angleB=30, ncurvA=0.5] {K2} {U2}
					
					\nccurve[angleA=-120, angleB=30, ncurvA=0.5] {C34} {a3}
					
					\nccurve[angleA=-60, angleB=25, ncurvA=0.5] {C34} {b4}
					
					\nccurve[angleA=-60, angleB=25, ncurvA=0.25] {K3} {U3}
					
					\nccurve[angleA=-30, angleB=15] {K3} {U4}
					
					\nccurve[angleA=-120, angleB=30, ncurvA=0.5] {C12} {a1}
					
					\nccurve[angleA=-60, angleB=25, ncurvA=0.5] {C12} {b2}
					
					\nccurve[angleA=-120, angleB=30] {K4} {U12}
					
					\nccurve[angleA=-60, angleB=25] {K4} {U34}
			\end{tabular}}

\\ \\ \\

\multicolumn{2}{c}{

\rnode {SLPF} {
\begin{tabular}{c}
\psset{nodesep=0.0cm, arrows=->, linewidth=1.5pt, arrowscale=2, rowsep=0.0cm, colsep=0.875cm, border=0.00cm}
\begin{psmatrix}
& & & & \ovalnode {223ab} {$)_2 \, _2( \, a_3$} & & & & \\
				
\ovalnode {123i} {$_1( \, _2( \, a_3$} & & \ovalnode {224a} {$)_2 \, _2( \, b_4$} & & & & \ovalnode {224aba} {$)_2 \, _2( \, b_4$} & & \ovalnode {21dashvabab} {$)_2 \, )_1 \dashv$} \\
				
& & & & \ovalnode {2256ab} {$)_2 \, _2( \, _5( \, a_6$} & & & & \\[0.25cm]
				
& & & & \ovalnode {5223ab} {$)_5 \, )_2 \, _2( \, a_3$} & & & & \\
				
\ovalnode {1256i} {$_1( \, _2( \,_5( \, a_6$} & & \ovalnode {7a} {$ \ b_7 \ $} & & & & \ovalnode {7aba} {$ \ b_7 \ $} & & \ovalnode {521dashvabab} {$)_5 \, )_2 \, )_1 \dashv$} \\
				
& & & & \ovalnode {52256ab} {$)_5 \, )_2 \, _2( \,_5( \, a_6$} & & & &
				
\psset{linewidth=1pt, border=0.05cm}
				
\ncline {123i} {224a}
				
\ncline {1256i} {7a}
				
\nccurve[angleA=30, angleB=-180] {224a} {223ab}
				
\nccurve[angleA=-30, angleB=180] {224a} {2256ab}
				
\nccurve[angleA=35, angleB=-180] {7a} {5223ab}
				
\nccurve[angleA=-35, angleB=180] {7a} {52256ab}
				
\nccurve[angleA=-0, angleB=150, ncurvA=0.125] {223ab} {224aba} \naput[labelsep=3pt, nrot=:U]{tree $1$}
				
\nccurve[angleA=-0, angleB=150, ncurvA=0.5] {2256ab} {7aba} \naput[labelsep=3pt, nrot=:U, npos=0.25]{tree $2$}
				
\nccurve[angleA=0, angleB=-150, ncurvA=0.5] {5223ab} {224aba} \nbput[labelsep=3pt, nrot=:U, npos=0.25]{tree $3$}
				
\nccurve[angleA=0, angleB=-150, ncurvA=0.25] {52256ab} {7aba} \nbput[labelsep=3pt, nrot=:U]{tree $4$}
				
\ncline {224aba} {21dashvabab}
				
\ncline {7aba} {521dashvabab}
\end{psmatrix}
\end{tabular}}}
\end{tabular}
\nput[labelsep=-1.25cm] {-25} {SPPF} {\shadowbox{\Large \textsc{SPPF $e_3$}}}
\nput[labelsep=-2.25cm] {0} {SLPF} {\shadowbox{\Large \textsc{SLPF $e_3$}}}
\nput[labelsep=-17.0cm] {0} {SLPF} {\shadowbox{\text{$1$: cross \quad $2$: union \quad $5$: concatenation}}}
\end{center}
\caption{(top) The four (graphic) syntax trees  of the string $x = a\,b\,a\,b$ of RE $e_3$ (left) and their SPPF forest~\cite{DBLP:journals/toplas/ScottJ06} (right). The square root node in the SPPF represents different expansions of the Kleene cross. Notice that in the SPPF some subtrees and leaves are shared.  (bottom) The corresponding (linearized) SLPF for comparison. } \label{fig:SPPF}
\end{figure}

In context-free grammar parsing, the set of  syntax trees of a string is represented in a so-called \emph{shared packed parse forest} (SPPF)~\cite{Tomita86,Johnson1991}. Though originally designed for grammars, an SPPF can  directly represent the trees of an RE, without any need of materializing a grammar equivalent to the RE~\cite{DBLP:journals/vlc/BorsottiBCM23}. For comparison, the four trees of string $x = a\,b\,a\,b$ of RE $e_3$ are represented in the SPPF of Fig.~\ref{fig:SPPF}, top. There are four alternative subtrees for expanding the cross operator, and several subtrees (or even leaf nodes) are shared.
Although the SPPF of a non-infinitely ambiguous string is also a directed acyclic graph, SLPF and SPPF are quite different data-structures: the former is based on the parser NFA and encodes the trees in a linearized form, while the latter has a hierarchical structure related to the RE. The SLPF shown in Fig.~\ref{fig:SPPF}, bottom, for the same string $x$ outlines the difference between SPPF and SLPF. The practically relevant difference between the two comes from their memory footprint, measured by the total number of nodes and edges.
\par
In general the SPPF of a string produced by a grammar has a polynomial size with respect to the string length~\cite{DBLP:journals/scp/ScottJ18}, where the polynomial degree depends on the grammar form, and for certain very ambiguous REs the size may even be non-polynomial~\cite{DBLP:journals/vlc/BorsottiBCM23}. On the other hand, an SLPF has a size linear with respect to the length of the input string. In fact, the SLPF consists of a number of columns equal to the string length plus one, each column has a size bounded by the number of NFA states, i.e., segments, which is independent of the string length, and the edges span consecutive columns only. Thus in general an SLPF scales better than an SPPF as the string length grows.
\section{-- Optimization of the SLPF} \label{app:SLPFoptimization}
\paragraph{SLPF encoding}
Basically the SLPF is an array of columns $C_0$, $C_r$, i.e., segment sets, with $C_0, C_r \subseteq Q$ ($Q$ is the set of all segments) for  $1 \leq r \leq n$ ($n$ is the text length). An
SLPF column can be naively represented as a bitstring of length $\ell = \vert \, Q \, \vert$ bits, in one or more memory words.
Unfortunately, for a long text, the SLPF memory footprint may be quite large and penalize execution. A significant memory optimization consists of
 encoding the columns more efficiently, by using shorter bitstrings. To  this end, we exploit the  property that all the segments
in a column are reached from one or more segments in the previous column. Consider an RE $e$, a valid text $x$, the SLPF of $x$ according to $e$, and a generic SLPF column $C_{r - 1}$ with text character $x_r = a$. Assume that $\rho = \mu_\# \, a_\#$ is any segment with an end-letter $a_\#$ that corresponds to character $a$. By construction the next column is $C_r = \underset {\rho \, \in \, C_{r - 1}} \bigcup \, \mathit{FolSeg} \, (e_\# \dashv, \, \rho)$ as of Eq.~\eqref{eq:folsegdef}. For instance, for the RE $e_2$ of Ex.~$2$ with text character $x_1 = a$, since there are two initial segments with an end-letter corresponding to $a$, namely the numbered characters $a_4$ and $a_6$ (see Tab.~\ref{tab:segments}), it holds (see the parser NFA in Fig.~\ref{fig:NFA-RE2}):
\[
C_1 = \mathit{FolSeg} \, \big( e_{2\#} \dashv, \, \text{``\,$_1( \, _2 \, _3( \, a_4$\,''} \big) \; \cup \; \mathit{FolSeg} \, \big( e_{2\#} \dashv, \, \text{``\,$_1( \, _2 \, a_6$\,''} \big) = \big\{ \; \underbracket[0.75pt]{\text{``\,$b_5$\,''}}_\text{follows $a_4$}, \; \underbracket[0.75pt]{\text{``\,$)_2 \, _2( \, _3( \, a_4$\,''}, \; \text{``\,$)_2 \, _2( \, a_6$\,''}, \; \text{``\,$)_2 \, )_1 \dashv$\,''}}_\text{all the three of them follow $a_6$} \; \big\}
\]
Thus an SLPF column can be equivalently represented as a set of pairs $\langle a_\#, \, \sigma \rangle$, where segment $\sigma$ is a follower of segment $\rho$ according to the numbered character $a_\#$, which is the end-letter of $\rho$; see also Eq.~\eqref{eq:folsegdef}. In each set, the numbered characters $a_\#$ of all the pairs correspond to the same input character $a$ when their numbering is canceled. For instance in the four pairs of the set below, which represents column $C_1$ above, both numbered characters $a_4$ and $a_6$ correspond to the input character $a$:
\[
\set{ \; \big\langle \, a_4, \, \text{``\,$b_5$\,''} \, \big\rangle, \; \big\langle \, a_6, \, \text{``\,$)_2 \, _2( \, _3( \, a_4$\,''} \, \big\rangle, \; \big\langle \, a_6, \, \text{``\,$)_2 \, _2( \, a_6$\,''} \, \big\rangle, \; \big\langle \, a_6, \, \text{``\,$)_2 \, )_1 \dashv$\,''} \, \big\rangle \; } \hspace{1.5cm} \text{pair set representing column $C_1$}
\]
Though the number of such sets of pairs may be large, most often their cardinality is small compared to the total segment count. For instance, the set above contains four pairs, while RE $e_2$ has ten segments (see Tab.~\ref{tab:segments}), i.e., over twice  as many. Indeed, in a large RE with many segments, a specific segment is likely to have only a small fraction of segments as  follower.
\par
This low cardinality is the key to save memory for the SLPF. We sort each such set of pairs in some way, e.g., lexicographically, and we number its elements. Each SLPF column can then be represented as a sequence of indices out of a certain set of pairs (as in the final clean SLPF some segments may be deleted). Accordingly the column can be encoded as a bitstring formed by the sequence of such indices. Therefore these bitstrings are likely to be shorter than the total number of segments $\ell$. Our tool uses this kind of encoding, and  we found  that most often one $64$-bit memory word suffices for a bitstring. When the bitstring is too long, the tool uses multiple words, organized in a hash table (as the same column may occur many times in the SLPF). In this case the column contains a one-word pointer to the hash table entry. The SLPF  footprint is  proportional to the text length, with a much smaller coefficient. This allows our tool to keep the whole SLPF in  main memory.
\paragraph{SLPF compression}
 The SLPF can be  encoded in compressed form by representing
it  as a deterministic finite-state automaton (SLPF-DFA), the states of which
coincide with the segment sets stored in the SLPF columns. Given an RE $e$, a valid text $x$ and the SLPF of $x$ according to $e$,
suppose $C_0$, $C_r$, with $1 \leq r \leq n$ (text length) are the columns. Assume that $\delta_\text{SLPF-DFA}$ is the transition function
of the SLPF-DFA, such that $\delta_\text{SLPF-DFA} \, (C_{r-1}, \, x_r) = C_r$, and represent the function as a look-up table.
The SLPF-DFA has at most $2^\ell$ states, since a column is a set of segments ($\ell$ total segment count),
independently of the text length $n$. Then starting from the initial column, the whole SLPF can be reconstructed by running the SLPF-DFA on text $x$.
This technique drastically reduces the memory footprint of the SLPF, of course at the cost of the reconstruction time. Note that the SLPF-DFA is similar to the DFA,
but is specific to text $x$, thus it cannot be obtained from the DFA (nor from the reverse DFA or their intersection). Generating the transition function $\delta_\text{SLPF-DFA}$ requires examining the whole SLPF and is a slow process, but it is done once and for all, and is useful for storing the SLPF on file. Our tool simply offers this compression option for archival purpose.
\par
A further improvement (currently not implemented) consists of reconstructing  only a section of interest of the SLPF starting from an intermediate column, or  the entire SLPF from a selection of evenly spaced columns stored
together with the SLPF-DFA, thus possibly accelerating the reconstruction by executing it in a parallel way.
\end{document}